\documentclass{emulateapj}
\usepackage{longtable}
\usepackage{lscape}

\newcommand{\cote}{C\^{o}t\'{e}\ }
\newcommand{\jordan}{Jord\'{a}n\ }
\newcommand{\etal}{et~al.\ }
\newcommand{\gz}{($g$--$z$)}

\newcommand{\sn}{$S_N$}
\newcommand{\snz}{$S_{N,z}$}
\newcommand{\sm}{$S_\mathcal{M}$}

\newcommand{\gsdss}{$g_{sdss}$}

\newcommand{\sersic}{S\'ersic}

\slugcomment{Accepted for publication in the Astrophysical Journal}

\shorttitle{Formation Efficiencies of Globular Clusters}
\shortauthors{Peng et al.}

\begin{document}


\title{The ACS Virgo Cluster Survey XV. The Formation Efficiencies of
  Globular Clusters in Early-Type Galaxies: The Effects of Mass
  and Environment\altaffilmark{1}}


\author{Eric W. Peng\altaffilmark{2,3,4}}
%\affil{Herzberg Institute of Astrophysics, 
%National Research Council of Canada, 
%5071 West Saanich Road, Victoria, BC  V9E 2E7, Canada} 
%\email{Eric.Peng@nrc-cnrc.gc.ca}

\author{Andr\'{e}s Jord\'{a}n\altaffilmark{5,6,7,8}}
%\affil{European Southern Observatory, Karl-Schwarzschild-Str. 2, 85748
%  Garching bei M\"{u}nchen, Germany}
%\email{ajordan@eso.org}
%
\author{Patrick C\^{o}t\'{e}\altaffilmark{2}}
%\affil{Herzberg Institute of Astrophysics, 
%National Research Council of Canada, 
%5071 West Saanich Road, Victoria, BC  V9E 2E7, Canada} 
%\email{Patrick.Cote@nrc-cnrc.gc.ca}
%
\author{Marianne Takamiya\altaffilmark{9}}
%\affil{Department of Physics \& Astronomy, University
%of Hawai'i, 200 W. Kawili Street, Hilo, HI 96720, USA}
%\email{takamiya@hawaii.edu}
%
\author{Michael J. West\altaffilmark{10,11,9}}
%\affil{Gemini Observatory, Casilla 603, La Serena, Chile}
%\email{mwest@gemini.edu}
%

\author{John P. Blakeslee\altaffilmark{2,12}}
%\affil{Department of Physics and Astronomy, Washington State University,
%Pullman, WA 99164-2814}
%\email{jblakes@wsu.edu}
%
\author{Chin-Wei Chen\altaffilmark{2,13}}

\author{Laura Ferrarese\altaffilmark{2}}
%\affil{Herzberg Institute of Astrophysics,
%National Research Council of Canada,
%5071 West Saanich Road, Victoria, BC  V9E 2E7, Canada}
%\email{Laura.Ferrarese@nrc-cnrc.gc.ca}
%
\author{Simona Mei\altaffilmark{14,15}}
%\affil{Department of Physics and Astronomy,
%  Johns Hopkins University, Baltimore, MD 21218, USA}
%\email{smei@pha.jhu.edu}
%
%\and
%
\author{John L. Tonry\altaffilmark{16}}
%\affil{Institute for Astronomy, University of Hawai'i, 2680 Woodlawn
%  Drive, Honolulu, HI 96822, USA}
%\email{jt@ifa.hawaii.edu}

\author{Andrew A. West\altaffilmark{17}}


\altaffiltext{1}{Based on observations with the NASA/ESA {\it Hubble
    Space Telescope} obtained at the Space Telescope Science Institute,
  which is operated by the Association of Universities for Research in
  Astronomy, Inc., under NASA contract NAS 5-26555.}
\altaffiltext{2}{Herzberg Institute of Astrophysics, 
  National Research Council of Canada, 
  5071 West Saanich Road, Victoria, BC  V9E 2E7, Canada; 
  Eric.Peng@nrc-cnrc.gc.ca}
\altaffiltext{3}{Space Telescope Science Institute, 3700 San Martin
  Drive, Baltimore, MD, 21218, USA}
\altaffiltext{4}{Department of Astronomy, Peking University, Beijing
  100871, China}
\altaffiltext{5}{Harvard-Smithsonian Center for Astrophysics,
60 Garden St., Cambridge, MA 02138}
\altaffiltext{6}{Clay Fellow}
\altaffiltext{7}{European Southern Observatory, Karl-Schwarzschild-Str. 2, 85748
  Garching bei M\"{u}nchen, Germany}
\altaffiltext{8}{Departamento de Astronom\'{\i}a y Astrof\'{\i}sica,
Pontificia Universidad Cat\'olica de Chile, Casilla 306, Santiago 22,
Chile}
\altaffiltext{9}{Department of Physics \& Astronomy, University
  of Hawai'i, 200 W. Kawili Street, Hilo, HI 96720, USA}
%; takamiya@hawaii.edu}
\altaffiltext{10}{European Southern Observatory, Alonso de Cordova 3107,
  Vitacura, Santiago, Chile}
\altaffiltext{11}{Gemini Observatory, Casilla 603, La Serena, Chile}
%;  mwest@gemini.edu}
\altaffiltext{12}{Department of Physics and Astronomy, Washington State 
  University, Pullman, WA 99164-2814}
%; jblakes@wsu.edu}
\altaffiltext{13}{Institute for Astronomy, National Central University
  Taiwan, Chung-Li 32054, Taiwan}
%; Chin-Wei.Chen@nrc-cnrc.gc.ca}
\altaffiltext{14}{University of Paris 7 Denis Diderot,  75205 Paris Cedex
  13, France}
\altaffiltext{15}{GEPI, Observatoire de Paris, Section de Meudon, 5 Place
  J. Janssen, 92195 Meudon Cedex, France}
%; Simona.Mei@obspm.fr}
\altaffiltext{16}{Institute for Astronomy, University of Hawai'i, 2680 Woodlawn
  Drive, Honolulu, HI 96822, USA}
%; jt@ifa.hawaii.edu}
\altaffiltext{17}{Department of Astronomy, University of California, 601
  Campbell Hall, Berkeley, CA 94720}
%; awest@astro.berkeley.edu}


\begin{abstract}

The fraction of stellar mass contained in globular clusters (GCs), also
measured by number as the specific frequency, is a fundamental
quantity that reflects both a galaxy's early star formation and its
entire merging history.  We present specific frequencies, luminosities, 
and mass fractions for the globular cluster systems of 100 early-type 
galaxies in the ACS Virgo Cluster Survey.  This catalog represents the 
largest homogeneous catalog of GC number and mass fractions across 
a wide range of galaxy luminosity ($-22<M_B<-15$). We find that 
1) GC mass fractions can be high in both giants and dwarfs, but 
are universally low in galaxies with intermediate luminosities 
($-20<M_B<-17$).  2) The fraction of red GCs increases with galaxy luminosity, 
but stays constant or decreases for galaxies brighter than $M_z=-22$.  As a 
result, although specific frequencies for blue and red GCs are both higher in 
massive galaxies, the behavior of specific frequency across galaxy mass is 
dominated by the blue GCs. 3) The GC fractions of low-mass galaxies exhibit 
a dependence on environment, where dwarf galaxies closer to the cluster center 
have higher GC  fractions.  Nearly all dwarfs with high GC fractions are within 
1~Mpc of the cD galaxy M87, presenting the first strong evidence that GC
formation in dwarf galaxies is biased toward dense environments.
4) GC formation in central dwarfs is biased because their stars form earliest 
and most intensely.  Comparisons to early-type dwarf galaxies in the Millennium 
Simulation show that central dwarfs are likely to have older
stellar populations and form more of their stars at higher star
formation rates (SFRs) and star formation rate surface densities.
In addition, the SFR surface density in simulated dwarfs peaks 
{\it before} the total SFR, naturally producing GC populations that are 
older and more 
metal-poor than the field stars. 5) Dwarfs within $\sim40$~kpc of the giant 
ellipticals M87 and M49 are red for their luminosities and have few or no GCs, 
suggesting that they have been tidally stripped and have contributed their GCs 
to the halos of their giant neighbors. The central dwarfs with high GC mass
fractions are thus likely to be the survivors most similar to the
protogalaxies that assembled the rich M87 globular cluster system.

\end{abstract}


%% subject header.  Objects should be in the appropriate "individual"
%% headers (e.g. quasars: individual, stars: individual, etc.) with the
%% additional provision that the total number of headers, including each
%% individual object, not exceed six.  The \objectname{} macro, and its
%% alias \object{}, is used to mark each object.  The macro takes the object
%% name as its primary argument.  This name will appear in the paper
%% and serve as the link's anchor in the electronic edition if the name
%% is recognized by the data centers.  The macro also takes an optional
%% argument in parentheses in cases where the data center identification
%% differs from what is to be printed in the paper.


\section{Introduction}

Globular clusters (GCs) constitute a small fraction of the stellar mass
in galaxies, but their ubiquity, relative simplicity, and old ages
make them the most prominent representatives of a bygone epoch of galaxy
formation.  GCs are made of stars that are among the oldest in
galaxies, and they can be observed at large distances (e.g., Blakeslee
\etal 2003a).  These old star clusters are thus unique, both intrinsically and
observationally, for understanding the early, intense star-forming
episodes that mark galaxy formation. 

In the local universe, we see massive star clusters forming wherever
there are high star formation rate surface densities (Larsen \& Richtler
2000), providing a
connection that suggests the properties of star cluster populations---age, 
metallicity, mass---should scale quite closely with field stars
formed in the same events.  The properties of globular
cluster systems do in fact correlate strongly with the properties of
the field stars of their host galaxies.  The mean
metallicities of GC systems have 
long been known to scale with the metallicity of their host (van den
Bergh 1975; Brodie \& Huchra 1991), and the mean metallicities of both
the metal-rich and metal-poor subpopulations also correlate with the
luminosity and mass of the host galaxy (Larsen \etal 2001; Peng \etal 2006a and
references therein).  However, if GC systems directly followed the
underlying field light in 
every way, they might be less interesting.  For
instance, although the metallicities of GC systems may track
those of galaxies, they are consistently offset to lower values
by $\sim0.5$--0.8~dex in [Fe/H] (e.g., \jordan \etal 2004a, Lotz \etal
2004).  Most conspicuously, even the most massive and metal-rich
galaxies have GC systems dominated by metal-poor star clusters
([Fe/H]$\lesssim -1$).  This suggests a disconnect between
the formation of ``halo'' stellar populations and the bulk of the
galaxy.  

One of the most studied aspects of this GC--galaxy
duality concerns the specific frequency of globular clusters, or
the number of GCs per unit stellar luminosity.  Specific frequency,
\sn, was introduced by Harris \& van den Bergh (1981) and is defined
as the number of GCs normalized to a galaxy luminosity of $M_V=-15$.
The purpose of studying $S_N$ across galaxies of different masses,
morphologies, and environments is, in the words of that initial paper, 
``to investigate whether there is in fact a `universal'
and uniform capability for globular cluster formation''.  
This simple quantity, and similar
ones related to it, turn out to be extremely interesting galaxy
diagnostics.  It appears true 
that for galaxies above a certain mass there is 
a nearly universal capability to form globular clusters, but this process
is not uniform across all galaxies, at least as seen in comparison to
the field stars.  Specific frequencies of spiral galaxies like the
Milky Way are generally 0.5--1 (Goudfrooij \etal 2003; Rhode \&
Zepf 2004; Chandar, Whitmore \& Lee 2004), although they have a mean of 
$\sim4$ when normalized only to bulge luminosity (\cote \etal 2000).  
The specific frequencies of massive
ellipticals are 2--6, and those of some cD galaxies such as M87 (N4486)
can be well in excess of 10.  These trends are apparent even when the
number of GCs is normalized to stellar mass as opposed to stellar
luminosity (Rhode, Zepf \& Santos 2005).  Dwarf elliptical galaxies
(dEs), whose GC systems are predominantly metal-poor, can also
have high $S_N$ similar to those of giant ellipticals (Durrell \etal 1996;
Miller \etal 1998; Lotz \etal 2004; Miller \& Lotz 2007; Puzia \&
Sharina 2007), 
as can some dwarf irregulars
(Seth \etal 2004), suggesting the possibility that
the halos of large galaxies
were formed mainly through the accretion of dwarf-like objects (Searle
\& Zinn 1978; C\^{o}t\'{e}, Marzke \& West 1998; \cote \etal 2000).

A central question in the study of GC systems is: How do we 
understand different GC fractions in the context of galaxy assembly?
The formation of globular cluster systems
has been particularly tied to the formation of massive elliptical
galaxies, in which GCs are often present in large numbers and where
GCs are most easily observed.  The mergers and accretion events
expected during the hierarchical assembly of these galaxies
must also be able to form their GC systems.
Observations and simulations of
elliptical galaxy formation are creating a picture
in which the stars form early and quickly (e.g., Kodama \etal 1998),
mimicking the traditional ``monolithic collapse'' scenario, 
but where the assembly of these stars into a single galaxy
continues until late times through largely dissipationless mergers
(De~Lucia \etal 2006; De~Lucia \& Blaizot 2007), and star formation at
late times is suppressed by energy feedback (Springel \etal 2005;
Croton \etal 2006).

Intertwined with the issue of galaxy formation is that of the
formation efficiency of the GCs themselves: Why do globular clusters
form with different efficiencies with respect to their light in different 
galaxies?  Blakeslee, Tonry \& Metzger (1997) and Blakeslee (1999)
studied the GC systems 
of brightest cluster galaxies (BCGs) in Abell galaxy clusters and found
that the number of GCs scaled with the velocity dispersion of the
galaxy cluster rather than with the luminosity of the BCGs, suggesting
that GC formation is closely linked to the
total mass of the system, i.e., $S_N\propto \mathcal{M}/L$.  In a similar vein,
McLaughlin (1999a) examined the high \sn\ in M87 and found
that the large number of GCs it possessed was not anomalous when
normalized to the total baryonic mass (including the hot X-ray gas)
rather than just to the stellar mass.  McLaughlin (1999a) defined a
``universal'' GC formation efficiency of $\hat{\epsilon}=0.26\%$, where
$\hat{\epsilon}$ is the fraction of the baryonic mass that ends up in
globular clusters.  Kravtsov
\& Gnedin (2005) studied the formation of the GC system in a high
resolution hydrodynamic simulation and also found that the mass in GCs
was directly proportional to the total halo mass of the
galaxy.  If total mass drives GC formation, then it is the variation in
converting baryons into field stars that drives trends in specific frequency.  

The connection between GCs and galactic mass (baryonic or total)
makes it tempting to try to explain them using simulations of dark
matter and galaxy assembly.  Beasley \etal (2002) simulated the color
distributions of GCs using semi-analytic models of galaxy formation,
while the aforementioned simulations of Kravtsov \& Gnedin (2005)
formed a Milky Way mass galaxy in detail.  Moore \etal (2006)
identified metal-poor GCs with early 
collapsing dark matter peaks in cosmological N-body simulations.
An empirical connection between the total mass of a system and the
mass contained in its globular clusters does not, however, explain why
the star formation histories of the GCs and the field should be different.

One clue is that in the local Universe, high mass fractions in massive
star clusters occur in regions of high star formation surface density
(Larsen \& Richtler 2000).  Thus,
one explanation for high-\sn\ galaxies might be that they formed more of
their stars in high efficiency events.  Given that the star
formation rate (SFR) in galaxies was most intense at very early times
($z\gtrsim2$), variations in \sn\ could result from different times of
formation, especially when coupled with a sharp star formation cutoff
at reionization (Santos 2003).  In this scenario, with \sn\ a function
of formation time, GC formation is biased towards the earliest
collapsing halos which can create a large fraction of their stars at
high efficiency before reionization.  Low mass halos in dense
environments collapse 
earlier (Gao, Springel \& White 2005; Diemand, Madau \& Moore 2005), and these
are also the ones most susceptible to heating through photoionization
and feedback.  One expectation of this scenario is that \sn\ will be
``biased'' towards dense environments (West 1993), especially in low mass
galaxies.  The simulations of Moore \etal (2006) suggested that 
GCs and satellites would be more highly clustered in dense regions for 
higher redshifts of reionization, and the same idea can be applied to
\sn\ at fixed reionization redshift.  

Previous observational studies have shown some evidence of an
environmental dependence for \sn.  West (1993) showed that the mean
\sn\ of elliptical galaxies correlates with the local galaxy density.
Blakeslee (1997, 1999) found that \sn\ in brightest cluster galaxies
(BCGs) scaled with properties that reflect cluster
density (cluster velocity dispersion, X-ray temperature, and X-ray
luminosity), and that cluster galaxies closer to their cluster's X-ray
center have higher \sn. West \etal (1995) also found a correlation
between the \sn\ of BCGs and the X-ray luminosity of the cluster,
interpreting it as evidence for a population of intergalactic globular
clusters.  On the other hand, Lotz \etal (2004) and Miller \& Lotz
(2007) did not find an obvious correlation between \sn\ and
clustercentric radius for dEs in the Virgo and Fornax clusters,
although the number of GCs in their galaxies were small and \sn\
errors were quite large.  Also, Spitler \etal (2007) recently found
that the relatively isolated elliptical NGC~821 has an \sn\ comparable
to cluster galaxies of like luminosities.

The specific frequencies (or formation efficiencies) of globular
cluster systems is clearly a fundamental property of galaxies.
However, accurately measuring \sn\ is traditionally fraught with
uncertainty despite the fact that at its most basic level it requires
only simple counting of GCs.  In practice, observations need to be
deep enough to observe past the mean of the GC luminosity function,
and GC selection needs to be efficient enough so that contaminants do
not overwhelm the GCs in lower mass galaxies.  The total magnitude
and distance of the galaxy also needs to be known to establish its
luminosity (or mass).  Studies of larger galaxies also benefit from wide-field
coverage so that enough of the GC system is sampled to minimize
extrapolation errors.  It is because of these limitations that homogeneous
surveys of \sn\ across a wide range of galaxy mass are difficult to conduct.
To put together a complete picture of globular cluster system
and galaxy formation, it is important to study galaxies at all masses
in the same way---at low masses we study the survivors of hierarchical
assembly, and at high masses we study its final products. 

With this in mind, we have undertaken a careful study of the formation
efficiencies for the GC systems of 100 early-type galaxies in the ACS
Virgo Cluster Survey (ACSVCS; \cote \etal 2004).  The 
deep, high-resolution, relatively wide-field imaging provides the most
complete census to date of GCs in early-type galaxies over a large
range in galaxy luminosity ($-22 < M_B < -15$).  In this paper, we
present the specific frequencies and other related quantities of the
ACSVCS galaxies and quantify their trends as a function of host galaxy
properties.  

\section{Observations and Data}
The ACS Virgo Cluster Survey (ACSVCS) is a large program to image 100
early-type galaxies in the Virgo Cluster with the HST Advanced Camera
for Surveys (ACS; Ford \etal 1998).  The survey is described in detail
in Paper I (\cote \etal 2004), and the data reduction techniques are
outlined in \jordan \etal 2004b (Paper II), but we briefly summarize
them here.  We obtained images in the F475W
($g$) and F850LP ($z$) filters of galaxies selected to be early-type
and with confirmed cluster membership in the Virgo Cluster Catalog
(VCC) of Binggeli, Sandage, \& Tammann (1985).  Our sample excludes
the Southern Extension, has $B_T<16$, and the galaxies are morphologically
classified as E, S0, dE, dE,N, dS0, or dS0,N.  This early-type galaxy
sample is magnitude limited for the brightest 26 galaxies ($B_T<12.15$
or $M_B< -19.10$), and contains a representative sample of galaxies
for $M_B<-15.11$.  The color distributions of the GC systems were
presented in Peng \etal (2006a; Paper IX), and their size distributions
in \jordan \etal (2005; Paper X).  We have measured distances 
for 84 galaxies from the method of surface brightness fluctuations
(Mei \etal 2005a, 2005b, 2007; Papers IV, V, and XIII).  For galaxies
where an SBF distance could not be measured, we adopt a
distance to the Virgo Cluster of $D=16.5~{\rm Mpc}$ with a distance
modulus of $31.09\pm0.03$~mag from Tonry \etal (2001), corrected by
the final results of the Key Project distances (Freedman \etal 2001;
see also discussion in Mei \etal 2005b).
The surface brightness profiles, total magnitudes, and colors of the
sample galaxies in $g$ and $z$ were described in Ferrarese \etal
(2006a; Paper VI), and the properties of the GC luminosity functions for 89
of our sample galaxies were presented in \jordan \etal (2007a,
2007b, Paper XII).  The galaxy nuclear properties were
presented in \cote \etal (2006; Paper VIII) and their connection to
supermassive black holes in Ferrarese \etal (2006b).  
These data have also been analyzed for 
ultra-compact dwarf galaxies (Ha\c{s}egan \etal 2005; Paper VII),
diffuse star clusters (Peng \etal 2006b; Paper XI), the connection
between GCs and low mass X-ray binaries (\jordan \etal 2004c, Sivakoff
\etal 2007), and color-magnitude relations in GC systems (Mieske \etal
2006; Paper XIV).  Together, these data create the best 
opportunity to date to study the formation efficiencies of GC systems
in early-type galaxies, and this paper will refer often to the
quantities measured in the preceding papers.

\subsection{Data Reduction and Control Fields}

Each galaxy was imaged with the Wide Field Channel (WFC) of the ACS.
We reduced the ACS/WFC images using a dedicated pipeline described in
\jordan \etal (2004b, Paper~II; see also Blakeslee \etal
2003b).  We produced the science 
images by combining and cleaning them of cosmic rays using the Pyraf
routine {\it multidrizzle} (Koekemoer \etal 2002).  We then subtracted
a model of the galaxy light and used the source detection program
SExtractor (Bertin \& Arnouts 1996) to detect and mask sources and
remove residual background.  Our final object detection
includes estimates of both the image noise and noise due to 
surface brightness fluctuations---ignoring the latter results in many
false detections in the bright central regions of the galaxy---and
objects are only included in the final catalog if they are detected in
both filters.  After rejecting very bright or elongated objects to eliminate
obvious foreground stars and background galaxies, and passing our
catalog through a generous color cut, we use the program KINGPHOT
(\jordan \etal 2005) to measure magnitudes and King model parameters.
KINGPHOT fits King (1966) model surface brightness profiles convolved
with the filter- and spatially-dependent point spread function
(PSF).  Magnitudes and colors are corrected for foreground extinction
using the reddening maps of Schlegel, Finkbeiner, \& Davis (1998) and
extinction ratios for a G2 star (Paper II; Sirianni \etal 2005).  For
the purposes of this paper, whenever we refer to $g$ and $z$, we mean
the HST/ACS magnitudes $g_{475}$ and $z_{850}$.  Magnitudes in the
Sloan Digital Sky Survey (SDSS) system will be designated explicitly,
e.g., as in \gsdss.  The low redshifts of the galaxies being studied 
($-575$ to $2284\ {\rm km\ s}^{-1}$) means that $K$-corrections are negligible
and thus we do not $K$-correct any of the magnitudes presented in this
paper.

In any study of extragalactic star clusters, it is important to
quantify and correct for background contamination.  To that end, we
have also reduced 16 blank, high-latitude control fields taken from
the ACS Pure Parallel Program (GO-9488, PI: Ratnatunga; and GO-9575,
PI: Sparks).  For each galaxy, we
have ``customized'' the control sample to mimic the spatially varying
detection efficiency that is a function of the surface brightness of
the unresolved galaxy light.  Details of this procedure are given in
Peng \etal (2006a,b; Papers IX and XI).

We select probable globular clusters using their sizes and
magnitudes.  We can assign a probability that any object is a GC based
on  position in the $r_h$--$z$ diagram and the locus of contaminants
in the same plane.  Objects with probabilities greater than 0.5 are
included in our GC sample, although the exact choice of the cutoff
value does not significantly affect our results.  We describe the
details of this selection in Peng \etal (2006a, Paper IX) and \jordan
\etal (2008, in preparation).

\subsection{WFPC2 Parallel Fields}

In addition to ACS/WFC imaging of our target galaxies, we have also
acquired parallel imaging of 100 ``blank'' fields using 
the Wide Field Planetary Camera 2 (WFPC2) in
the F606W (wide $V$) and F814W ($I$) filters.  The separation between
ACS and WFPC2 in the HST focal plane is approximately $5\farcm8$,
which is 29~kpc at the mean distance to the Virgo cluster (16.5~Mpc).
For most of the
ACSVCS galaxies, this distance puts the WFPC2 field well outside of
the target galaxy, and thus one motivation for taking them is to
search for intergalactic GCs (e.g., West \etal 1995, \jordan \etal
2003, Williams \etal 2007).  In this paper, we use the WFPC2 parallel
images instead
to constrain the total number of GCs in the larger, more luminous
galaxies, where the GC system is still detectable 29~kpc from the
center.  The WFPC2 images were reduced using a modified form of the
PyRAF pipeline written by  Alasdair 
Allan\footnote{http://www.astro.ex.ac.uk/people/aa/pages/computing/pyraf\_pipeline.html}.
We implemented the cosmic ray cleaning algorithm LACOSMIC
(van Dokkum 2001) and a geometric distortion correction 
(Bagget \etal 2002) in the PyRAF pipeline. Sources were
selected from a SExtractor catalog having magnitudes and colors of
GCs. The sizes and magnitudes of GC candidates were also measured with
KINGPHOT. To account for sample contamination, 
we selected 10 blank WFPC2 fields from the HST archive of similar
depth and observed with the same filters as our WFPC2 data. We
analysed them in the same manner as our data, using WFPC2 point spread
functions provided by P.\ B. Stetson.  We find that there are
on average $5.8\pm2.5$ contaminant sources per WFPC2 field.

\section{Calculating Specific Frequencies, Luminosities, and Masses}

\subsection{Total Numbers of GCs}
\label{sec:profs}
The determination of specific frequency requires knowledge of the
total numbers of GCs, the total apparent magnitude of the galaxy
(traditionally in the $V$ band), and the distance.  For the distances, we
use the SBF-determined distances presented in Mei \etal (2007;
Paper~XIII), using their polynomial calibration.
The measurement of the total number of GCs is in principle a simple
task of counting, but 
is not quite so straightforward in practice.  Our images are deep
enough that we generally detect the brightest $\sim90\%$ of the GC
luminosity function (GCLF), limiting the uncertainties from
extrapolations in luminosity.  Nevertheless, a full count of
the GCs within the ACS 
field of view involves knowing our level of completeness, which is a
function of the GC magnitude, size, and the local background flux.  We
also need to know the form of the GCLF so that we can estimate the
number of GCs missed.  

Fortunately, both the level of completeness and the GCLF
can be determined for our data. 
We empirically derived the completeness of our data as a function of
magnitude, background flux, and GC size, using simulations of nearly 5
million GCs placed in actual ACSVCS images for galaxies of different
surface brightnesses.  Using these data, we can nonparametrically estimate the
expected detection probability of any GC in the survey.  In addition,
\jordan \etal (2007; Paper XII) was able to determine the form of the
GCLF for 89 galaxies in the survey.  For galaxies where we were unable
to measure the form of the GCLF we assume the mean to be the same as
that of the cD galaxy VCC~1316, and a Gaussian distribution with 
sigma derived from Equation~2 in
\jordan \etal (2006).  In this paper, we will use the Gaussian
parameterization of the GCLF.

Most of our galaxies are small enough that the ACS/WFC field of
view is sufficient to 
encompass all of the galaxy's GC system.  In these cases, we
determine the total number of GCs by counting GC candidates within our
imaged field, correcting for incompleteness using the known mean and
sigma of the GCLF.  Calculating the completeness correction, however,
is not straightforward because the surface brightness of the galaxy
varies spatially.  If, for instance, there were strong radial gradients in
the GCLF or size distribution, we could be over- or under-correcting for
unseen GCs.  Our previous work on these galaxies (\jordan \etal 2005,
2007) shows that this is not likely to be the case.
We use bright GCs ($z<22.5$)---those that are bright enough as to be
complete over the entire range of galaxy surface brightness and GC
sizes---to sample the ``true'' distribution of background surface
brightnesses and GC sizes.  We then make the assumption that the
fainter GCs sample the same distributions to derive a mean
completeness correction for the entire population.  

% fig:radial1316
%%%%%%%%%%%%%%%%%%%%%%%%%%%%%%%%%%%%%%%%%%%%%%%%%%%%%%%%%%%%%%%%%%%%%%%%%
\begin{figure}
\epsscale{1.22}
% gcradialplot1316.ps
\plotone{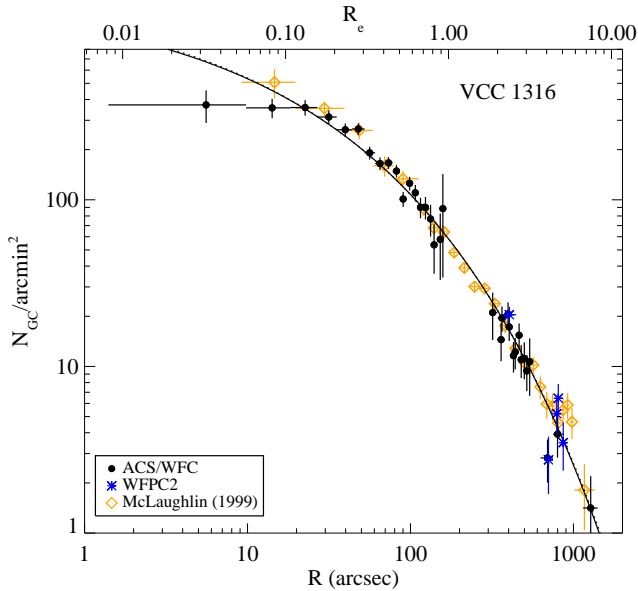}
\caption{Surface density of globular clusters in VCC~1316 (M87/N4486)
  as a function of projected galactocentric radius.  We use data from
  three sources to fit the spatial density profile: ACS pointings of
  the central field and companion fields (black dots), WFPC2 parallel
  pointings in adjacent blank fields (blue asterisks), and
  ground-based data from McLaughlin (1999a) (orange diamonds).  All
  data are in good agreement and are well fit by a S\'ersic profile (black
  line).  We integrate the S\'ersic profile to obtain the total number
  of GCs.
  \label{fig:radial1316}}
\end{figure}
%%%%%%%%%%%%%%%%%%%%%%%%%%%%%%%%%%%%%%%%%%%%%%%%%%%%%%%%%%%%%%%%%%%%%%%%%

In the more luminous galaxies, the ACS/WFC field of view is
insufficient to image the entire GC system.  To correct for this, 
we determine the radial spatial density profile of the GC system, fit a
S\'ersic profile, and integrate over all radii to estimate the total
number of GCs.  For the largest galaxies, we supplemented the density
profiles at larger radii with
data from three sources: 1) ACSVCS imaging of companion galaxies whose 
GC systems are dominated by the halo population of the
neighboring giant, 2) Our WFPC2 parallel imaging of nearby halo
fields (Takamiya et~al., in prep), and 3) Ground-based data from the
literature. In cases where there is no wide-field ground-based data, our
additional pencil-beam HST/WFC and WFPC2 observations still do a
comparably good job because the spatial resolution of HST minimizes
the noise from background contamination.
We use the same method described above for
completeness correction, except we do so in bins of galactocentric 
radius instead of for the whole field of view.  

Errors for $N_{GC}$
were determined using a Monte Carlo technique where we introduced the
appropriate random Poisson noise to our density profiles, then
fit and integrated the S\'ersic (1968) profile.  After doing this 1000
times, we used the distribution
of estimated $N_{GC}$ to determine the intervals that contained the
closest 68\% of the measurements to the mean.  
In Figure~\ref{fig:radial1316}, we show the combined radial profile
data for VCC~1316 (M87/N4486, hereafter M87), which includes the
central ACS pointing, four ACS pointings targeted at nearby
companions, WFPC2 pointings, and the ground-based data of McLaughlin
(1999a).  All are in good agreement and are well fit by a S\'ersic profile,
except perhaps in the very central regions which are difficult to measure
(especially from the ground) and do not contribute large numbers of GCs.
The spatial density profiles of the GC systems of
ACSVCS galaxies will be presented in Peng \etal (in prep), which
will contain a more detailed description of the techniques.
Below, we outline our use of supplementary data in the larger
galaxies.

\noindent {\bf VCC 1226} (M49/N4472, hereafter M49).  
For the most luminous galaxy in
the Virgo cluster, we used the GC counts in the fields of the
neighboring compact galaxies VCC 1192 and 1199, excluding a region of
$4~R_e$ around the galaxies themselves, where $R_e$ is the effective
radius.  We also used four WFPC2
parallel fields in the halo, and the ground-based density profiles of
McLaughlin (1999a) and Rhode \& Zepf (2001).  All these data are in
excellent agreement.

\noindent {\bf VCC 1316} (M87/N4486).  Four ACSVCS galaxies are
near enough to this cluster cD that the GCs observed in their images
are dominated by those of the giant: VCC~1327 (NGC~4486A), VCC~1297
(NGC~4486B), VCC~1279 and VCC~1250. 
VCC~1250 also appears to have ongoing star formation.
As described above for M49, we masked a region of $4~R_e$ around these
nearby neighbors and used the remaining GCs to constrain the density
profile of the larger galaxy.
In addition, we used five WFPC2 parallel fields, and the ground-based density
profiles from McLaughlin (1999a) (see Figure~\ref{fig:radial1316}).

\noindent {\bf VCC 881} (N4406).  We supplemented the profile with two
WFPC2 fields and the ground-based density profile of Rhode \& Zepf (2004).

\noindent {\bf VCC 763} (N4374).  We supplemented the profile with two
WFPC2 fields and the ground-based density profile of G\'omez \& Richtler
(2004).

\noindent {\bf VCC~798, 731, 1535, 1903, 1632,
1231, 2095, 1154, 1062, 1030, and 1664}. For these galaxies, 
we extended the measured
density profile using between 1 and 3 WFPC2 fields per galaxy, which
were at galactocentric radii between 5\arcmin and 17\arcmin.

When we compare the total number of GCs derived using the integrated
GC radial density profiles against the total number counted within the ACS/WFC
field of view, the numbers converge for galaxies with $M_B>-18$.
Because integrated radial density profiles are more uncertain and
not necessary in low
luminosity galaxies, we have adopted $M_B=-18$ as a cutoff; for
galaxies brighter than this, we use the S\'ersic-integrated GC number
counts, and for galaxies fainter than this, we use the corrected
number of GCs directly counted within the ACS/WFC image.

VCC~1938 has a close companion dE, VCC~1941, whose GC system
complicates an accurate count of the larger galaxy's GC system.  We
mask a $R=70\arcsec$ region around VCC~1941, and count only GCs in the
ACS/WFC field of view, so our count of the VCC~1938 GC system is
likely to be a lower limit.

All GC counts are corrected for foreground and background
contamination using control fields (except in the eight cases
described below).  In many other studies, the
background is taken from an annulus around the target galaxy.  The
advantages of using separate control fields are: 1) We can sample a much larger
area of sky to determine the mean surface density of contaminants, thus
greatly reducing the Poisson errors introduced in contaminant
subtraction, 2) We can use the full field of view for the target
galaxy to measure as much of the GC system as we can, 3) We can still
study systems where the GC system fills the entire field of view, 4)
We can naturally incorporate spatially varying completeness as a
function of position within the galaxy.  The only potential
disadvantage is the addition of cosmic variance into the error.  

The advantage in using multiple control fields over a local background
approach is most clear for the dwarf galaxies that have few GCs, where
we need to minimize the error in the mean expected background.
We determine the error in the background by measuring the number of
GC-like objects selected in each of the 16 custom control fields, and
taking the standard deviation of these counts.  This number combines
both Poisson errors and cosmic variance, and takes into account the varying
selection and completeness from galaxy-to-galaxy.
We find that the contribution of cosmic variance to the errors is on
the order of the Poisson noise or less, and that the use of control fields is
superior to the use of a local background.

For eight galaxies---VCC~1327, 1297, 1279, 1250, 1185, 1192, 1199,
1178---control fields are insufficient, and we measure the total
number of GCs by counting 
candidates in a $R=70\arcsec$ aperture and subtracting a local
background.  The first five are
nearby neighbors to the giant elliptical M87 and so some (and
sometimes all) GCs detected in their vicinity belong to the halos of
the giants.  The next three are neighbors of M49.  The size of this
aperture was chosen to fit the ACS/WFC field of view, and extends 11,
30, 4, 4, 10, 20, 11, and 5~$R_e$ in radius for these galaxies, 
respectively.  These numbers are, in principle, lower limits to the
total numbers of GCs in each system.

In total, we present a homogeneous catalog of GC number counts for 100
ACSVCS galaxies.
We have taken great care in this paper to produce our best estimate
for the total number of GCs in the ACSVCS galaxies.  Peng \etal
(2006a) listed numbers of GCs used in their analysis of color
distributions, but {\it we emphasize that those numbers were not
  corrected for field of view or completeness} and are best used
only to evaluate the signal-to-noise ratio of the color distributions.  
Forbes (2005) and Miller \& Lotz (2007) both derive \sn\ values from
Peng \etal (2006a) (the Forbes (2005) paper uses them exclusively), and so
the appropriate warnings apply.

\subsection{Total Luminosities and Masses of GCs}
\label{sec:slm}

Another quantity of interest is the total luminosity (or stellar mass)
in GCs.  We take a straightforward empirical approach to measuring the
total luminosity.  For each galaxy, we add up the $z$-band luminosity
in observed GCs down to 1~mag fainter than the mean of the GCLF, with
adjustments for contaminants and completeness.  For galaxies with $M_B
< -18$, we apply an aperture correction to the total luminosity in
GCs, which is the ratio of the total number of GCs determined from the
integrated radial profile to the total number
observed in the ACS/WFC image.  At this depth, we are
directly counting 84\% of the GCs in a Gaussian GCLF, but are sampling
99\% of the luminosity in GCs.  Thus, completeness corrections are
small, and it has been advocated (e.g. Harris 1991) that the
luminosity in GCs is a more robust quantity than the number of GCs.
To obtain the total mass in GCs, we assume that each GC is a 13~Gyr
simple stellar population (SSP) (Chaboyer \etal 1996) and use its \gz\ color in
conjunction with the Bruzual \& Charlot (2003; BC03) models with a
Chabrier (2003) initial mass function (IMF) to obtain
a mass-to-light ratio in the $z$ bandpass.  The range of
$\mathcal{M}/L_z$ variation with [Fe/H] for globular clusters is no
more than $\pm20\%$.

\subsection{Galaxy Magnitudes: \\ A Consistent Catalog of $M_V$ And $M_z$}

Globular cluster formation efficiencies can be measured against the
total luminosity or mass of a galaxy.  Specific frequency is typically
calculated using the absolute $V$ magnitude.  The Virgo Cluster
Catalog lists $M_B$ for all of our sample galaxies, but these
magnitudes were not derived from CCD photometry.  We do not have 
$V$ imaging for our galaxies, so we predict $V$ using
optical colors measured from SDSS imaging.  We use
$(ugriz)_{sdss}$ photometry measured directly from images in the Sloan
Digital Sky Survey Data Release 5 (Adelman-McCarthy \etal 2007).  
In many cases the large
sizes of the galaxies compared to the size of the SDSS CCDs required
careful stitching of neighboring runs and camera columns to produce
flat images with matching sky (West \etal 2007).  
Using these specially prepared images, we measured total
magnitudes in the five SDSS bands for all the ACSVCS galaxies using a
growth curve analysis.  Details of this analysis and the full
catalog of magnitudes and colors will appear in Chen et~al., in prep.

For the purposes of this paper, we only use the SDSS photometry for
two objectives, deriving a $V$ magnitude, and
supplementing the total $z$ magnitudes given in Paper VI derived from
ACS imaging.  For determining each galaxy's $V$ magnitude, we fit
SSPs from BC03 to the four optical colors---$(u-g)_{sdss}$, $(g-r)_{sdss}$,
$(r-i)_{sdss}$, $(i-z)_{sdss}$---and use the $g_{sdss}-V$ color of the
best fit model.  Across the entire sample, the $g_{sdss}-V$ color
ranges from 0.18 to 0.48 with a mean of 0.37.  In practice, this
approach is extremely robust because 
$g_{sdss}-V$ is almost entirely a function of $(g-r)_{sdss}$.  We
subtract this color from the measured \gsdss\ magnitude to obtain $V$.
The mean relationship between $V$ and $(g-r)_{sdss}$ matches the
empirical relation of Blanton \& Roweis (2007), except that our magnitudes 
are fainter in the mean by $0.03$~mag.

Because redder wavelengths are a better tracer of stellar mass than
the traditionally used $B$ and $V$, we will often use the total integrated
$z$ luminosities of galaxies in this paper.  For most of these
galaxies, we use the $z$ magnitudes measured from the ACS images as
presented in Paper~VI.  However, for the brightest 10 galaxies, the
ACS field of view was substantially smaller than the extent of the
galaxy, and we prefer to use the wide-field SDSS photometry (after
introducing a small ($<0.04$~mag) color-dependent correction between
the SDSS to the ACS photometric systems).  We also use the SDSS
photometry for VCC~1030, 575, and 1512, which have suspect integrated
magnitudes from their ACS surface brightness
profiles (see notes on these galaxies in Paper~VI).
Otherwise, the two independent measures of the total 
luminosity are in good agreement.  We choose to use the ACS magnitudes
for the remaining 87 galaxies because they have higher signal-to-noise
than the SDSS photometry, which were taken from the ground with
much higher sky backgrounds and shorter exposures.

\subsection{Stellar Mass}
Stellar mass is a better basis for comparison when studying
galaxies of different morphological types and star formation
histories.  The $V$ band light in star-forming galaxies is affected
significantly by young stars and is not the best tracer of the total mass.
Using the \gz\ colors from Paper VI, $J-K_s$ colors from the Two
Micron All Sky Survey (2MASS; Skrutskie \etal 2006), and the model
SSPs from BC03 with a Chabrier (2003) initial mass function, we obtain mean 
mass-to-light ratios in the $z$ bandpass ($\mathcal{M}/L_z$) from which we can
derive a total stellar masses.  
In the near-infrared, we measure the $J$ and $K_s$ galaxy magnitudes
in images taken by 2MASS, and supplement them with those in 
the 2MASS Extended Source Catalog (XSC) and Large Galaxy Atlas (Jarrett
\etal 2003).  For 9 galaxies with unreliable
$K_s$ photometry from 2MASS (those not included in the XSC), we use only the
\gz\ color assuming a 10~Gyr SSP to determine $\mathcal{M}/L_z$.  We assume a
younger age than for GCs because dEs are measured to have younger ages
than GCs or massive ellipticals (Geha \etal 2003). 
Although $\mathcal{M}/L$ is related to both age
and metallicity, it is easier to determine $\mathcal{M}/L$ than age or metallicity
individually.  Most of the
sample have $\mathcal{M}/L_z$ consistent with old stellar populations
(age $>5$~Gyr), though a few galaxies are dE/dI transition objects and have
noticeably younger mean ages and correspondingly lower $\mathcal{M}/L_z$ (the
most extreme example being VCC~1499).  For one of these galaxies,
VCC~1030, we derive a low mean age, high metallicity, and a
correspondingly low $\mathcal{M}/L_z$.  While this galaxy may be interacting, 
the colors may also be suspect due to a large-scale central dust
disk.  Until we get more data, we consider $\mathcal{M}/L$ for this galaxy uncertain.

% fig:sn_mv
%%%%%%%%%%%%%%%%%%%%%%%%%%%%%%%%%%%%%%%%%%%%%%%%%%%%%%%%%%%%%%%%%%%%%%%%%
\begin{figure}
\epsscale{1.22}
% snpf00.ps
\plotone{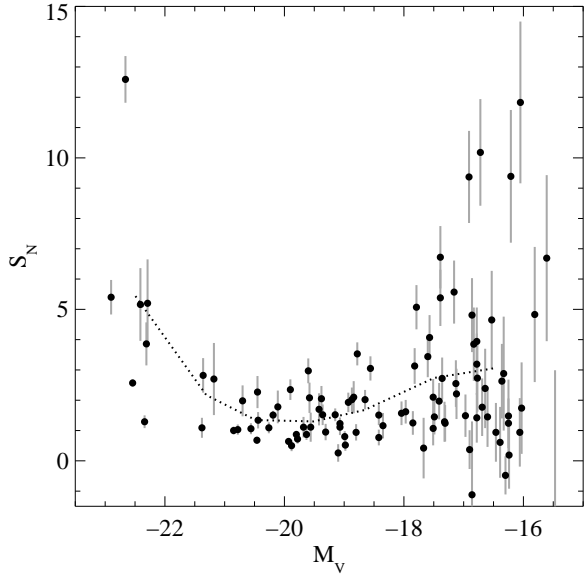}
\caption{$S_N$ versus galaxy $M_V$ for 100 ACSVCS galaxies.  M87, 
  the giant elliptical with $S_N\sim13$ is a well-known
  outlier.  Luminous early-type galaxies have higher $S_N$ than
  intermediate-luminosity early-type galaxies by a factor $\sim2$--3.
  Galaxies with intermediate luminosities ($-20.5 < M_V < -18$)
  generally have $S_N\sim1.5$.  Early-type dwarf galaxies have a large
  spread in $S_N$, with some having zero GCs, and others having among
  the highest measured $S_N$ in our sample.  The dotted line shows the
  mean trend (including M87) whose values are listed in Table~\ref{table:binmv}.
  \label{fig:sn_mv}}
\end{figure}
%%%%%%%%%%%%%%%%%%%%%%%%%%%%%%%%%%%%%%%%%%%%%%%%%%%%%%%%%%%%%%%%%%%%%%%%%

Fortunately, the $z$ bandpass is less sensitive to recent star
formation than bluer bandpasses, and the range of $\mathcal{M}/L_z$ is only 0.4
to 2.1, a dynamic range which is a factor of $\sim3$ smaller than that for
$\mathcal{M}/L_B$.  The error in $\mathcal{M}/L_z$ is dominated by the quality of the
infrared photometry.  We estimate errors in $\mathcal{M}/L_z$ with a Monte Carlo
procedure where the $J-K_s$ color is perturbed by a random amount
drawn from a Gaussian distribution with sigma equal to the claimed photometric
error.  The $\mathcal{M}/L_z$ is then recalculated in this fashion 1000 times
and the error is the half width of the middle 68\% of the
distribution.  The typical error in $\mathcal{M}/L_z$ is $\pm0.4$, or
$\sim25\%$.  These errors are propagated forward into mass-normalized
quantities, although any possible variations in the IMF are not included.

All of the global galaxy properties discussed above are listed in
Table~\ref{table:galxtable}, and all the quantities related to the
globular cluster systems are presented in Table~\ref{table:gctable}.

\section{Results}

\subsection{$S_N$ versus Galaxy Magnitude}

The historical definition of specific frequency, $S_N$, is
the number of GCs per unit $M_V=-15$ of galaxy luminosity, or
\begin{equation}
  S_N = N_{GC} \times 10^{0.4(M_V + 15)}
\end{equation}
(Harris \& van den Bergh 1981).  This number is approximately unity
for the Milky Way.  One of the motivations for calculating $S_N$ is to
understand whether GC formation scales with the bulk of star formation
in the same way across galaxy types and masses.
In Figure~\ref{fig:sn_mv} we show the behavior of $S_N$ as a function of
$M_V$ in our sample of 100 ACSVCS galaxies.  The luminous early-type
galaxies, mostly giant ellipticals, are well known to have $S_N\sim2$--$5$.
The cD galaxy, M87, has $S_N=12.6\pm0.8$ which is
consistent with other measures of its $S_N$ (e.g., Harris \etal 1998).  
Early-type galaxies of
intermediate luminosity ($-22 < M_V < -18$) have a nearly uniformly
low $\langle S_N\rangle \sim1.5$.  Galaxies in this luminosity range
also have a tendency to be lenticular, with the VCC classifying 70\%
of these ACSVCS intermediate-luminosity galaxies as S0, E/S0 or S0/E.
By contrast, the fainter galaxies ($M_V > -18$) exhibit a wide range
of $S_N$, with values as low as zero and as high as those of
M87\footnote{Strader \etal (2006) analyzed a subset of these data
and claimed a bimodal distribution of \sn\ in the dEs.  A histogram of
the \sn\ distribution shows that it is better described as strongly
peaked with a tail to higher \sn, especially when one includes dS0s.}.

% Table 3
%%%%%%%%%%%%%%%%%%%%%%%%%%%%%%%%%%
% Table 3
\setcounter{table}{2}
\begin{center}
\begin{deluxetable}{ccl}
\tablewidth{0pt}
\tablecaption{Specific Frequency in bins of $M_V$\label{table:binmv}}
\tablehead{
\colhead{$M_V$ range} & 
\colhead{$\langle M_V \rangle$} & 
\colhead{$S_N$}
}
\startdata
($-24,-22$)$^1$ & $-22.5$ & $  4.0$ \\
($-24,-22$) & $-22.5$ & $  5.4$ \\
($-22,-21$) & $-21.3$ & $  2.2$ \\
($-21,-20$) & $-20.5$ & $  1.3$ \\
($-20,-19$) & $-19.5$ & $  1.3$ \\
($-19,-18$) & $-18.7$ & $  1.7$ \\
($-18,-17$) & $-17.5$ & $  2.7$ \\
($-17,-15$) & $-16.4$ & $  3.1$ \\
\enddata
\tablenotetext{1}{Not including VCC 1316 (M87)}
\end{deluxetable}

\end{center}
%%%%%%%%%%%%%%%%%%%%%%%%%%%%%%%%%%

\begin{figure}
\epsscale{1.22}
% snpf01.ps
\plotone{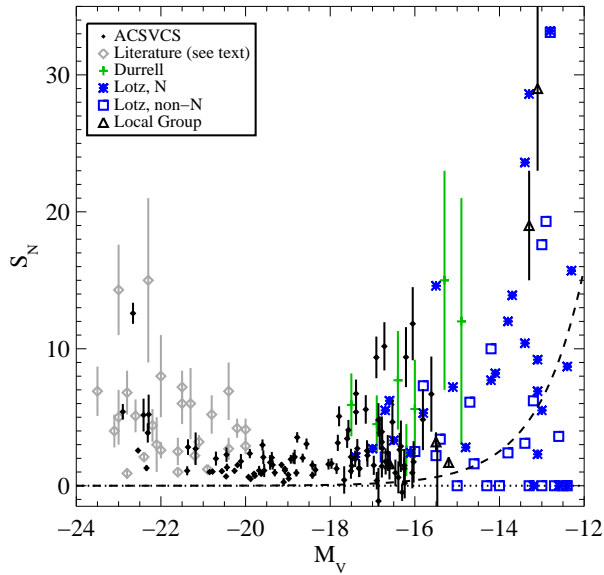}
\caption{$S_N$ versus galaxy $M_V$ for 100 ACSVCS galaxies (black
  circles), and early-type galaxy data from the literature (references
  in the text).  Literature values for $S_N$ follow and extend trends
  visible in the ACSVCS galaxies.  In particular, dwarf galaxy $S_N$
  values from the work of Durrell \etal (1996) and Lotz \etal (2004)
  show that fainter dwarf ellipticals can have an even larger range of
  specific frequency. For clarity, error bars are not plotted for the
  HST/WFPC2 dEs, but the uncertainty can be very large.  The dashed
  line shows the $S_N$ value if a galaxy at that magnitude had 1~GC.
  \label{fig:sn_mv_lit}}
\end{figure}

This trend has been hinted at previously from studies of giant
ellipticals, lenticulars, and dwarf ellipticals.  In
Figure~\ref{fig:sn_mv_lit}, we show the ACSVCS
sample combined with data compiled from the literature.  At the high
luminosity end, we take $S_N$ values from the compilation of Ashman \&
Zepf (1998), using ones that were reliably determined from CCD
data (gray diamond points). These include estimates from Kissler-Patig
\etal (1996, 1997), 
Dirsch \etal (2003a, 2003b), Dirsch, Schuberth \& Richtler (2005), 
Forbes \etal (1996), Rhode \& Zepf
(2004), Zepf \etal (1995), and Harris, Harris \& Geisler (2004).  At
the faint end, we include data from three sources which extend the
range in galaxy luminosity by many magnitudes.  We include the 7 dEs
studied by Durrell \etal (1996a, 1996b) that do not overlap with our
sample, the Virgo and Fornax dEs in the HST/WFPC2 study of Miller
\etal (1998) and Lotz \etal (2004), and the five Local Group dwarfs
listed in Lotz et~al.

% fig:sn_mv_lit
%%%%%%%%%%%%%%%%%%%%%%%%%%%%%%%%%%%%%%%%%%%%%%%%%%%%%%%%%%%%%%%%%%%%%%%%%
%%%%%%%%%%%%%%%%%%%%%%%%%%%%%%%%%%%%%%%%%%%%%%%%%%%%%%%%%%%%%%%%%%%%%%%%%

At the high luminosity end, the literature values have large scatter,
possibly due to their heterogeneous nature.  One of the benefits of
the ACSVCS is that relative distances between Virgo cluster galaxies
are both small and measured in a homogeneous fashion.  At low
luminosities, the studies are mainly in Virgo and Fornax so relative
distances are less of a problem.  At luminosities fainter than the
ACSVCS, the trend of higher $S_N$ coupled with a larger range of $S_N$
continues to the limit of the data.  The ACSVCS sample not only has
smaller errors, it fills the important luminosity regime between
giants and dwarfs.  Kundu \& Whitmore (2001a,b) studied galaxies in
the magnitude range $-23 < M_B < -16$, similar to the ACSVCS, 
and found that ellipticals
generally had higher \sn\ than lenticulars, but they were only able to
measure ``local'' \sn\ within the WFPC2 field of view.

\begin{figure}
\epsscale{1.22}
% snpf02.ps
\plotone{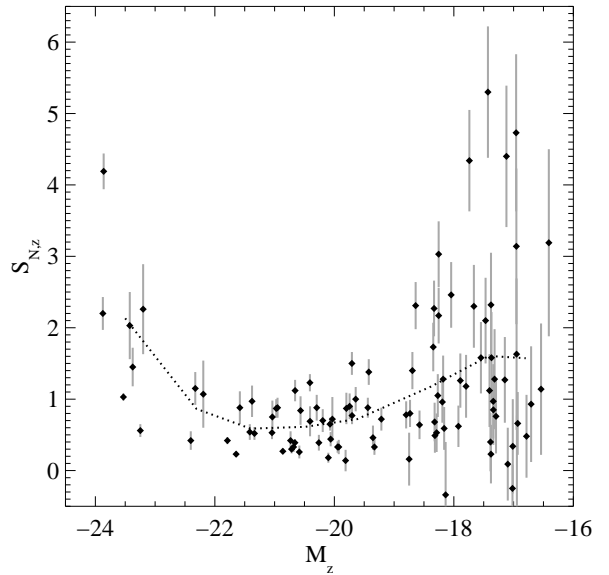}
\caption{$S_{N,z}$ versus galaxy $M_z$.  This figure shows similar
  quantities to those in Figure~\ref{fig:sn_mv} except the number of
  GCs is normalized to the galaxy absolute magnitude in the $z$-band, 
  with $M_z$ is plotted along the x-axis.  This far-red bandpass
  is a better tracer of the total stellar mass than the traditionally
  used $V$-band.  The
  dotted line shows the mean trend with values in Table~\ref{table:binmz}.
  \label{fig:snz_mz}}
\end{figure}

Ultimately, studies of GC specific frequency are about how GC
formation scales with galaxy {\it mass}, using $M_V$ as a proxy.  
However, a redder bandpass more faithfully traces stellar
mass, and in Figure~\ref{fig:snz_mz}, we introduce \snz, which is defined
identically to $S_N$ except that the number of GCs is normalized to
an absolute $z$ magnitude of $M_z=-15$, and plotted against $M_z$.
\snz\ values are 1.5 to 3 times smaller than \sn\ because early-type 
galaxies are
red (and thus more luminous in $z$).  The specific frequencies of
giant ellipticals are adjusted more than those of the dwarfs because
massive galaxies are redder in color.  While the trends in this figure
are the same as those in Figure~\ref{fig:sn_mv}, one interesting
difference is that the highest \snz\ values for the dwarfs now equal
or exceed that of M87, with the change being due to M87's redder color.
For the rest of the paper, we will use quantities based on $M_z$.

% fig:snz_mz
%%%%%%%%%%%%%%%%%%%%%%%%%%%%%%%%%%%%%%%%%%%%%%%%%%%%%%%%%%%%%%%%%%%%%%%%%
%%%%%%%%%%%%%%%%%%%%%%%%%%%%%%%%%%%%%%%%%%%%%%%%%%%%%%%%%%%%%%%%%%%%%%%%%

\subsection{$S_{N,z}$ and Bulge Luminosity}

In most scenarios of GC system formation, the GCs are associated with
the stellar spheroids---either halo or bulge---and not with the
formation of present day disks.  If this is the case, then perhaps
normalizing GC numbers to total spheroid luminosity would be more
fundamental than normalizing to total luminosity.  In our ACSVCS
sample, over half of our galaxies are morphologically classified in the VCC as
either E/S0, S0/E, S0, or dS0.  Most of these galaxies are at
intermediate luminosities, and make up a substantial fraction of the
galaxies that have low \snz.  Could it be the case that these low
\snz\ values are due to the inclusion of a stellar disk in the total
luminosity?  Moreover, the scatter in the \snz\ of the intermediate
luminosity galaxies in our sample exceeds the scatter expected from
the errors, implying that a parameter such as bulge fraction might be
important.  

\begin{figure}
\epsscale{1.22}
% snpf19.ps
\plotone{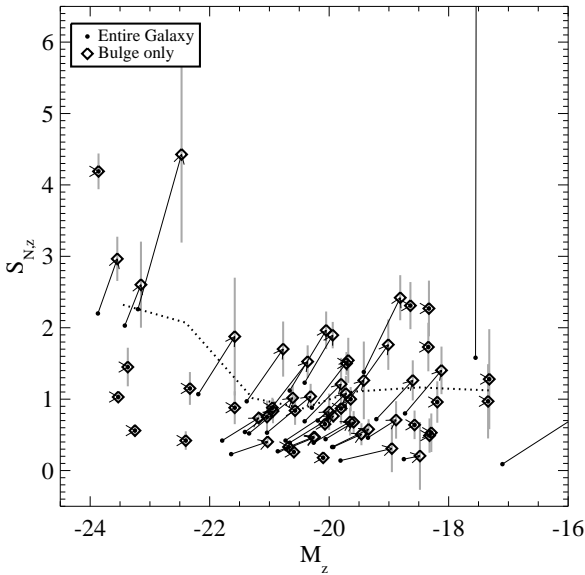}
\caption{\snz\ versus galaxy $M_z$ for ACSVCS galaxies with VCC
  morphological types of E/S0, S0/E, S0, and dS0, and which have
  morphological T-types in the RC3.  We plot the total \snz\ (same as
  Figure~\ref{fig:snz_mz}) as well as an arrow connecting to the
  galaxy's \snz\ normalized to bulge luminosity alone.  Some galaxies
  have \snz\ similar to those in luminous ellipticals, but others are
  relatively unchanged.
  \label{fig:bulge}}
\end{figure}

We can test whether normalizing by bulge luminosity is more
fundamental with quantitative bulge-disk decompositions (Chen et~al.,
in prep).  In this paper, however, we use the morphological T-types
given in the Third Reference Catalog of Bright Galaxies (RC3)
(de~Vaucouleurs \etal 1991).  These T-types are correlated with the
ratio of bulge to total luminosity using the relation of Simien \&
de~Vaucouleurs (1986).  We can do this for 55 galaxies in the ACSVCS
sample and we show how their \snz\ values change in
Figure~\ref{fig:bulge}.  This figure shows \snz\ as derived using
total luminosity compared to \snz\ as derived from bulge luminosity
alone with the two values connected by an arrow.  For some galaxies,
their specific frequencies do increase to values typical of the
luminous ellipticals (\snz$\sim1$--2), but for others whose disks are
less prominent, their \snz\ stay relatively unchanged.  For two
galaxies, their RC3 morphological types values imply that they have
little or no bulge, which give them very high \snz.

Based on the current morphological classifications, it is not apparent
that GCs form at a constant efficiency with respect to bulge
luminosity.  Normalizing to bulge luminosity increases
the scatter in \snz\ in the intermediate luminosity range.  However,
a more thorough analysis based on modern bulge-to-disk
decompositions is necessary before we can reach any strong conclusions.

% fig:bulge

%%%%%%%%%%%%%%%%%%%%%%%%%%%%%%%%%%%%%%%%%%%%%%%%%%%%%%%%%%%%%%%%%%%%%%%%%
%%%%%%%%%%%%%%%%%%%%%%%%%%%%%%%%%%%%%%%%%%%%%%%%%%%%%%%%%%%%%%%%%%%%%%%%%

% Table 4
%%%%%%%%%%%%%%%%%%%%%%%%%%%%%%%%%%
% Table 4
%\input{sntable_binmz.tex}
\begin{center}
\begin{deluxetable*}{ccrrccccccc}
\tablewidth{0pt}
\tablecaption{Bins of $M_z$\label{table:binmz}}
\tablehead{
\colhead{$M_z$ range} & 
\colhead{$\langle M_z \rangle$} & 
\colhead{$S_{N,z}$} & 
\colhead{$T$} & 
\colhead{$S_L$} & 
\colhead{$S_M$} & 
\colhead{$S_{N,z,blue}$} & 
\colhead{$S_{N,z,red}$} & 
\colhead{$f_{red}$} & 
\colhead{$S_{N,z,close}^1$} & 
\colhead{$S_{N,z,far}^2$}
}
\startdata
($-25,-23$)$^3$ & $-23.4$ & $  1.6$ & $ 13.2$ & $ 0.90$ & $ 0.74$ & $ 1.17$ & $ 0.47$ & $0.29$ & $ 1.72$ & $ 1.57$ \\
($-25,-23$) & $-23.5$ & $  2.1$ & $ 18.2$ & $ 1.12$ & $ 1.04$ & $ 1.53$ & $ 0.60$ & $0.28$ & $ 2.55$ & $ 1.57$ \\
($-23,-22$) & $-22.3$ & $  0.9$ & $  8.5$ & $ 0.51$ & $ 0.50$ & $ 0.49$ & $ 0.37$ & $0.43$ & $ 1.11$ & $ 0.43$ \\
($-22,-21$) & $-21.4$ & $  0.6$ & $  4.1$ & $ 0.21$ & $ 0.17$ & $ 0.40$ & $ 0.19$ & $0.32$ & $ 0.65$ & $ 0.45$ \\
($-21,-20$) & $-20.5$ & $  0.6$ & $  4.7$ & $ 0.18$ & $ 0.16$ & $ 0.43$ & $ 0.19$ & $0.31$ & $ 0.57$ & $ 0.68$ \\
($-20,-19$) & $-19.6$ & $  0.7$ & $  5.8$ & $ 0.21$ & $ 0.17$ & $ 0.57$ & $ 0.15$ & $0.21$ & $ 0.75$ & $ 0.67$ \\
($-19,-18$) & $-18.4$ & $  1.2$ & $ 11.4$ & $ 0.38$ & $ 0.34$ & $ 0.99$ & $ 0.18$ & $0.16$ & $ 1.41$ & $ 0.80$ \\
($-18,-17$) & $-17.4$ & $  1.6$ & $ 17.8$ & $ 0.36$ & $ 0.37$ & $ 1.46$ & $ 0.15$ & $0.09$ & $ 2.04$ & $ 1.02$ \\
($-17,-16$) & $-16.8$ & $  1.6$ & $ 17.3$ & $ 0.51$ & $ 0.35$ & $ 1.36$ & $ 0.21$ & $0.13$ & $ 2.03$ & \nodata \\
\enddata
\tablenotetext{1}{Within 1~Mpc of VCC 1316 (M87), in projection}
\tablenotetext{2}{Outside 1~Mpc of VCC 1316 (M87), in projection}
\tablenotetext{3}{Not including VCC 1316 (M87)}
\end{deluxetable*}

\end{center}
%%%%%%%%%%%%%%%%%%%%%%%%%%%%%%%%%%

\subsection{Normalizing to Stellar Mass}
\begin{figure}
\epsscale{1.22}
% snpf03.ps
\plotone{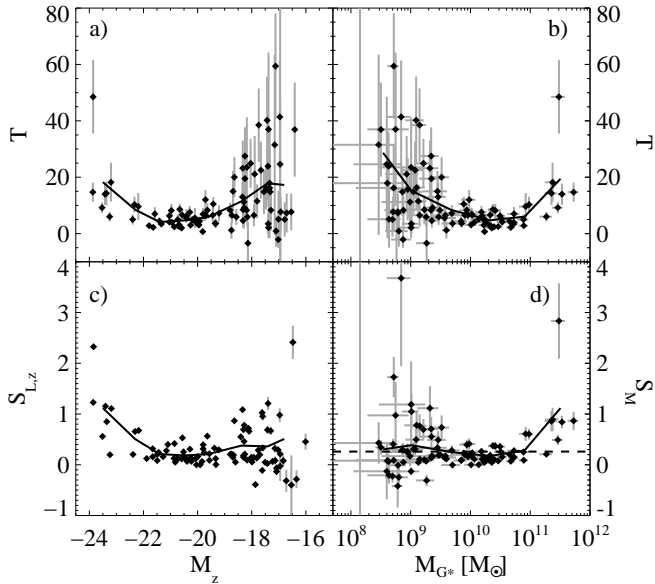}
\caption{$T_N$ versus $M_z$ (a, upper left), and versus galaxy stellar
  mass (b, upper right).  $T_N$ is the number of GCs per $10^9
  \mathcal{M}_\odot$.  In the bottom two plots we show the specific $z$
  luminosity, $S_L$, versus $M_z$ (c, lower left) and the specific
  mass, $S_\mathcal{M}$, versus stellar mass (d, lower right). The dashed line
  in (d) marks $S_\mathcal{M} = 0.26$, the ``universal'' GC formation
  efficiency from McLaughlin (1999a).  Because the
  ACSVCS sample is all early-types, $\mathcal{M}/L$ does not vary much, and
  hence in all four panels we see trends similar to those for $S_N$
  and $S_{N,z}$ (see text for exceptions).  The difference between
  dwarfs and giants in $S_L$ and $S_\mathcal{M}$ is not as large as it is in
  $S_N$ or $T_N$.  This reflects the changing GCLF across galaxy
  luminosity (see Figure~\ref{fig:lgczmean}). In all figures, the
  solid line shows the mean trend, values in Table~\ref{table:binmass}.
  \label{fig:t_m}}
\end{figure}

Using the stellar masses for the ACSVCS galaxies, we calculate the
$T$ parameter introduced by Zepf \& Ashman (1993), which is the
number of GCs ($N_{GC}$) per $10^9 \mathcal{M}_\odot$,
\begin{equation}
T = N_{GC} / (\mathcal{M}_{G\star}/10^9 \mathcal{M}_\odot)
\end{equation}
where $\mathcal{M}_{G\star}$ is the stellar mass of the galaxy.  The advantage of using
$T$ instead of \sn\ is that it allows comparisons across galaxies with
different mass-to-light ratios.  In
Figure~\ref{fig:t_m}a and b, we show $T$ plotted against $M_z$ and
$\mathcal{M}_{G\star}$.  Although the errors are larger than in
Figure~\ref{fig:snz_mz}, the same trends are evident.  Our values for
$T$ are higher than those in previous studies, such as Rhode
\& Zepf (2004).  This is due to differences in the mass-to-light
ratios used.  Previous studies have assumed $\mathcal{M}/L_V=10$ for
elliptical galaxies, whereas we have estimated $\mathcal{M}/L$ using galaxy
colors and the BC03 models with a Chabrier (2003) initial mass function
(IMF).  Even accounting for the different 
bandpasses used, our $\mathcal{M}/L$ are systematically lower.  For example, for
M49, the most luminous elliptical in the sample, 
$(V-z)=0.97$ and its $\mathcal{M}/L_z=2.0$ translates to $\mathcal{M}/L_V=5$, 
which is a factor of two lower than the canonical Zepf \& Ashman values.
We feel that although the $\mathcal{M}/L$ values in Zepf \& Ashman (1993) have
been valuable as standard conversions, they are too high, and that our values
of $\mathcal{M}/L$ are more reasonable given recent
dynamical mass measurements of elliptical
galaxies (Kronawitter \etal 2000; Cappellari \etal 2006).  The use of a
Salpeter (1955) IMF increases $\mathcal{M}/L$ by a factor of 1.8, but is not
observationally motivated for stars below 1~$\mathcal{M}_\odot$ and would
produce stellar $\mathcal{M}/L$ values higher than the total $\mathcal{M}/L$ measured in
some early-type galaxies.  For the early-type galaxies
in our sample, $\mathcal{M}/L_V$ ranges from 1 to 5, although some galaxies with
obvious star formation have $\mathcal{M}/L_V<1$.  We reiterate, though, that a
using redder bandpass is better, and all our masses are derived using
$\mathcal{M}/L_z$, and we discuss the $V$ bandpass only for comparing
with previous work.

For the purposes of this paper, however, the absolute scale of $T$ is
not as important as the relative scale within the sample.
Figure~\ref{fig:t_m}b shows that $T$ spans a wide range at masses 
$\mathcal{M}_{G\star}<4\times10^9 \mathcal{M}_\odot$, stays constant until $10^{11} \mathcal{M}_\odot$, and
then increases again at higher mass.

% fig:t_m
%%%%%%%%%%%%%%%%%%%%%%%%%%%%%%%%%%%%%%%%%%%%%%%%%%%%%%%%%%%%%%%%%%%%%%%%%
%%%%%%%%%%%%%%%%%%%%%%%%%%%%%%%%%%%%%%%%%%%%%%%%%%%%%%%%%%%%%%%%%%%%%%%%%

\subsection{Specific Luminosity and Mass}
If we want to know what fraction of a galaxy's luminosity (or mass) is
in GCs, we can total it up directly, as described in
Section~\ref{sec:slm}.  We use the definition of specific luminosity,
$S_L$ presented in Harris (1991) except that we use the $z$ bandpass
instead of $V$:
\begin{equation}
S_{L,z} = 100 \times L_{GC,z} / L_{galaxy,z}
\end{equation}
where $L_{GC,z}$ and $L_{galaxy,z}$ are the total $z$-band
luminosities of the GCs and the galaxy.  This quantity has two
advantages over $S_N$ in that it is independent of distance and that
the completeness corrections for unobserved faint GCs are extremely
small.  We also plot the specific mass, $S_\mathcal{M}$, defined as
\begin{equation}
S_\mathcal{M} = 100 \times \mathcal{M}_{GC} / \mathcal{M}_{G\star}
\end{equation}
where $\mathcal{M}_{GC}$ is the total stellar mass in GCs, calculated as
described in Section~\ref{sec:slm}.

Figure~\ref{fig:t_m}c shows the specific luminosity $S_L$ of GCs
plotted against $M_z$, and Figure~\ref{fig:t_m}d shows $S_\mathcal{M}$ versus
$\mathcal{M}_{G\star}$.  The dashed line marks the 0.26\%
``universal'' GC formation efficiency from McLaughlin (1999a), and is
a reasonable description of the GC mass fraction in intermediate mass
galaxies.  For the ACSVCS galaxies with $-22<M_z<-19$, $\langle S_L
\rangle = 0.20$, $\sigma_{S_L}=0.14$.  Similarly, for galaxies with
$\mathcal{M}_{G\star} = 0.4$--$6\times 10^{10}\mathcal{M}_\odot$,  
$\langle S_\mathcal{M}\rangle = 0.17$, $\sigma_{S_\mathcal{M}}=0.11$.
That the mass fraction is slightly lower than the luminosity fraction
can be explained by the bluer colors and hence lower $\mathcal{M}/L$
of the GCs as compared to their host galaxies.

% fig:lgczmean
%%%%%%%%%%%%%%%%%%%%%%%%%%%%%%%%%%%%%%%%%%%%%%%%%%%%%%%%%%%%%%%%%%%%%%%%%
\begin{figure}
\epsscale{1.22}
% snpf04.ps
\plotone{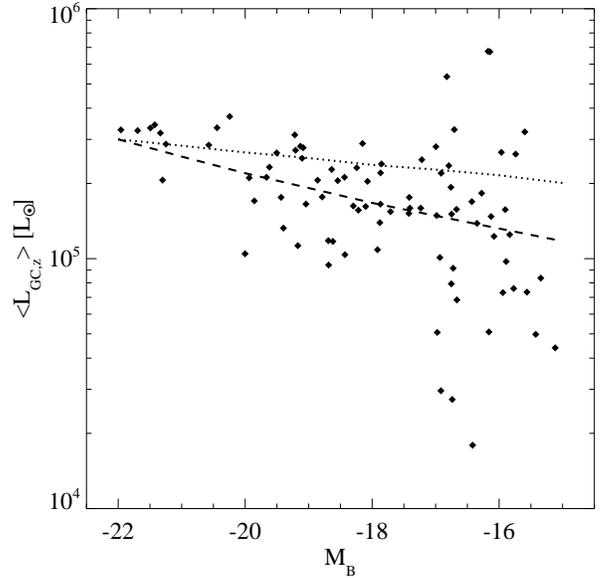}
\caption{Mean $z$ luminosity of GCs, $\langle L_{GC,z} \rangle$, in
  ACSVCS galaxies versus $M_z$.  If the GCLF was constant across all
  galaxies, $\langle L_{GC,z} \rangle$ would also be constant.  The
  affect of a fainter GCLF turnover in dwarf galaxies (dotted line)
  only partially explains the lower mean luminosities.  A combination
  of fainter turnovers and narrower GCLFs (dashed line) reproduces the
  observed trend.
  \label{fig:lgczmean}}
\end{figure}
%%%%%%%%%%%%%%%%%%%%%%%%%%%%%%%%%%%%%%%%%%%%%%%%%%%%%%%%%%%%%%%%%%%%%%%%%

% Table binmass
%%%%%%%%%%%%%%%%%%%%%%%%%%%%%%
% Table 5
\begin{center}
\begin{deluxetable}{ccrrr}
\tablewidth{0pt}
\tablecaption{Bins of $M_{G\star}$\label{table:binmass}}
\tablehead{
\colhead{$M_\star$ range$^1$} & 
\colhead{$\langle M_\star \rangle^1$} & 
\colhead{$T$} & 
\colhead{$S_L$} & 
\colhead{$S_M$}
}
\startdata
($   0.1,   0.5$) & $   0.3$ & $ 18.2$ & $ 1.12$ & $ 0.30$ \\
($   0.5,   2.2$) & $   1.0$ & $  8.5$ & $ 0.51$ & $ 0.39$ \\
($   2.2,  10.0$) & $   4.8$ & $  4.1$ & $ 0.21$ & $ 0.25$ \\
($  10.0,  46.4$) & $  21.1$ & $  4.7$ & $ 0.18$ & $ 0.17$ \\
($  46.4, 215.4$) & $  80.8$ & $  5.8$ & $ 0.21$ & $ 0.29$ \\
($ 215.4,1000.0$) & $ 321.0$ & $ 11.4$ & $ 0.38$ & $ 1.12$ \\
($ 215.4,1000.0$)$^2$ & $ 324.7$ & $ 14.0$ & $ 1.00$ & $ 0.80$ \\
\enddata
\tablenotetext{1}{$M_\star/10^9 M_\odot$}
\tablenotetext{2}{Not including VCC 1316 (M87)}
\end{deluxetable}

\end{center}
%%%%%%%%%%%%%%%%%%%%%%%%%%%%%%

The major difference in Figure~\ref{fig:t_m} between the top two
panels showing $T$ (a and b) and the bottom two showing $S_{L,z}$ and
$S_\mathcal{M}$ (c and d) is the comparison between the giants and dwarfs.
While the dwarfs have very high number fractions, their luminosity and
mass fractions are substantially lower compared to the giants.  Why
should this be the case?  The reason for this has to do with the
GCLF.  If the GCLF was constant across all galaxies (i.e., same mean
$\mu$ and width $\sigma$), then $S_L$ would mirror $T$.  However,
\jordan \etal (2006, 2007) showed that the GCLF varies as a
function of galaxy mass in the sense that less massive galaxies host
GC systems with fainter $\mu$ and smaller $\sigma$.  This has the
effect of lowering the mean luminosity (and mass) of GCs in the dwarf
galaxies.  We illustrate this in Figure~\ref{fig:lgczmean}, where we
plot the mean $z$ luminosity of GCs in each galaxy, $\langle
L_{GC,z}\rangle$, against galaxy luminosity (in this case $M_B$
because that is how we presented the data in \jordan et al.).  It is
clear that a constant GCLF does not describe the data well.  The
dotted line shows the effect of varying $\mu$, and the dashed line
shows the relation predicted by the combination of fainter $\mu$ and a
narrower $\sigma$ in the dwarfs.  Although the scatter is large, the
expected change in the GCLF can account for the changes in mean GC
luminosity that we see.  This trend is actually more
pronounced in mass than in luminosity because GCs in dwarfs are more
metal-poor and have lower $\mathcal{M}/L$ than those in giants.

The result is that some of the more extreme $S_N$ or $T$ values seen
in the dwarfs are somewhat less extreme when expressed as a mass
fraction.  Nevertheless, many low-luminosity galaxies still have $S_L$
and $S_\mathcal{M}$ that are significantly higher than those in the
intermediate-$L$ galaxies.  Because the global trends are similar in
$S_N$, $T$, $S_L$, and $S_\mathcal{M}$, we will refer to these
quantities collectively as ``GC fractions''.

\subsection{Specific Frequencies of Red and Blue GCs}

% fig:rfrac
%%%%%%%%%%%%%%%%%%%%%%%%%%%%%%%%%%%%%%%%%%%%%%%%%%%%%%%%%%%%%%%%%%%%%%%%%
\begin{figure}
\epsscale{1.22}
% snpf12.ps
\plotone{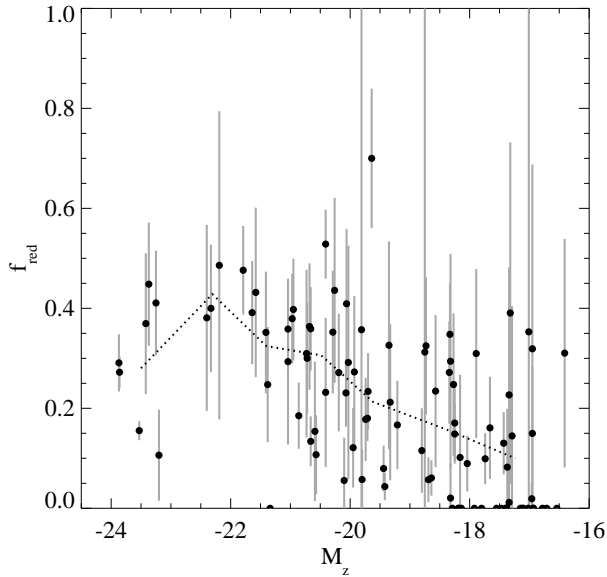}
\caption{Fraction of red GCs ($f_{red}$) versus $M_z$.  The dotted
  line represents the mean trend (Table~\ref{table:binmz}).  More
  luminous galaxies have higher fractions of red GCs up to
  $M_z\sim-22$, but at higher luminosities
  there is a flattening or turnover in $f_{red}$, and the most luminous
  galaxies do not have the highest red GC fractions.
  \label{fig:rfrac}}
\end{figure}
%%%%%%%%%%%%%%%%%%%%%%%%%%%%%%%%%%%%%%%%%%%%%%%%%%%%%%%%%%%%%%%%%%%%%%%%%

Our previous study of GC color and metallicity distributions in the ACSVCS
galaxies show that they are, on average, either bimodal or asymmetric
across the entire luminosity range of the sample, and we use the
products of that analysis (Peng \etal 2006a) in the current study.
The blue (metal-poor) and red (metal-rich) are believed to trace
either different epochs of formation or different progenitor halos
(however, see Yoon, Yi \& Lee 2006 and Cantiello \& Blakeslee 2007 
for arguments that the bimodality may
be an observational consequence of a nonlinear metallicity-color
relation).  It
is an interesting question to ask how the specific frequencies of each
GC population scale with galaxy properties.  For example, Rhode \& Zepf (2004)
showed that the mass-normalized number of blue GCs, $T_{blue}$,
increases as one goes from spirals to
elliptical galaxies, although their spirals were less massive than
ellipticals so the trend could also have been one in galaxy mass.
Assuming that mergers of spirals only produce 
new red GCs, they argued that the GC systems of ellipticals cannot be formed
purely by spiral mergers.

We determine the fraction of blue and red GCs in a hybrid approach
similar to what we do for total numbers.  For the brightest 21
galaxies in the sample (a complete magnitude-limited sample, as ranked
by $B_T$ in \cote \etal 2004; 
Paper~I) there are sufficient numbers of GCs that the KMM two-Gaussian
fits to the \gz\ distribution performed by Peng \etal (2006a; Paper IX)
are reliable enough that we can use the ``dip'' between the two
Gaussians---the color at which a GC is equally likely to belong to the
blue or red GC distribution---as the dividing color between the two
populations. For the remaining galaxies, we assume a fixed dividing
color of \gz$=1.16$.  Although the colors of the individual peaks vary as a
function of galaxy luminosity, the dip color is relatively invariant
(see Figure~5 in Peng \etal 2006a).

The brighter galaxies may have better defined color distributions, but
we only observe a fraction of their entire GC system, and the
red-to-blue ratio is observed to decrease as a function of galactocentric
radius.  We correct for this bias in the brightest 14 galaxies by
fitting \sersic\ models to the surface density profiles of the
red GCs separately, and integrating the best model to obtain their total
numbers.  We fit to profiles derived from our ACS and WFPC2 data, as
described in \S\ref{sec:profs}.  In the most luminous galaxies
such as M49 and M87, the 
fraction of red GCs within the ACS/WFC is as high as 0.6, but the
red fraction of the entire GC system is more like 0.3.  For the remainder of the
galaxies, we find that the ACS/WFC encompasses nearly all of the red
GC system and use our corrected counts to determine the red GC fraction.

% fig:snz_mz_br
%%%%%%%%%%%%%%%%%%%%%%%%%%%%%%%%%%%%%%%%%%%%%%%%%%%%%%%%%%%%%%%%%%%%%%%%%
\begin{figure}
\epsscale{1.22}
% snpf09.ps
\plotone{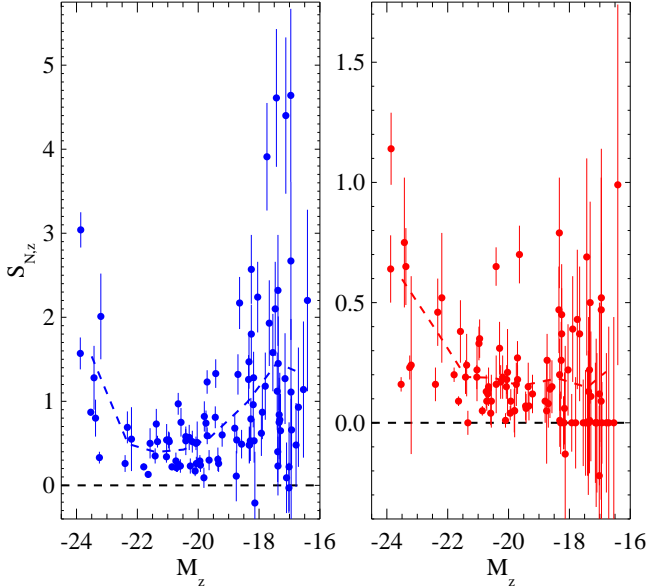}
\caption{$S_{N,z}$ versus $M_z$ for blue (left) and red (right)
  globular clusters.  Both red and blue GCs show enhanced \snz\ in
  massive galaxies, but the variation in $S_{N,z}$ across
  galaxy mass is dominated by the blue GCs.  This is true even in
  the massive galaxies. The dashed
  lines show the mean trends with values listed in Table~\ref{table:binmz}.
  \label{fig:snz_mz_br}}
\end{figure}
%%%%%%%%%%%%%%%%%%%%%%%%%%%%%%%%%%%%%%%%%%%%%%%%%%%%%%%%%%%%%%%%%%%%%%%%%

Figure~\ref{fig:rfrac} shows the fraction of red GCs as a function of
$M_z$.  The fraction of red GCs, $f_{red}$, generally increases with galaxy
luminosity, going from $\sim0.1$ to $\sim0.5$, but at $M_z\lesssim-22$,
the trend in $f_{red}$ appears to either flatten or perhaps turn over.  This
is in contrast to the results of Paper~IX which only quantified the
red GC fraction within the ACS/WFC images, and thus were biased to
detecting the more centrally concentrated red GCs.  With the proper
aperture corrections, we can see that the most luminous galaxies in
the sample do not have increasingly higher fractions of red GCs.  Is this
because there are fewer red GCs in these galaxies or more blue GCs?

Figure~\ref{fig:snz_mz_br} shows the \snz\ for blue and red GCs as
function of $M_z$.  The trend in the blue GCs mirrors the overall
trend seen in Figure~\ref{fig:snz_mz}, with the massive and dwarf
galaxies having the highest \snz.  This is not surprising on the faint
end since most of the GCs in the fainter galaxies are blue.  Even for
the massive galaxies, the high $S_{N,z}$ values are
dominated by blue GCs.  However, the specific frequencies of red GCs
also exhibit an increase for $M_z<-21$, especially for the cD
galaxy M87, which is a $4\sigma$ outlier.  The elevated
$S_{N,z,blue}$ {\it and} $S_{N,z,red}$ for the most massive galaxies,
suggests that massive galaxies are not underproducing red GCs, which
is one possible interpretation of
Figure~\ref{fig:rfrac}.  Instead, these massive galaxies have more red
GCs, but even more blue GCs.

\subsection{$S_{N,z}$ and Nucleation}

Observations by the HST/WFPC2 of dEs in the Virgo and Fornax Clusters
(Miller \etal 1998; Lotz \etal 2004; Miller \& Lotz 2007) have found
that dEs with stellar nuclei have a higher mean specific frequency,
possibly implying that a higher past star formation efficiency
resulted in the formation of both nuclei and GCs.  The dEs in our
sample are more luminous than the ones studied in the WFPC2 snapshot
survey, and almost all of the galaxies are nucleated (\cote \etal
2006; Paper~VIII).  Therefore, it is difficult to test if there is any
correlation between \snz\ and nucleation.  Only four dwarf galaxies
are definitely non-nucleated (Type~II in Table~1 of
Paper~VIII)---VCC~1049, 1833, 1499, and 1512.  Of these galaxies, VCC~1499
and 1512 are dE/dI transition objects with young stars, and VCC~1049 has
bluer colors toward the center.  The \snz\ values of these galaxies
range from 0.97 (VCC~1049) to 3.14 (VCC~1499) and do not appear
different from the rest of the sample.  When normalized to stellar
mass, however, the young stellar populations have lower
$\mathcal{M}/L$ and thus much higher $T$ and \sm\ values.

% fig:snz_rp
%%%%%%%%%%%%%%%%%%%%%%%%%%%%%%%%%%%%%%%%%%%%%%%%%%%%%%%%%%%%%%%%%%%%%%%%%
\begin{figure}
\epsscale{1.22}
% snpf05.ps
\plotone{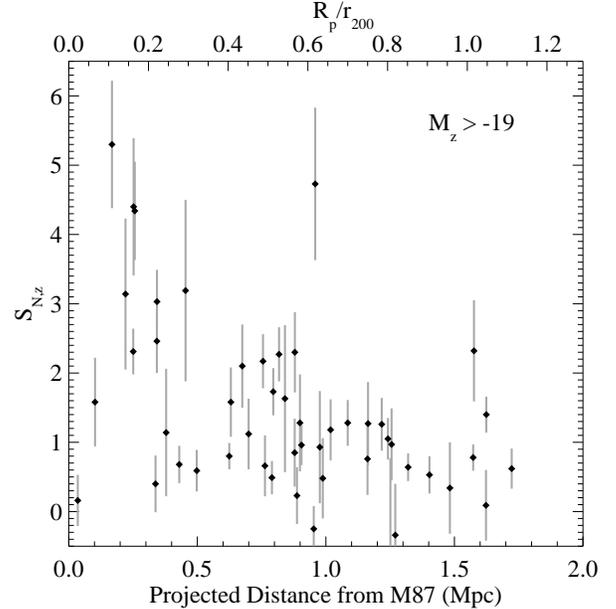}
\caption{$S_{N,z}$ vs projected clustercentric distance for low-luminosity
  galaxies ($M_z > -19$), excluding VCC~571 and 538 which are known to
  be $\sim6$--7~Mpc behind the Virgo core (Mei \etal 2007).  The
  center of the Virgo cluster is taken to be the location of the cD
  galaxy, M87, and the top axis shows $R_p/r_{200}$ where
  $r_{200}=1.55$~Mpc.  There is a notable trend in specific
  frequency with clustercentric radius.  All but one galaxy with
  $S_{N,z} > 2$ are within $R_p\sim 1$~Mpc of the cD.  
  \label{fig:snz_rp}}
\end{figure}
%%%%%%%%%%%%%%%%%%%%%%%%%%%%%%%%%%%%%%%%%%%%%%%%%%%%%%%%%%%%%%%%%%%%%%%%%

% table binrp
%%%%%%%%%%%%%%%%%%%%%%%%%%%%%%
% Table 6
\begin{center}
\begin{deluxetable}{cccrcc}
\tablewidth{0pt}
\tablecaption{Bins of $R_p$ for Galaxies with $M_z>-19$\label{table:binrp}}
\tablehead{
\colhead{$R_p$ range} & 
\colhead{$\langle R_p \rangle$} & 
\colhead{$S_{N,z}$} & 
\colhead{$T$} & 
\colhead{$S_{N,z,blue}$} & 
\colhead{$S_{N,z,red}$} \\
\colhead{Mpc} & \colhead{Mpc} & 
\colhead{} & \colhead{} & \colhead{} & \colhead{} 
}
\startdata
($0.00,0.15$) & $ 0.07$ & $ 0.47$ & $  3.9$ & $ 0.40$ & $ 0.07$ \\
($0.15,0.30$) & $ 0.23$ & $ 3.46$ & $ 33.3$ & $ 3.17$ & $ 0.29$ \\
($0.30,0.50$) & $ 0.40$ & $ 1.58$ & $ 15.7$ & $ 1.34$ & $ 0.23$ \\
($0.50,1.00$) & $ 0.83$ & $ 1.38$ & $ 13.0$ & $ 1.14$ & $ 0.23$ \\
($1.00,1.50$) & $ 1.24$ & $ 0.70$ & $  8.4$ & $ 0.59$ & $ 0.11$ \\
($1.50,2.00$) & $ 1.62$ & $ 1.05$ & $ 11.6$ & $ 0.99$ & $ 0.06$ \\
\enddata
\end{deluxetable}

\end{center}
%%%%%%%%%%%%%%%%%%%%%%%%%%%%%%

\subsection{$S_{N,z}$ and Environment}

One of the more intriguing questions presented by plots such as
Figure~\ref{fig:snz_mz} is the nature of \snz\ in the low-luminosity
galaxies ($M_z > -19$), which we will loosely refer to as ``dwarfs'',
although some of the galaxies in our sample are on the more massive
end of the spectrum of dwarf galaxies.  Our relatively small errors
show that the large spread in specific frequency for dwarf
galaxies is not simply due to the expected observational
scatter. Given our errors, the observed distribution of \snz\ is 2.4
times broader than we would expect if all dwarfs had a single \snz.
This implies that there is at least one other parameter besides galaxy
mass that governs the GC formation efficiency.

% fig:snz_r3d
%%%%%%%%%%%%%%%%%%%%%%%%%%%%%%%%%%%%%%%%%%%%%%%%%%%%%%%%%%%%%%%%%%%%%%%%%
\begin{figure}
\epsscale{1.22}
% snpf06.ps
\plotone{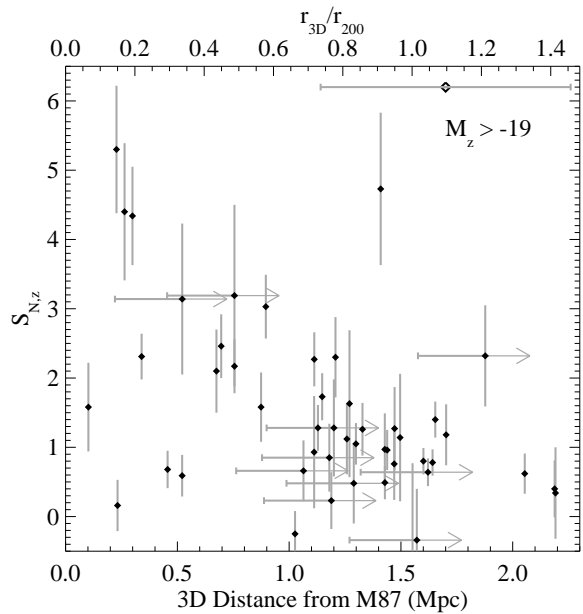}
\caption{$S_N$ versus three dimensional clustercentric distance for
  low-luminosity galaxies ($M_z > -19$).  The top axis shows
  $R_{3D}/r_{200}$ where $r_{200}=1.55$~Mpc. The error bar at top right
  shows the mean distance error.  Points with arrows do not have SBF
  distances so we have used their projected distances and added
  0.3~Mpc (the median $(R_{3D}-R_p)$ for the rest of the sample).
  The left error bar is at their projected radii, and are 
  thus lower limits on their 3-d radii.
  \label{fig:snz_r3d}}
\end{figure}
%%%%%%%%%%%%%%%%%%%%%%%%%%%%%%%%%%%%%%%%%%%%%%%%%%%%%%%%%%%%%%%%%%%%%%%%%

In this section, we investigate the relationship between specific
frequency and the galaxy's environment---specifically, its distance
from the cluster center, which is taken to be the location of the cD
galaxy, M87 (Binggeli \etal 1985).  Environment clearly plays a role in
galaxy evolution, as is evidenced by the morphology-density relation,
and this can be through gravitational and hydrodynamic processes such
as tidal and ram pressure stripping, or through initial conditions
where halos in denser regions collapse earlier.  Dwarf galaxies, being
the most vulnerable, are more likely to express the
effects of their environment.

Figure~\ref{fig:snz_rp} shows \snz\ for dwarf galaxies against their
projected distance from the cluster center ($R_p$), ignoring VCC 571
and 538 which are known from their SBF distances to lie 6--7~Mpc
behind the cluster core (Mei \etal 2007).  There is a remarkably
clear correlation in which dwarfs closer to M87 have higher \snz.  Of
the 14 galaxies with $S_{N,z}>2$, 13 lie within a projected radius of
1~Mpc (the exception being VCC~21, a possible dE/dI transition
galaxy).  We note that this is roughly half of $r_{200}$ for
the Virgo~A subcluster centered on M87 (McLaughlin 1999b; \cote
\etal 2001), where $r_{200}=1.55$~Mpc is the radius at which the mass density
of M87 is 200 times the critical density.  This value has been recalculated for
$D_{M87}=16.5$~Mpc using the McLaughlin (1999a) model (D. McLaughlin,
priv.\ comm.).  When we plot $S_{N,z}$ against three-dimensional
clustercentric distance (Figure~\ref{fig:snz_r3d}), we see the
same trend although the error in the line-of-sight distance
introduces more scatter.  Tables~\ref{table:binrp} and
\ref{table:binr3d} list the values for the mean trends in these Figures.

% Table binr3d
%%%%%%%%%%%%%%%%%%%%%%%%%%%%%%
% Table 7
%\begin{center}
%\input{sntable_binr3d.tex}
\begin{deluxetable}{cccrcc}
\tablewidth{0pt}
\tablecaption{Bins of $R_{3d}$ for Galaxies with $M_z>-19$\label{table:binr3d}}
\tablehead{
\colhead{$R_{3d}$ range} & 
\colhead{$\langle R_{3d} \rangle$} & 
\colhead{$S_{N,z}$} & 
\colhead{$T$} & 
\colhead{$S_{N,z,blue}$} & 
\colhead{$S_{N,z,red}$} \\
\colhead{Mpc} & \colhead{Mpc} & 
\colhead{} & \colhead{} & \colhead{} & \colhead{} 
}
\startdata
($0.00,0.25$) & $ 0.19$ & $ 1.38$ & $ 11.3$ & $ 1.19$ & $ 0.18$ \\
($0.25,0.50$) & $ 0.34$ & $ 2.39$ & $ 23.9$ & $ 2.19$ & $ 0.20$ \\
($0.50,1.00$) & $ 0.71$ & $ 2.13$ & $ 20.8$ & $ 1.87$ & $ 0.26$ \\
($1.00,1.50$) & $ 1.27$ & $ 1.27$ & $ 13.5$ & $ 1.05$ & $ 0.22$ \\
($1.50,2.00$) & $ 1.65$ & $ 0.78$ & $  7.7$ & $ 0.68$ & $ 0.10$ \\
($2.00,2.50$) & $ 2.14$ & $ 0.50$ & $  5.2$ & $ 0.47$ & $ 0.03$ \\
\enddata
\end{deluxetable}

%\end{center}
%%%%%%%%%%%%%%%%%%%%%%%%%%%%%%

\begin{figure}
\epsscale{1.0}
% dE_hisn_gray.eps
\plotone{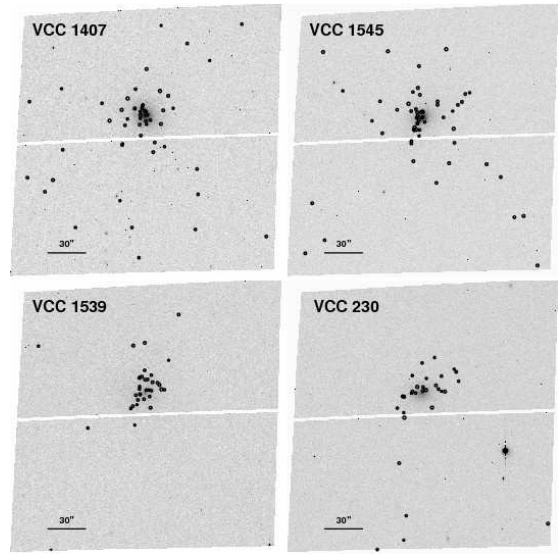}
\caption{Spatial distributions of GC candidates (circles) in 4 dEs with
  the highest specific frequencies (VCC~1407, 1545, 1539, and 230).  
  In all cases, there is evidence of a rich GC
  system that is centrally clustered around the galaxy, showing that
  their elevated $S_N$ is not due to an enhanced level of interloping
  GCs from the cD galaxy or an intracluster population.  These
  galaxies have $\sim5\times$ more GCs than ``normal'' galaxies with
  \snz$=1$.  The images show the entire ACS field of view, and
  the scale bar in the lower left of each image has a
  length of $30\arcsec$.  The blank diagonal strip is the gap
  between the two ACS CCDs.  Catalogs have been statistically cleaned
  using expected contamination determined from control fields.
  \label{fig:dEhisn}}
\end{figure}

The M87 globular cluster system is one of the most extreme in the
local supercluster and it is possible that some of the M87 halo
GCs, or a population of intracluster GCs, 
are contaminating the GC systems of the nearby dwarfs.  In fact,
the five galaxies closest to the cD, VCC 1297, 1327, 1279, 1250, and 1185
have been treated differently in our analyses in this paper because it
is obvious that some to all the GCs detected are part of the M87 halo.
One of the 
telltale signatures of this is whether the GCs are uniformly distributed
across the ACS/WFC image rather than centrally concentrated around the
targeted galaxy.  Figure~\ref{fig:dEhisn} shows the locations of GC
candidates around the 4 dEs with the highest \snz.  In all cases, the
star clusters are concentrated toward the center of the galaxy
and not in a uniform spatial distribution.  An extension of the M87
GC radial density profile using a \sersic\ model fit to the ACSVCS
data and the ground-based data of McLaughlin (1999) (see
Figure~\ref{fig:radial1316})
predicts only 1--3 M87 GCs over the entire ACS/WFC
field of view at the distances of VCC~1407, 1545, and 1539.  The GC radial
profile from Tamura \etal (2006) gives slightly higher values, with an
expected $9^{+3}_{-6}$ M87 GCs in the ACS field at the projected
distance of VCC~1407, a dE which has 50~GCs.  Given the difficulties
in measuring the M87 GC density from ground-based data, especially
given uncertainties in background subtraction, these independent
estimates are consistent.  Extrapolations of a de~Vaucouleurs profile
fit to the Tamura \etal data predict $\sim2$ M87 GCs in the ACS fields
for VCC~1545 and 1539.  We also have a parallel WFPC2 observation at a
distance of 40\arcmin\ from M87 in which we find a number of GCs
consistent with zero.  Finally, as these galaxies have 31--54 GCs
each, this contamination from M87 GCs is expected to be {\it at most} 6--18\%.

% fig:dEhisn
%%%%%%%%%%%%%%%%%%%%%%%%%%%%%%%%%%%%%%%%%%%%%%%%%%%%%%%%%%%%%%%%%%%%%%%%%
%%%%%%%%%%%%%%%%%%%%%%%%%%%%%%%%%%%%%%%%%%%%%%%%%%%%%%%%%%%%%%%%%%%%%%%%%

% fig:spatial_snz
%%%%%%%%%%%%%%%%%%%%%%%%%%%%%%%%%%%%%%%%%%%%%%%%%%%%%%%%%%%%%%%%%%%%%%%%%
\begin{figure}
\epsscale{1.22}
% snpf07.ps
\plotone{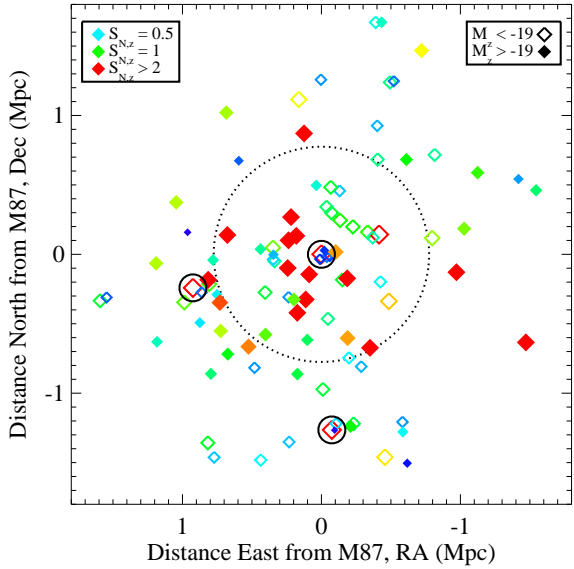}
\caption{Spatial distribution of ACSVCS galaxies color coded by
  $S_{N,z}$.  Specific frequency increases as colors
  change from blue to red.  Filled points are dwarfs ($M_z>-19$) 
  and open points are the more luminous galaxies.  The three large
  solid circles represent the 
  three massive ellipticals in the Virgo cluster: the cD galaxy M87
  center, M49 bottom, and M60
  (VCC1978/N4649) left.  The large dotted circle represents
  $r_{200}/2=775$~kpc for the M87/Virgo~A subcluster.  The high
  $S_{N,z}$ galaxies are preferentially around the cD,
  while no enhancement appears around the cluster's most
  luminous galaxy M49.  Dwarfs immediately in the vicinity of the
  giants have low $S_{N,z}$ and may be tidally stripped.
  \label{fig:spatial_snz}}
\end{figure}
%%%%%%%%%%%%%%%%%%%%%%%%%%%%%%%%%%%%%%%%%%%%%%%%%%%%%%%%%%%%%%%%%%%%%%%%%

The general trend of dEs having higher \snz\ is reversed for the two
dwarfs (VCC 1297 and 1185) with $R_p\lesssim100$~kpc.  These galaxies
have low intrinsic numbers of GCs and we hypothesize that they have
had their GCs stripped from them by M87.  Both VCC~1297 and the
more luminous VCC~1327 are within $R_p\approx40$~kpc, and have GC
numbers consistent with zero.  The 
galaxies VCC~1192 and 1199, which are at comparable distances from
M49, also have undetectable intrinsic GC systems.

Figure~\ref{fig:spatial_snz} shows the positions of the ACSVCS
galaxies on the sky, color and size coded by \snz, with open points
representing giants and filled points for dwarfs.  The high-\snz\
dwarfs are clustered around M87, and there is no similar effect around
the brightest galaxy in the cluster, M49, nor around
VCC~1978 (M60/N4649).  Interesting exceptions are the aforementioned galaxies
within the immediate vicinities of the giants ($R\lesssim40$~kpc),
which have few or no GCs and may have been tidally stripped of their
GC systems.  The specific frequencies of the more massive
galaxies are generally uniform, as shown in
Figure~\ref{fig:snz_mz}, and thus are relatively unaffected by their
distance from M87.  

% fig:snz_mz_rp
%%%%%%%%%%%%%%%%%%%%%%%%%%%%%%%%%%%%%%%%%%%%%%%%%%%%%%%%%%%%%%%%%%%%%%%%%
\begin{figure}
\epsscale{1.22}
% \plotone{snpf08.pdf}
\plotone{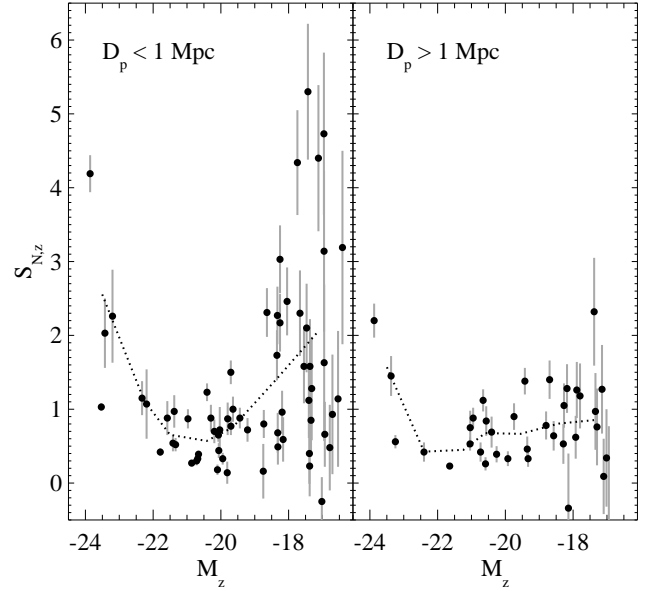}
\caption{$S_{N,z}$ versus galaxy $M_z$, as in Figure~\ref{fig:snz_mz},
  except divided by clustercentric distance.  Galaxies within 1~Mpc
  of the cluster center exhibit the full range of $S_{N,z}$, but nearly
  all the dwarfs outside of 1~Mpc have low specific frequencies 
  ($S_{N,z} \leq 1.5$).  The only exception is a dE/dI transition
  object (VCC~21).
  \label{fig:snz_mz_rp}}
\end{figure}
%%%%%%%%%%%%%%%%%%%%%%%%%%%%%%%%%%%%%%%%%%%%%%%%%%%%%%%%%%%%%%%%%%%%%%%%%

% fig:snz_rp_br

%%%%%%%%%%%%%%%%%%%%%%%%%%%%%%%%%%%%%%%%%%%%%%%%%%%%%%%%%%%%%%%%%%%%%%%%%
\begin{figure}
\epsscale{1.22}
%\plotone{snpf10.pdf}
\plotone{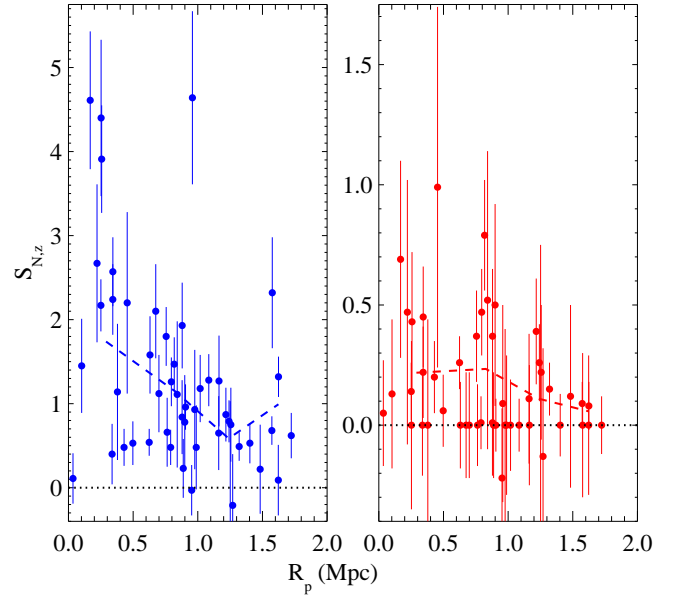}
\caption{$S_{N,z}$ versus projected clustercentric distance for
  blue GCs (left) and red GCs (right) in low-luminosity galaxies ($M_z
  > -19$).  Given that most GCs in dwarfs are metal-poor, it is not
  surprising that the trend for blue GCs mirrors that in
  Figure~\ref{fig:snz_rp}. The $S_{N,z}$ for the red GCs may also show
  a tendency to be enhanced at small clustercentric radii, but the
  numbers involved are small.
  \label{fig:snz_rp_br}}
\end{figure}
%%%%%%%%%%%%%%%%%%%%%%%%%%%%%%%%%%%%%%%%%%%%%%%%%%%%%%%%%%%%%%%%%%%%%%%%%

We can re-plot Figure~\ref{fig:snz_mz}, except divided into two
samples, one inside and one outside a projected clustercentric distance of
1~Mpc.  Figure~\ref{fig:snz_mz_rp} shows the difference between the
two samples.  Galaxies within 1~Mpc mirror the trends seen in
Figure~\ref{fig:snz_mz}, with both giant and dwarf galaxies having
high \snz.  However, the sample of galaxies outside of 1~Mpc
entirely lacks high-\snz\ dwarfs.  The division is not perfect, as
there are still low-\snz\ dwarfs within 1~Mpc (some of which 
have larger 3-D distances), but the lack of high-\snz\ dwarfs on the
outskirts of the cluster or around the other massive ellipticals is
very clear.
The one possible exception to the general trend is VCC~230, which lies
at $R_p = 0.96$~Mpc and has an \snz\ higher by a factor 2 than other
galaxies at comparable distances.

The GCs in dwarf galaxies are predominantly from the blue (metal-poor)
subpopulation, but many in the ACSVCS sample also possess small
numbers of red GCs.  Are both of these subpopulations affected by
environment?  In Figure~\ref{fig:snz_rp_br}, we plot \snz\ in dwarf
galaxies against $R_p$ for the blue and red GCs separately.  As
expected, the blue GCs mirror the overall trends since they dominate
the GC budget.  For the red GCs, although the errors are large, there
is a slight hint that dwarfs with smaller clustercentric
distances tend to have elevated red GC specific frequencies.  The mean
$S_{N,z,red}$ of dwarfs within 1~Mpc is 0.2, compared to 0.1 for those
outside.

\begin{figure}
\epsscale{1.22}
%\plotone{snpf11.pdf}
\plotone{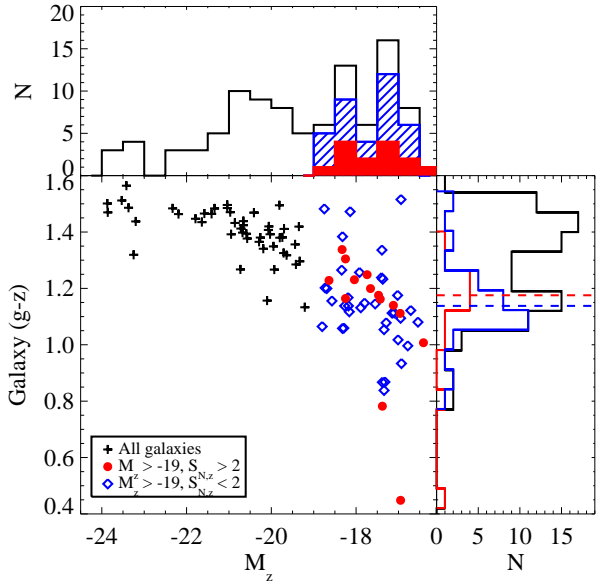}
\caption{Color-magnitude properties of dwarfs with high and low
  $S_{N,z}$.  We show \gz\ vs. $M_z$ and accompanying histograms for the ACSVCS
  sample of galaxies.  Galaxies with $M_z > -19$ are separated at
  $S_{N,z} = 2$.  The luminosity distribution of the two subsets are
  nearly identical, while the high-$S_{N,z}$ dwarfs have a median color
  that is formally redder than that of the low-$S_{N,z}$ dwarfs
  (shown by dashed lines in right hand histogram).
  \label{fig:cmd}}
\end{figure}

Do the dwarfs with higher \snz\ exhibit any intrinsic characteristics that
differentiate them from the others?  Figure~\ref{fig:cmd} shows the
\gz-$M_z$ color-magnitude diagram for the ACSVCS galaxies, singling
out the dwarfs with both low and high \snz, and dividing the sample at
\snz$=2$.  The top histogram plots the $M_z$ distribution of the
total, high-\snz, and low-\snz\ samples and shows how the luminosity
distributions of the two groups are nearly identical.  On the
right-hand histogram, we plot the color distribution with the median
colors marked by the dashed lines.  Although the high-\snz\ dwarfs
formally have redder colors, $\Delta (g-z)=0.05$~mag, the difference
is not significant.  We have also investigated possible differences in
their structural properties (\sersic\ $n$, $r_e$) as well as in the
colors of their globular clusters, and for none of these properties are
high-\snz\ dwarfs significantly different from low-\snz\ dwarfs.

% fig:cmd
%%%%%%%%%%%%%%%%%%%%%%%%%%%%%%%%%%%%%%%%%%%%%%%%%%%%%%%%%%%%%%%%%%%%%%%%%
%%%%%%%%%%%%%%%%%%%%%%%%%%%%%%%%%%%%%%%%%%%%%%%%%%%%%%%%%%%%%%%%%%%%%%%%%

\section{Discussion}

\subsection{Mass Dependence of the GC Mass Fraction}
Globular cluster specific frequency and its related quantities are
clearly dependent on the mass of the host galaxy.  However, GC
fraction does {\it not} 
vary monotonically with galaxy mass, making it
unlike most other properties measured for the ACSVCS galaxies---e.g.,
colors, structural
parameters (Ferrarese \etal 2006a), core deficit and excess (\cote
\etal 2007), GC mean colors (Peng \etal 2006a), GC mean sizes (\jordan
\etal 2005), and GC luminosity function parameters (\jordan \etal 2007).
These papers (particularly Papers VI and VIII) show that there is no
clear distinction between ``dwarfs'' and ``giants'', and that early-type galaxy
properties have a smooth, monotonic dependence on galaxy mass.  For GC
fractions, however, a transition does appear to exist.

It is possible that GC fraction is telling us more about the
variable formation efficiency of stars in the field than it is about
the formation of GCs themselves.  Blakeslee \etal (1997), Blakeslee (1999) and 
McLaughlin (1999a) have shown that, at least on the high mass end, the
number of GCs appears to scale directly to the baryonic, or even the total
mass.  Kravtsov and Gnedin (2005) show a similar relation in their
simulations of GC formation.  High $S_N$ values may then be
interpreted as a lower fraction of baryons that form field stars.  

The mismatch between total mass and stellar mass across the galaxy
mass function is now
well known in the study of galaxy formation.  Observationally,
dynamical mass estimates of galaxies across a
wide range of mass require high mass-to-light ratios for 
both high (\cote \etal 2001, 2003) and low mass galaxies (Mateo 1998),
with low $\mathcal{M}/L$ measured for galaxies around $L^*$ (Romanowsky
\etal 2003; Peng \etal 2004; Napolitano \etal 2005).  Mass-to-light
ratios for ensembles of galaxies derived from weak lensing have also produced
similar trends in $\mathcal{M}/L$ (e.g., Guzik \& Seljak 2002;
Hoekstra \etal 2005; Mandelbaum \etal 2006).
One can also take a statistical approach and match 
the expected dark matter halo mass distribution from simulations
and the observed galaxy luminosity function
(Berlind \& Weinberg 2002; van den Bosch \etal 2003, 2007; Vale \&
Ostriker 2007), which does not
require us to know the details of galaxy formation.  These halo
occupation studies
infer a maximum conversion efficiency of baryons to stars at halo mass
$\mathcal{M}_h\sim2\times10^{11} \mathcal{M}_\odot$, or a stellar mass of 
$\mathcal{M}_{G\star}\sim7.5\times 10^9 \mathcal{M}_\odot$.  
In simulations of galaxy formation, this $\mathcal{M}_h/L$--$L$
relation (or alternatively, $\mathcal{M}_h/\mathcal{M}_{G\star}$--$\mathcal{M}_{G\star}$) requires
various heating mechanisms (photoionization, stellar
and AGN feedback) to prevent the formation of more stars in galaxies
than are observed (Benson \etal 2002; Croton \etal 2006).  

What happens if we make the simple assumption that
$N_{GC}\propto\mathcal{M}_h$, or equivalently 
$S_\mathcal{M}\propto \mathcal{M}_h/\mathcal{M}_{G\star}$?
Ideally, we would know the total dynamical mass of each galaxy in our
sample.  Without this information, we can apply 
average halo mass-to-light ratios derived from the halo occupation
studies, in this case using the parameterization of van den Bosch \etal
(2007) and a WMAP3 cosmology (Spergel \etal 2007) .  We transform
their $L_B$ to $\mathcal{M}_{G\star}$ using a 
range of $\mathcal{M}_{G\star}/L_B$, monotonically increasing from 1.8
to 4.1 as a
function of galaxy luminosity, derived from a polynomial fit to the
inferred $\mathcal{M}_{G\star}/L_B$ and measured $L_z$ of the ACSVCS galaxies. 
This results in a simple prediction
for the behavior of $S_\mathcal{M}$ as function of galaxy stellar mass.  

% fig:vdb
%%%%%%%%%%%%%%%%%%%%%%%%%%%%%%%%%%%%%%%%%%%%%%%%%%%%%%%%%%%%%%%%%%%%%%%%%
\begin{figure}
\epsscale{1.22}
%\plotone{snpf15.pdf}
\plotone{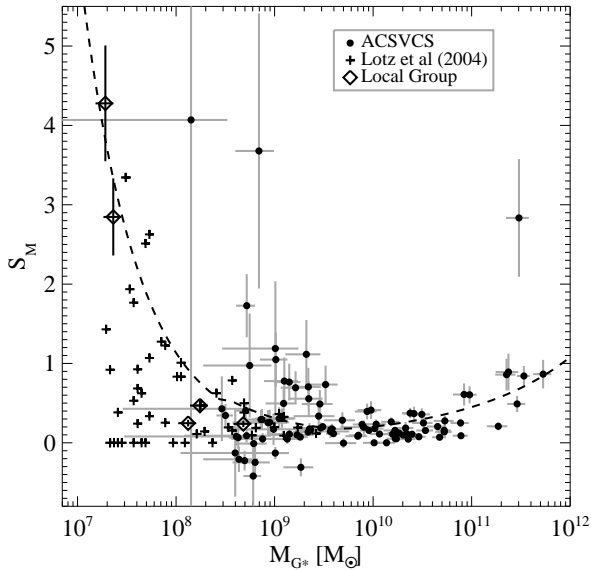}
\caption{Mass fraction of GCs, $S_\mathcal{M}$, versus galaxy stellar
  mass $\mathcal{M}_{G\star}$ for ACSVCS galaxies (filled dots), Virgo and Fornax
  dEs from Lotz \etal (2004) (crosses), and Local Group early-type
  dwarfs (diamonds). We overplot the expected behavior of $S_\mathcal{M}$
  assuming that the mass in GCs follows the total halo mass as
  inferred from the $\mathcal{M}_h/L$--$\mathcal{M}_h$ relation of
  van den Bosch (2007).  This
  assumption can explain the rise in $S_\mathcal{M}$ for luminous
  galaxies, although the cD galaxy M87 is off the relation because its
  stellar mass is not representative of its halo mass.
  At the low mass end, we also expect a sharp
  rise in $S_\mathcal{M}$ which is traced by some cluster and Local Group dEs.
  At masses of $\sim10^9 \mathcal{M}_\odot$, however,
  many dEs in the ACSVCS sample have higher $S_\mathcal{M}$ than might
  be expected.  The
  high-$S_\mathcal{M}$ and low-$S_\mathcal{M}$ dwarfs cannot be
  explained simultaneously if GC mass fraction is solely a function of
  galaxy stellar mass. 
  \label{fig:vdb}}
\end{figure}
%%%%%%%%%%%%%%%%%%%%%%%%%%%%%%%%%%%%%%%%%%%%%%%%%%%%%%%%%%%%%%%%%%%%%%%%%

In Figure~\ref{fig:vdb}, we plot the ACSVCS data for $S_\mathcal{M}$
against $\mathcal{M}_{G\star}$ with errors for both $S_\mathcal{M}$ and
$\mathcal{M}_{G\star}$.  We extend the data to lower 
mass galaxies by including the Virgo and Fornax dEs and five Local Group
dEs from Lotz \etal (2004).  For these latter data, we assume 
$\mathcal{M}_{G\star}/L_B\sim2.5$ (an extension of the mass-to-light
ratios fitted to the ACSVCS dwarfs) and 
$\langle \mathcal{M}_{GC} \rangle = 1.6\times10^5 \mathcal{M}_\odot$,
the mean GC mass in our ACSVCS dwarfs.  The Virgo 
and Fornax dE points do not include errors in their stellar masses, and for
the Local Group dEs, we have assumed an error in their luminosities of 0.25~mag.
Overplotted on the data is the predicted behavior of
$S_\mathcal{M}$ assuming $S_\mathcal{M}\propto
\mathcal{M}_h/\mathcal{M}_{G\star}$. The only truly free parameter in 
the plot is the vertical normalization, which is chosen to match both
the massive elliptical M49 and the Local Group dwarf spheroidal
galaxies Fornax and Sagittarius. 
The assumption that the GC mass fraction is proportional to the inverse of
the stellar mass fraction, $S_\mathcal{M}\propto
\mathcal{M}_h/\mathcal{M}_{G\star}$ predicts that 
$S_\mathcal{M}$ should be high at both high and low masses.  At high stellar
masses, the rise reflects the increasing $\mathcal{M}/L$ of massive elliptical
galaxies.  One galaxy, M87, is well off the predicted relation, but
this is because its GC population is more likely representative of the
entire Virgo~A subcluster within which M87 resides (McLaughlin 1999a).
The sharp cutoff in the galaxy 
luminosity function and the nearly universal luminosity of brightest
cluster galaxies means that $S_\mathcal{M}$ can vary widely
depending on whether 
the luminosity is associated with a galaxy halo or the cluster halo
(Blakeslee 1999, \jordan \etal 2004a).  At low masses, $S_\mathcal{M}$ is also
predicted to rise steeply, but 
this is at masses {\it lower} than those in our sample.  
Many of the ACSVCS dwarfs around $\mathcal{M}_{G\star}\sim10^9
\mathcal{M}_\odot$ have $S_\mathcal{M}$ 
higher than we would infer by a factor of 4--8, and these are the same
ones that are in the inner regions of the Virgo cluster.
At masses below $\mathcal{M}_{G\star}\sim10^8 \mathcal{M}_\odot$, the dEs
from Lotz \etal 
(2004) are either consistent with or below the expected rise, although the
observational error is quite large (and not shown for clarity).  
So, while the linear scaling of $\mathcal{M}_{GC}\propto \mathcal{M}_h$
does seem to be relevant to $S_\mathcal{M}$  across a large galaxy mass range,
the $S_\mathcal{M}$ for the ACSVCS dwarf galaxies in this 
diagram is not explained by a dependence on galaxy mass alone.  The
scatter in \sm\ for dwarfs (and also possibly for intermediate mass
galaxies) also cannot be explained by the intrinsic scatter in galaxy
$\mathcal{M}/L$ (from van den Bosch \etal 2007).
The scatter is even harder to explain for $S_L$ where the errors are
considerably smaller.

Both Forbes (2005) and Bekki, Yahagi \& Forbes (2006)
have also attempted to explain the increase in \sn\ for
dwarf galaxies by assuming a relation between GCs and galaxy mass.
These attempts followed similar arguments by Durrell \etal (1996) and
McLaughlin (1999a).
Forbes (2005) used the data for ACSVCS galaxies presented in Peng
\etal (2006a, Paper~IX) to compare the approximate $S_N$ of blue GCs in
dwarf galaxies with the scaling relation of Dekel \&
Birnboim (2006), where $\mathcal{M}/L_V\propto \mathcal{M}^{-\alpha}$.
Forbes (2005) use $\alpha = 2/3$ from Dekel \& Birnboim (2006) for
masses below the critical galaxy stellar mass of $3\times10^{10}
\mathcal{M}_\odot$ where galactic winds should be more efficient at
blowing gas and metals from the galaxy  
(Kauffmann \etal 2003; Tremonti \etal 2004).  They also assume a
constant number of GCs as a function of galaxy mass, giving 
$S_N\propto \mathcal{M}^{-5/3}$. Although this formulation predicts a rapid rise
in \sn\ and $S_\mathcal{M}$ for fainter galaxies, it is much too steep
to accommodate the data in the way that they present it.  The
normalization of this relation as 
presented in Forbes (2005) is not constrained, but because the
relation is steep, the normalization is critical for comparing to
the data.  If we fix to the mean $S_\mathcal{M}$ at the critical mass,
the relation overpredicts $S_\mathcal{M}$ for all the dwarfs.  This is
due to the fact that dwarf galaxies do not have a constant number of GCs.
Bekki \etal (2006) simulate GC
systems and test various values for $\alpha$, concluding that the
observations are not consistent with $\alpha=0$ (a constant
$\mathcal{M}/L$), and it would be interesting to determine $\alpha$
again with this new data set.  Whatever the best value of $\alpha$,
however, a simple relation where the specific frequency is determined
solely by the present day stellar mass of a galaxy is unable to describe
all of the data on the low mass end. 

\subsection{Environment and Biased GC Formation}
\subsubsection{Cluster Dwarfs in the Millennium Simulation}

As we showed in Figures~\ref{fig:snz_rp}-\ref{fig:snz_rp_br}, GC
specific frequency is not purely a function of galaxy mass, but can
also depend on environment.  We do not see environmental effects on
\sn\ for massive galaxies with the exception of M87, whose location defines
the cluster center.  For intermediate-mass galaxies in particular,
\sn\ does not appear to depend on environment. For dwarf galaxies,
however, specific frequency
is a strong function of proximity to the center of the Virgo
Cluster. Nearly all dwarfs with high \sn\ are within $R_p=1$~Mpc.  

One of the main
problems with building up the blue GC sub-population in the halos of massive
galaxies is that the ratio of metal-poor GCs to metal-poor field stars
in galaxies is very high (Harris \& Harris 2002).  In other words,
if halos are built up through the accretion of dwarf galaxies or
dwarf-mass fragments then their specific frequencies need to be high
in order to keep the main body of the galaxy metal-rich.  
These central dwarfs with high \sn\ may then hold the key to the
formation of GC systems, as they are likely to be most similar to
the progenitors of any dwarfs that have since merged into the halo of M87.
Figure~\ref{fig:snz_mz_br} shows that the $S_{N,z,blue}$ values for the
innermost dwarfs are as high or higher than that of M87 or any of the
other giant ellipticals.  
These dwarfs could be the survivors of an accreted population
of protogalaxies that may have had even higher GC mass fractions and redder
GCs.  In fact, the two galaxies closest
to M87---VCC~1327 
and 1297---have GC systems that are either entirely stripped or
undetectable against the M87 GC system, and their $g-z$ colors are
much redder than those of other galaxies with the same luminosity (see the
red outliers in Figure~\ref{fig:cmd}).

What is the cause of the enhanced GC fractions in these central
dwarfs?  Are they more efficient at producing GCs for their mass,
less efficient at forming field stars, or better at keeping the GCs
they form?  As discussed earlier, one
possible way for GC production to be biased towards the central dense
regions of the galaxy cluster is for these galaxies to form a larger
fraction of their stars earlier and at a higher star formation rate
density. Although we do not currently have the ability to determine
the detailed star formation histories of the ACSVCS galaxies at these
early times, we can use simulations to study global trends in
the star formation histories of low mass early-type galaxies in the
cluster environment.  By using simulations, we can test the
consistency of the hypothesis that the central dwarfs must be both
older and have higher SFR densities.

% fig:meanage
%%%%%%%%%%%%%%%%%%%%%%%%%%%%%%%%%%%%%%%%%%%%%%%%%%%%%%%%%%%%%%%%%%%%%%%%%
\begin{figure}
\epsscale{1.22}
%\plotone{milfig20.pdf}
\plotone{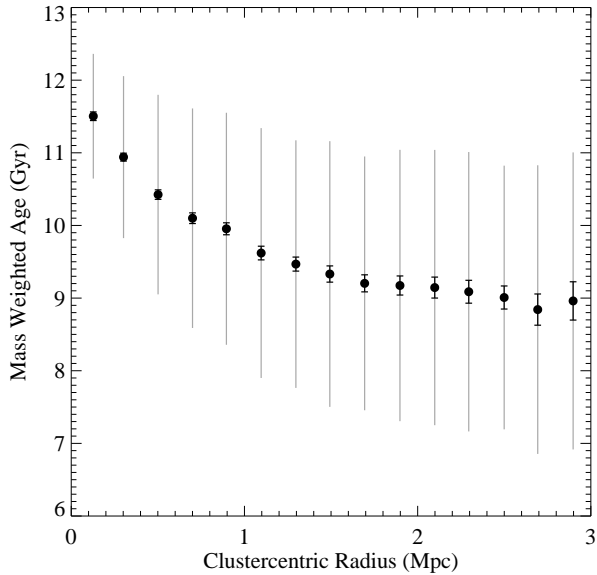}
\caption{The mass-weighted stellar age of low mass ($M_z>-19$)
  early-type cluster
  galaxies in the Millennium Simulation.  Galaxies in the inner regions
  of the galaxy cluster have older mean ages.  The black error bars
  with hats depict the $2\sigma$ error in the mean, while the gray
  error bars indicate
  the $1\sigma$ width of the age distribution in each bin.
  \label{fig:meanage}}
\end{figure}
%%%%%%%%%%%%%%%%%%%%%%%%%%%%%%%%%%%%%%%%%%%%%%%%%%%%%%%%%%%%%%%%%%%%%%%%%

The simulation we use for comparison is the ``Millennium Simulation''
carried out by the Virgo Consortium (Springel \etal 2005) coupled with the
semi-analytic models presented by De~Lucia \etal (2006).  This simulation
consisted of $2160^3$ dark matter particles followed from $z=127$ to
the present day in a volume $500h^{-1}$~Mpc on a side.  The spatial
resolution is $5h^{-1}$~Mpc, and the simulation is essentially
complete for all galaxies with stellar mass greater than $3\times10^8
\mathcal{M}_\odot$.  This mass limit is comparable to the lowest mass
galaxies in the ACSVCS.  From this simulation, we selected a sample of low mass
early-type cluster galaxies and their progenitors.  We identified 126
massive galaxy clusters using the same criteria as in De~Lucia \etal
(2007), requiring a halo mass greater than $7\times10^{14} \mathcal{M}_\odot$
at $z=0$.  This mass limit also corresponds roughly to the 
mass of Virgo (B\"{o}hringer \etal 1994).  
Within each of these clusters, we then selected all
galaxies with $M_{z,sdss} > -19$ that have early-type morphologies,
matching the low luminosity sample we have in the ACSVCS.
Early-type galaxies were defined by the bulge-to-total luminosity
ratio with $\Delta M < 1.56$ ($\Delta M = M_{bulge} - M_{total}$),
using the empirical criteria of Simien \& de~Vaucouleurs (1986).
In addition, we selected only galaxies with $g_{sdss}-z_{sdss} > 0.5$
at $z=0$ in order to best match the colors of the ACSVCS galaxies.  In
total, we selected 15,506 simulated galaxies at $z=0$ and all of their
progenitors over 63 snapshots back to $z\sim12$.

In Figure~\ref{fig:meanage}, we show the mass-weighted age of
the stellar populations in the selected early-type dwarfs as a
function of distance from the center of the cluster's dark matter
halo.  The error bars show the $1\sigma$ width of the age distribution
in each bin of radius.  This figure clearly shows that the stellar age
of the simulated dwarfs have a strong dependence on clustercentric
radius, at least within a distance of 1.5~Mpc, with the central dwarfs
having mean ages of 11.5~Gyr decreasing to 9~Gyr at the cluster
outskirts.  The radius within which the age gradient is evident is
also the radius in Virgo within which we see elevated GC fractions.

% fig:sfr

%%%%%%%%%%%%%%%%%%%%%%%%%%%%%%%%%%%%%%%%%%%%%%%%%%%%%%%%%%%%%%%%%%%%%%%%%
\begin{figure}
\epsscale{1.22}
%\plotone{milfig21.pdf}
\plotone{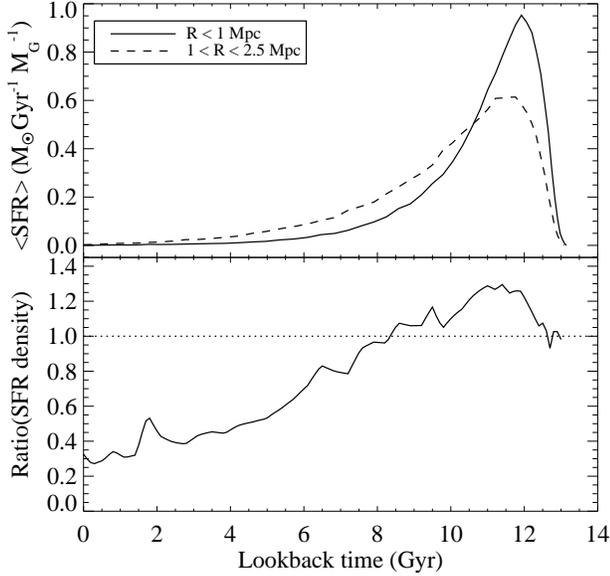}
\caption{(a, top) Average normalized star formation rate versus lookback 
  time for low mass
  early-type cluster galaxies in the Millennium simulation.  The solid
  line shows the average SFR history for ``central'' galaxies within 1~Mpc of the
  cluster center, and the dashed line shows ``outer'' galaxies.  Low
  mass early-type galaxies in the inner cluster regions formed their
  stars earlier and with higher peak star formation rate (SFR) than those
  in the outskirts.  Higher peak SFRs could result in higher globular
  cluster mass fractions. (b, bottom) Average ratio in SFR surface
  density between inner and outer 
  low mass early-type cluster galaxies as a function of lookback
  time.  Galaxies in the inner regions have a higher SFR surface
  density during the period of highest total star formation
  ($t_{lookback} > 8$~Gyr).  This results in central
  galaxies having a higher mass fraction in massive star clusters.
  \label{fig:sfr}}
\end{figure}
%%%%%%%%%%%%%%%%%%%%%%%%%%%%%%%%%%%%%%%%%%%%%%%%%%%%%%%%%%%%%%%%%%%%%%%%%

The central dwarfs in the simulation
may be older, but they also need to form their
stars more intensely (higher peak star formation rates and densities)
in order to have more of their stars in massive clusters.  
Figure~\ref{fig:sfr}a shows the
average normalized star formation rate of early-type cluster dwarfs as
a function of lookback time.  The star formation rate is normalized to
the final stellar mass of the galaxy.  The two curves represent SFR
histories for dwarfs divided into two bins of clustercentric radius
at 1~Mpc from the cluster center.  We note that these are average star
formation histories, and that for any individual galaxy, the bursts of
star formation are more intense, brief, and stochastic.  Combining
galaxies allows us to see that the central dwarfs
have a more peaked SFR at earlier times with a rapid falloff,
whereas the outer dwarfs not only have a lower peak but also more
star formation extending to later times.  

It has been suggested that massive
star clusters form preferentially in high-pressure environments
(Harris \& Pudritz 1994; Elmegreen \&
Efremov 1997; Ashman \& Zepf 2001), of which
SFR surface density is one possible indication.  
Figure~\ref{fig:sfr}b shows the ratio of SFR surface density in central dwarfs
to that in outer dwarfs. We calculate the SFR surface density using the SFR
and the disk radius, which the semi-analytic models calculate using the
analytic model of Mo, Mao \& White (1998). 
Not only is the peak star formation rate higher in central
dwarfs, but the {\it intensity} of star formation, as measured by the
SFR surface density, is also higher during the epoch when 
these galaxies are forming most of their stars.  At later times, the
SFR surface density in the outer dwarfs is higher on average than that in
central dwarfs, but this
is at much lower absolute star formation rates and SFR surface
densities.  If GCs are formed in the same events that produced the
bulk of these low mass galaxies, then we would expect that the
central dwarfs would retain a higher fraction of their stellar mass in
massive star clusters.

% fig:sfrd
%%%%%%%%%%%%%%%%%%%%%%%%%%%%%%%%%%%%%%%%%%%%%%%%%%%%%%%%%%%%%%%%%%%%%%%%%
\begin{figure}
\epsscale{1.22}
%\plotone{milfig22.pdf}
\plotone{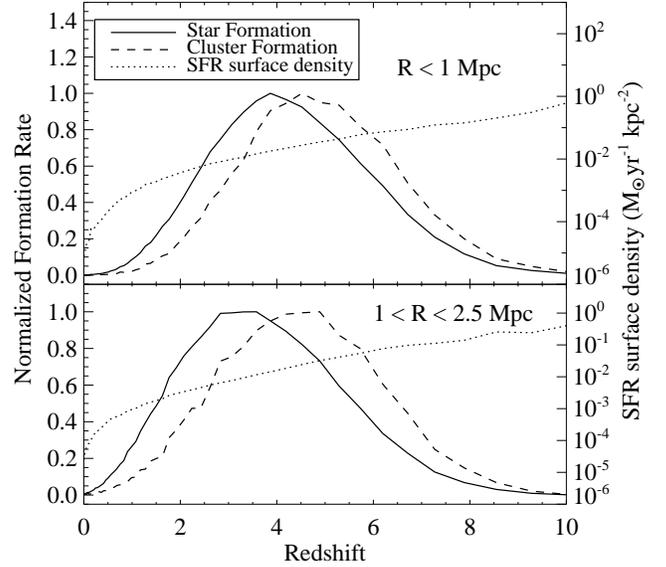}
\caption{For central (top) and outer (bottom) early-type dwarfs in the
  Millennium  Simulation, a comparison of the average star formation
  rate (solid), the average star formation rate surface density
  (dotted), and the inferred average cluster formation rate (dashed).
  The highest SFR surface densities occur at earlier times than the 
  highest SFRs because the disks within which stars form are smaller
  at high redshift.  If massive star clusters preferentially form at
  high SFR surface densities, then GCs in early-type dwarfs will, on
  average, have older ages and lower metallicities than the field stars.
  Formation rates are normalized to their maximum values.
  \label{fig:sfrd}}
\end{figure}
%%%%%%%%%%%%%%%%%%%%%%%%%%%%%%%%%%%%%%%%%%%%%%%%%%%%%%%%%%%%%%%%%%%%%%%%%

\subsubsection{Inferring Cluster Formation Histories}

We can quantify these effects more directly 
by inferring the cluster formation rate (CFR)
history, using an empirical relationship between
$S_\mathcal{M}$, the fraction of stellar mass formed in massive star
clusters, and the SFR surface density, $\Sigma_{SFR}$.  Fitting the
data of Larsen \& Richtler 
(2000), we adopt $S_\mathcal{M}\propto (\Sigma_{SFR})^{0.8}$, up to a
maximum efficiency of 100\%.  The
cluster formation rate then scales with a combination of the SFR and the SFR
surface density as $CFR \propto SFR\times(\Sigma_{SFR})^{0.8}$.  
The Larsen \& Richtler (2000) data
measured the fraction of luminosity in young massive clusters (YMCs),
which had ages between 10 and 500~Myr and more massive than
$\sim3\times10^4 \mathcal{M}_\odot$.  Given that most clusters will be destroyed
over a Hubble time, we need to estimate what fraction of these
clusters will survive.  We adopt the two-stage disruption law of
Whitmore, Chandar \& Fall (2007) where ``infant mortality'' causes
clusters to disrupt in constant numbers, $dN/d\tau\propto \tau^{-1}$,
for ages less than a few $\times 10^8$~Gyr (Fall, Chandar \& Whitmore
2005), and a constant mass loss over a Hubble time of
$\mu_{ev}=1.9\times10^4 \mathcal{M}_\odot Gyr^{-1}$ due to two-body relaxation
(Fall \& Zhang 2001; \jordan \etal 2007).  Starting with the data of
Larsen \& Richtler (2000), we estimate that $\sim1.8\%$ of their young
massive clusters survive to the present day.  Using this formalism, we
produce the corresponding cluster formation rate for central
early-type dwarfs, normalized to the final mass in GCs.

Figure~\ref{fig:sfrd} compares the
average SFR and SFR surface density as a function of redshift for the sample
of central dwarfs (top) and outer dwarfs (bottom).  While the SFR
peaks at $z=3.5-4$, $\Sigma_{SFR}$ 
is highest at $z=10$ and falls a factor of 35 by $z=4$.  Because
the formation rate of massive star clusters depends on both
quantities, their peak epoch of formation will be {\it earlier}
than that of the stars.  The result is that the CFR 
peaks at earlier redshift, $z\sim4.5$~to~5, than the total star formation
rate, a difference of $\sim350$--$500$~Myr.  
In all dwarfs, the dependence of massive star cluster formation on the SFR
surface density naturally produces GCs that are older and more
metal-poor than the stars.    

Depending on how massive star cluster formation scales with SFR and
SFR density, we also expect that the mean age difference between
the GCs and the field would be different in the central and outer
dwarfs. Because GCs in these dwarfs will mostly form at or before the
SFR peak, they will always be old, but the field stars can continue to form at
later times.  This
is evident when comparing the difference in lookback time for the
SFR peaks for the central and outer simulated dwarfs in Figure~\ref{fig:sfr}a
and the difference in their mean ages in Figure~\ref{fig:meanage}.
The outer dwarfs have mass-weighted mean ages that are younger by
$\sim2.5$~Gyr, but the times of their peak star formation rates differ by only
0.1~Gyr.  Figure~\ref{fig:sfrd} shows how the CFR
peak happens at similar redshifts for all dwarfs, but the peak in star formation
happens later in outer dwarfs.
One prediction is then that, on average, the
difference between the mean age of the metal-poor GCs and the mean age of the
stars should be larger in dwarf galaxies with lower GC mass fractions
(i.e., the mean age of the galaxy is proportional to $S_\mathcal{M}$).

\begin{figure}
\epsscale{1.22}
%\plotone{milsim_sm_plot.pdf}
\plotone{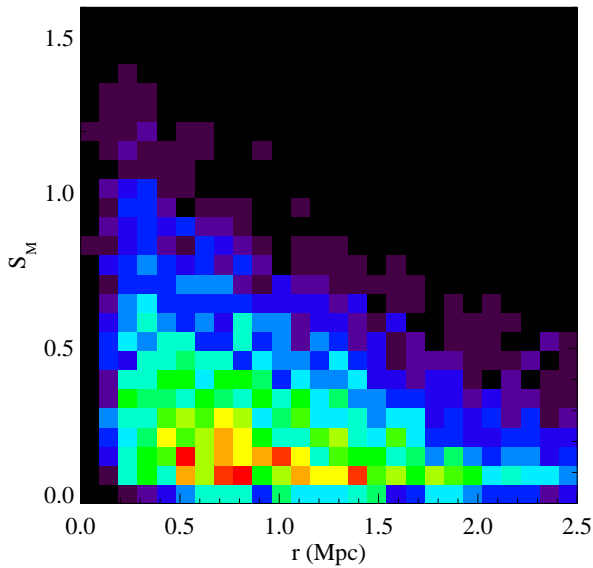}
\caption{Simulated $S_M$ versus clustercentric radius for early-type
  dwarfs in the Millennium simulation.  Colors from purple to red
  (or black to white)
  symbolize lower to higher numbers of simulated dwarfs in a particular cell in
  this diagram.  Like in the observations (Figures~\ref{fig:snz_rp}
  and \ref{fig:snz_r3d}), there is a tendency for
  dwarf galaxies with higher $S_M$ to be at the center of the cluster.
  \label{fig:sm_sim}}
\end{figure}

Lastly, we test whether this framework can produce higher GC mass
fraction in the central dwarfs. Figure~\ref{fig:sm_sim} shows \sm\
against clustercentric radius for simulated dwarfs.
The absolute scale for \sm\ is highly sensitive to the assumed
destruction but is not important for our purposes.  It is the
relative comparison between central and outer simulated dwarfs that is
of interest, and the high-\sm\ simulated dwarfs do appear
preferentially in the central regions, similar to what the data shows
in Figures~\ref{fig:snz_rp} and \ref{fig:snz_r3d}.  

The simulations do
not match the data exactly, as there also appears to be many low-\sm\
dEs within 1~Mpc, and the metallicities of the GCs as derived from the
simulation are generally too high, but it is encouraging that this
simple scaling for 
cluster formation in the Millennium Simulation can reproduce many of the
observed trends.  A more detailed treatment of ram pressure and tidal
stripping in cluster cores in a higher resolution simulation may
help resolve some discrepancies.

% fig:sm_sim
%%%%%%%%%%%%%%%%%%%%%%%%%%%%%%%%%%%%%%%%%%%%%%%%%%%%%%%%%%%%%%%%%%%%%%%%%
%%%%%%%%%%%%%%%%%%%%%%%%%%%%%%%%%%%%%%%%%%%%%%%%%%%%%%%%%%%%%%%%%%%%%%%%%

\subsection{Possible Mechanisms for Quenching Star Formation}

The central dwarfs may produce a higher
fraction of GCs at or before their peak SFR, but it is also important that their
subsequent star formation is rapidly quenched or kept at a low level.
At these and lower masses, photoionization heating of the ISM by
the UV background 
is believed to be an important mechanism for suppressing star
formation in dwarf galaxies (Bullock, Kravtsov \& Weinberg 2000;
Benson \etal 2002), and this could universally truncate the epoch of
efficient formation of metal-poor GCs, although detailed work still needs
to be done to show that reionization is a plausible mechanism for
halting GC formation.  At least one scenario (Cen 2001) has
reionization as the cause for GC formation, another suggests that
starburst-driven shocks may trigger GC formation (Scannapieco,
Weisheit \& Harlow 2004), and yet another shows that GCs themselves
could plausibly be the source of reionizing photons (Ricotti 2004).
So, the relationship between GC formation and reionization, if any, is
still very much undetermined.

One potential problem with the scenario where reionization halts GC
formation is that in the simulations of
Bekki \etal (2006), the bias introduced by the truncation redshift is
not very strong.  Changing the truncation redshift from $z=15$ to
$z=6$, for example, does not alter the \sn\ of galaxies by very much,
and at least in these simulations, reionization-induced biased GC formation by
itself does not seem to be able to explain the scatter in \sn\ for
dEs.  However, it may be sufficient to explain the general trend towards
higher GC fractions at lower masses seen in Figure~\ref{fig:vdb}.

Another plausible explanation is that the environment plays an
important role
in quenching the later, lower level star formation that builds up the field.
Ram pressure stripping of gas will be more efficient in low mass
galaxies at the center of the cluster.  Since these halos are the
earliest to fall into the cluster, they will have their star formation
quenched earlier and more efficiently than their counterparts in the
outer cluster regions.
Moore, Lake and Katz (1998) also showed through
numerical simulations that ``galaxy harassment'' in clusters is
efficient at transforming small spirals into low mass spheroidal
galaxies.  More recently, Mayer \etal (2007) used
hydrodynamic simulations to demonstrate that the environmental effects of tidal
shocks and ram pressure stripping can combine to create the most dark
matter-dominated dwarf spheroidals (dSphs) in the Local Group.  Although the
Virgo dEs are much more massive than dSphs such as Draco or Ursa
Minor, it is possible that similar processes caused them to
fail in their conversion of gas to field stars.  Ram pressure
stripping in Virgo has been observed out to a cluster radius of 0.8~Mpc
(Kenney, van Gorkom \& Vollmer 2004), similar to the radius within which
we see enhanced GC fractions.  The comparison of
central dEs and those on the outskirts, however, shows no strong
differences between their observed properties.  If harassment was
important, we might expect different morphologies or surface
brightness profiles (\sersic\ $n$ or $r_e$) in the two groups, but this
is not the case. The cumulative GC color distributions of these two
populations of dEs are also indistinguishable.  If the central dEs were
once much more luminous, we would expect them to be outliers on the
color-magnitude relation, but the ones with high \sn\
are not. Only the innermost galaxies that are well within the M87 GC
system and have no GCs {\it are} outliers and are red for their luminosities.
These galaxies are good candidates for having been harassed.

\subsection{Globular Cluster Destruction}

Up until now, we have only discussed the possible variation in
formation histories to explain the observed trends in GC fraction.
However, star clusters can also be destroyed,
and we expect that a substantial fraction of clusters initially
formed will not survive a Hubble time.  Observations of young star
clusters in nearby galaxies show that most star clusters are disrupted
very early (Fall \etal 2005; Bastian \etal 2005).  In addition,
subsequent evolution 
through two-body relaxation, stellar mass loss, and tidal shocks will
destroy even more low mass clusters (Fall \& Rees 1977; Vesperini
1998; Fall \& Zhang 2001).  Globular clusters are thus the survivors
of what was once a much larger population of star clusters.  Any
differences in the survival rates between different galaxies could
produce different GC fractions in the present day.

Unfortunately, the cosmological simulations used to study galaxy
formation cannot at the present time
model GC destruction.  One benefit of studying trends in terms of
$S_\mathcal{M}$, or GC mass fraction, is that while most of the destroyed star
clusters after the initial 1~Gyr are preferentially low mass, most of
the mass remains in the 
higher mass objects.  Thus, GC mass fraction is more
robust against disruption processes that preferentially destroy low
mass clusters.  If, 
however, there was a mechanism that could affect the survival
efficiency of massive clusters---perhaps variable infant mortality or
dynamical friction as 
influenced by the dark matter halo profile---then
destruction could play a role in 
driving the trends in GC mass fraction.  This mechanism would
require central dwarfs to be able to retain a larger fraction of the
massive star clusters than their counterparts at larger cluster
radii.  This is perhaps counter to what one might expect if the
cluster tidal field had a role in stripping or destroying GCs.
We also see no signs in our data that central dwarfs have higher
mean GC masses or different GC luminosity functions, things that might
point to internal destruction being a dominant mechanism driving GC fractions.

For a real test of the mechanisms that drive GC formation efficiency,
it would be extremely beneficial to have more and better simulations of galaxies
and their star cluster populations.  Both hydrodynamic simulations and
dark matter simulations coupled with semi-analytic models can provide
useful quantitative predictions.  These would ideally be able to
produce a $z=0$ volume comparable to the Virgo cluster with a stellar mass
resolution that would allow the resolution of GC mass objects
($\sim10^5\mathcal{M}_\odot$).  

\subsection{\sn\ Variation in Metal-poor and Metal-rich GCs}

Another result of this study is the
behavior of the specific frequencies of blue and red GCs. 
The variation in \sn\ is dominated by the blue, metal-poor GCs, although the
red GCs do exhibit similar trends with fewer GCs and lower
significance.  Figure~\ref{fig:rfrac} shows that the fraction of red
GCs rarely exceeds $50\%$ and that the trend of increasing red fraction
with galaxy mass either flattens or turns over for the most massive
galaxies.  This suggests that massive ellipticals have accreted
the most dwarf-like galaxies, the presumed original hosts
of metal-poor GCs, or that the dwarfs which they accreted had the
highest \sn, or both. Also, the rise
in red GC fraction from $M_z=-19$ to $M_z=-22$ requires that
$S_{N,z,blue}$ and $S_{N,z,red}$ respectively fall and rise in equal
proportions to produce the nearly constant \snz\ over this range.  The
fraction of blue GCs is lowest where the total \snz\ is at a
minimum, implying again that they are driving the overall trends.  

The higher values of $S_{N,z,red}$ for the massive galaxies, in
particular M87, means that the formation of metal-rich GCs also occurs
at different efficiencies.  For massive galaxies, these results are
consistent with the data of Rhode \etal (2005), who found that the
mass-normalized number of GCs, $T_{red}$, is positively correlated
with galaxy mass (shown in Figure~4 of Brodie \& Strader 2006).
Could these metal-rich GCs have been accreted from dwarfs in a
scenario similar to that described for the metal-poor GCs?  
Figure~\ref{fig:snz_rp_br} shows that the
number fraction of red GCs may also be enhanced in central dwarfs,
like the metal-poor GCs, but it is unclear whether a biasing scenario
similar to the one described above provide an explanation since we
would expect that the formation of metal-rich GCs at a later time might actually
be suppressed.  However, the numbers we observe are very small and
uncertain and may be sensitive to the dividing color
between blue and red GCs, which is chosen to be the same in all
galaxies at these luminosities.  Another possibility is that the
trend in the red GCs is determined more by the mass of the host, with red
GCs forming in the most massive progenitor, and the more massive
protogalaxies forming their stars earlier and with higher star formation rates.
This is consistent with observations of the stellar populations of
elliptical galaxies, which find that the stars in massive ellipticals
($L>L^*$) form have old ages and high [$\alpha$/Fe]---i.e., they
formed early and rapidly---while lower mass early-type galaxies have younger
mean ages and lower [$\alpha$/Fe] (Thomas \etal 2005).  Similar trends
in [$\alpha$/Fe] are seen in the GCs of these galaxies (Puzia \etal 2006).

The fact that we see the same trends for both the blue and red GC
sub-populations suggests the possibility that the two populations are
not very distinct, and that their separation is merely the result of 
placing a dividing line on a continuum of GC properties.  This possibility
has been raised by Yoon, Yi, \& Lee (2006), who suggest that the
GC metallicity bimodality in early-type galaxies is a result of a
nonlinear metallicity-color relation.  Although our data does not
constrain this hypothesis, we do not observe a clear
distinction between blue and red GC fractions across galaxy mass or
environment.

\section{Conclusions}

We have measured globular cluster specific frequencies (\sn), luminosity
fractions ($S_L$), and stellar mass fractions ($S_\mathcal{M}$) for 100
early-type galaxies in the ACS Virgo Cluster Survey.  These galaxies
span the mass range from giants to dwarfs ($-22 < M_B < -15$) and
these represent the largest homogeneous catalog of GC number and mass
fractions to date.  We have studied these quantities as a function of
galaxy mass and environment and find that:
\begin{enumerate}

\item Globular cluster fractions can be high ($S_{N,z}>2$) for both
  high and low luminosity early-type galaxies, but are universally low
  for intermediate luminosity galaxies ($-22<M_z<-19$).

\item There is a large spread in GC fraction in early-type dwarfs
  ($0<S_{N,z}<5.5$) which can be understood as an underlying dependence on
  environment. Almost all dwarfs ($M_z>-19$) with $S_{N,z}>2$ are
  within a projected radius of 1~Mpc from M87 and the cluster center
  (or 1.5~Mpc in three-dimensional cluster radius).  The spatial
  distributions of GCs 
  in the high-\sn\ dwarfs are centrally concentrated, showing that
  they are intrinsic to the host dwarf and not interloping GCs from
  M87 or an intracluster population.  We do not detect higher GC
  fractions in dwarfs around the other massive ellipticals in the cluster.
  We present this as evidence that
  GC formation in low mass galaxies is biased towards the densest environments.

\item Galaxies within $R_p\sim40$~kpc of M87 and M49 have few or no GCs,
  and are likely to have had their GC systems tidally stripped by
  their giant neighbors.  Galaxies out to $R_p\sim100$~kpc from M87
  (VCC~1185) also appear to be affected.

\item Analyzing the blue and red GC populations separately, we
  find that the fraction of red GCs increases with galaxy luminosity until
  $M_z\sim-22$ at which point it flattens or declines.  Trends in GC
  fraction are dominated by the blue GCs, although the \snz\ of red GCs does
  exhibit similar relative enhancement for massive galaxies, and
  possibly shows a weak enhancement for central dwarfs.

\item We use a globally averaged
  $\mathcal{M}_h/\mathcal{M}_{G\star}$--$\mathcal{M}_{G\star}$ relation 
  from halo occupation studies to test whether trends in GC fraction
  can be explained by the assumption that the mass in GCs is directly
  proportional to the halo mass, $\mathcal{M}_{GC}\propto
  \mathcal{M}_h$.  While this may be able to explain the mean trend in
  $S_\mathcal{M}$, it is unable to account 
  simultaneously for all dwarfs, particularly 
  the dwarfs with high GC fractions in the cluster's central regions.

\item Comparisons with semi-analytic models of galaxy formation in the
  Millennium Simulation show that early-type dwarfs in the central
  1~Mpc of massive galaxy clusters are expected to be
  older and have higher peak star formation rates and SFR surface densities than
  their counterparts on the outskirts of the cluster. The higher
  stellar mass fractions in globular cluster for central dwarfs can be
  explained if higher SFR surface densities are responsible for more
  efficient formation of massive star clusters.

\item The peak SFR surface density in simulated dwarfs
  occurs before the peak SFR, which we propose as an explanation for why
  GCs are, on average, both older and more metal-poor than the field stars in
  their host galaxies.

\end{enumerate}

We present a picture of globular cluster system formation in the Virgo
Cluster where
the highest GC mass fractions are formed in the oldest systems.  These
progenitors formed a larger fraction of their stars at higher
peak SFR surface densities, and also have star formation suppressed at
later times.
In regions of high density, halos collapse and star formation
starts earlier, and the mechanisms for truncating and suppressing
subsequent star formation are stronger.  For all dwarfs, the highest
SFR surface densities occur earlier than the peak SFR, naturally
producing the metal-poor GC populations we see today---old
($>10$~Gyr) and more metal-poor than the bulk of the field stars. Merging of
many low mass 
progenitors with high GC fractions produce the extremely high \sn\
for blue GCs seen in the cD galaxy, M87.  They may also, at a lower
level, produce the elevated blue GC fractions seen in other massive
ellipticals.   Future detailed 
simulations of GC system formation will be crucial to test, in a
quantitative way, the scenarios of galaxy and GC system formation that
we are beginning to assemble.

\acknowledgments
E.\ W.\ P. gratefully acknowledges the support of the National
Research Council of Canada's Plaskett Research Fellowship at the
Herzberg Institute of Astrophysics.  We thank
Gabriella De Lucia, Gerard Lemson, and Darren Croton for their help with
the Millennium Simulation, and thank Rupali Chandar, Dean McLaughlin, 
Anil Seth, and Peter Stetson for useful discussions and comments on
the manuscript. 
We also thank Peter Stetson for providing point spread functions used
in the analysis of our WFPC2 data.  We thank Naoyuki Tamura, who
provided the numbers for their M87 GC radial profile.  We are grateful
to the anonymous referee who provided comments that improved the manuscript.
M.\ J.\ W.\ and M.\ T.\ acknowledge support through NSF grant AST 02-05960.

This publication makes use of data products from the Sloan Digital
Sky Survey (SDSS).  Funding for SDSS and SDSS-II has
been provided by the Alfred P. Sloan Foundation, the Participating
Institutions, the National Science Foundation, the U.S. Department
of Energy, the National Aeronautics and Space Administration, the
Japanese Monbukagakusho, the Max Planck Society, and the
Higher Education Funding Council for England. The SDSS Web site is
http://www.sdss.org/. 

The SDSS is managed by the Astrophysical Research Consortium (ARC)
for the Participating Institutions. The Participating Institutions
are the American Museum of Natural History, Astrophysical
Institute Potsdam, University of Basel, University of Cambridge,
Case Western Reserve University, The University of Chicago, Drexel
University, Fermilab, the Institute for Advanced Study, the Japan
Participation Group, The Johns Hopkins University, the Joint
Institute for Nuclear Astrophysics, the Kavli Institute for
Particle Astrophysics and Cosmology, the Korean Scientist Group,
the Chinese Academy of Sciences (LAMOST), Los Alamos National
Laboratory, the Max-Planck-Institute for Astronomy (MPIA), the
Max-Planck-Institute for Astrophysics (MPA), New Mexico State
University, Ohio State University, University of Pittsburgh,
University of Portsmouth, Princeton University, the United States
Naval Observatory, and the University of Washington. 

This publication makes use of data products from the Two Micron All
Sky Survey, which is a joint project of the University of
Massachusetts and the Infrared Processing and Analysis
Center/California Institute of Technology, funded by the National
Aeronautics and Space Administration and the National Science
Foundation. 

This research has made use of the NASA/IPAC Extragalactic Database
(NED) which is operated by the Jet Propulsion Laboratory, California
Institute of Technology, under contract with the National Aeronautics
and Space Administration.

Facilities: \facility{HST(ACS,WFPC2)}

{}

\clearpage

\clearpage

%% Tables may also be prepared as separate files. See the accompanying
%% sample file table.tex for an example of an external table file.
%% To include an external file in your main document, use the \input
%% command. Uncomment the line below to include table.tex in this
%% sample file. (Note that you will need to comment out the \documentclass,
%% \begin{document}, and \end{document} commands from table.tex if you want
%% to include it in this document.)

%%%%%%%%%%%%%%%%%%%%%%%%%%%%%%%%%%
\setcounter{table}{0}
\LongTables
\begin{deluxetable}{crcccccc}
\tablewidth{0pt}
\tablecaption{Global Properties of ACS Virgo Cluster Survey Galaxies\label{table:galxtable}}
\tablehead{
\colhead{No.} & 
\colhead{VCC} & 
\colhead{$M_V$} & 
\colhead{$M_z$} & 
\colhead{$R_p$} & 
\colhead{$R_{3D}$} & 
\colhead{$L_{z,\star}$} & 
\colhead{$M_\star$} \\ 
\colhead{(1)} &
\colhead{(2)} &
\colhead{(3)} &
\colhead{(4)} &
\colhead{(5)} &
\colhead{(6)} &
\colhead{(7)} &
\colhead{(8)} 
}
\startdata
1 & 1226 & $ -22.90$ & $ -23.87$ & 1.27 & $1.28 \pm 0.39$ & $263.04$ & $531.84 \pm 110.48$ \\
2 & 1316 & $ -22.66$ & $ -23.86$ & 0.00 & $0.00$ & $188.40$ & $302.27 \pm  79.13$ \\
3 & 1978 & $ -22.41$ & $ -23.42$ & 0.94 & $1.04 \pm 0.46$ & $167.62$ & $339.32 \pm  50.28$ \\
4 &  881 & $ -22.54$ & $ -23.53$ & 0.36 & $1.49 \pm 0.42$ & $181.42$ & $289.63 \pm  59.87$ \\
5 &  798 & $ -22.34$ & $ -23.25$ & 1.71 & $2.09 \pm 0.42$ & $132.84$ & $186.50 \pm  43.84$ \\
6 &  763 & $ -22.29$ & $ -23.20$ & 0.43 & $2.09 \pm 0.44$ & $141.95$ & $236.49 \pm  61.04$ \\
7 &  731 & $ -22.31$ & $ -23.37$ & 1.53 & $6.76 \pm 0.54$ & $141.76$ & $226.21 \pm  52.45$ \\
8 & 1535 & $ -21.38$ & $ -22.40$ & 1.37 & \nodata & $ 57.88$ & $ 77.21 \pm  16.21$ \\
9 & 1903 & $ -21.18$ & $ -22.19$ & 0.82 & $1.87 \pm 0.35$ & $ 50.26$ & $ 83.87 \pm  19.10$ \\
10 & 1632 & $ -21.36$ & $ -22.33$ & 0.34 & $0.63 \pm 0.38$ & $ 60.33$ & $ 95.35 \pm  16.89$ \\
11 & 1231 & $ -20.70$ & $ -21.58$ & 0.31 & $1.30 \pm 0.36$ & $ 32.37$ & $ 53.46 \pm  12.30$ \\
12 & 2095 & $ -20.78$ & $ -20.95$ & 1.59 & \nodata & $ 33.74$ & $ 52.75 \pm  10.12$ \\
13 & 1154 & $ -20.85$ & $ -21.79$ & 0.47 & $0.66 \pm 0.38$ & $ 36.86$ & $ 77.95 \pm  13.64$ \\
14 & 1062 & $ -20.56$ & $ -21.34$ & 0.77 & $1.42 \pm 0.36$ & $ 28.65$ & $ 52.69 \pm   8.88$ \\
15 & 2092 & $ -20.46$ & $ -21.64$ & 1.54 & $1.61 \pm 0.38$ & $ 25.55$ & $ 47.32 \pm   7.66$ \\
16 &  369 & $ -20.19$ & $ -20.41$ & 0.79 & $1.04 \pm 0.37$ & $ 20.17$ & $ 32.46 \pm   7.67$ \\
17 &  759 & $ -20.44$ & $ -21.41$ & 0.46 & $0.60 \pm 0.40$ & $ 25.74$ & $ 45.15 \pm  10.04$ \\
18 & 1692 & $ -20.26$ & $ -21.04$ & 1.54 & $1.64 \pm 0.40$ & $ 21.85$ & $ 33.97 \pm   6.55$ \\
19 & 1030 & $ -20.45$ & $ -21.38$ & 0.30 & $0.33 \pm 0.39$ & $ 22.63$ & $ 13.32 \pm   9.05$ \\
20 & 2000 & $ -19.60$ & $ -20.66$ & 1.03 & $2.03 \pm 0.35$ & $ 11.71$ & $ 23.76 \pm   4.10$ \\
21 &  685 & $ -20.11$ & $ -21.04$ & 1.33 & \nodata & $ 19.12$ & $ 31.23 \pm   6.69$ \\
22 & 1664 & $ -19.90$ & $ -20.97$ & 0.48 & $0.77 \pm 0.38$ & $ 15.39$ & $ 26.00 \pm   6.15$ \\
23 &  654 & $ -19.88$ & $ -20.59$ & 1.35 & \nodata & $ 14.22$ & $ 22.74 \pm   4.12$ \\
24 &  944 & $ -19.80$ & $ -20.66$ & 0.86 & $1.05 \pm 0.38$ & $ 13.94$ & $ 29.14 \pm   4.88$ \\
25 & 1938 & $ -19.93$ & $ -20.86$ & 0.89 & $1.00 \pm 0.40$ & $ 15.71$ & $ 24.86 \pm   5.50$ \\
26 & 1279 & $ -19.78$ & $ -20.72$ & 0.04 & $0.47 \pm 0.40$ & $ 13.42$ & $ 22.03 \pm   4.03$ \\
27 & 1720 & $ -19.63$ & $ -20.68$ & 0.94 & $0.97 \pm 0.38$ & $ 11.55$ & $ 20.21 \pm   5.31$ \\
28 &  355 & $ -19.42$ & $ -20.41$ & 1.07 & $1.55 \pm 0.36$ & $ 10.06$ & $ 15.78 \pm   3.62$ \\
29 & 1619 & $ -19.36$ & $ -20.19$ & 0.33 & $1.51 \pm 0.36$ & $  8.65$ & $ 17.24 \pm   3.12$ \\
30 & 1883 & $ -19.68$ & $ -20.73$ & 1.65 & $1.69 \pm 0.38$ & $ 11.19$ & $ 16.66 \pm   3.92$ \\
31 & 1242 & $ -19.38$ & $ -20.29$ & 0.49 & $1.42 \pm 0.43$ & $  9.07$ & $ 15.27 \pm   3.17$ \\
32 &  784 & $ -19.31$ & $ -20.26$ & 1.00 & $1.30 \pm 0.37$ & $  8.48$ & $ 16.79 \pm   3.14$ \\
33 & 1537 & $ -18.99$ & $ -19.95$ & 0.39 & $1.18 \pm 0.36$ & $  6.18$ & $ 10.24 \pm   2.35$ \\
34 &  778 & $ -19.56$ & $ -20.03$ & 0.79 & $1.17 \pm 0.41$ & $ 10.73$ & $ 18.18 \pm   4.19$ \\
35 & 1321 & $ -18.80$ & $ -19.93$ & 1.26 & $1.99 \pm 0.35$ & $  4.86$ & $  6.98 \pm   1.75$ \\
36 &  828 & $ -19.15$ & $ -20.07$ & 0.38 & $1.42 \pm 0.42$ & $  7.51$ & $ 13.69 \pm   3.00$ \\
37 & 1250 & $ -18.98$ & $ -20.10$ & 0.06 & $0.79 \pm 0.57$ & $  5.23$ & $  3.68 \pm   2.04$ \\
38 & 1630 & $ -19.07$ & $ -20.06$ & 0.34 & $0.51 \pm 0.38$ & $  7.03$ & $ 11.61 \pm   2.39$ \\
39 & 1146 & $ -18.93$ & $ -19.64$ & 0.28 & $0.60 \pm 0.38$ & $  5.71$ & $  8.69 \pm   2.17$ \\
40 & 1025 & $ -19.58$ & $ -20.57$ & 1.24 & $5.98 \pm 0.53$ & $ 10.90$ & $ 21.39 \pm   4.25$ \\
41 & 1303 & $ -18.84$ & $ -19.80$ & 0.97 & $0.98 \pm 0.39$ & $  5.55$ & $ 10.39 \pm   2.33$ \\
42 & 1913 & $ -18.87$ & $ -19.74$ & 1.58 & $1.67 \pm 0.40$ & $  5.67$ & $ 10.74 \pm   2.72$ \\
43 & 1327 & $ -19.10$ & $ -19.81$ & 0.04 & $1.45 \pm 0.51$ & $  8.36$ & $ 17.24 \pm   2.34$ \\
44 & 1125 & $ -19.07$ & $ -19.44$ & 0.24 & \nodata & $  6.80$ & $  8.06 \pm   3.40$ \\
45 & 1475 & $ -18.56$ & $ -19.42$ & 1.13 & $1.15 \pm 0.38$ & $  4.07$ & $  7.73 \pm   1.50$ \\
46 & 1178 & $ -18.35$ & $ -19.35$ & 1.22 & $1.48 \pm 0.37$ & $  3.64$ & $  7.09 \pm   1.42$ \\
47 & 1283 & $ -18.65$ & $ -19.70$ & 0.34 & $0.86 \pm 0.41$ & $  4.83$ & $  9.08 \pm   1.79$ \\
48 & 1261 & $ -18.42$ & $ -19.21$ & 0.47 & $1.44 \pm 0.51$ & $  3.14$ & $  4.87 \pm   1.69$ \\
49 &  698 & $ -18.78$ & $ -19.71$ & 0.58 & $1.80 \pm 0.43$ & $  5.03$ & $  9.52 \pm   2.56$ \\
50 & 1422 & $ -17.97$ & $ -18.73$ & 0.62 & $1.60 \pm 0.43$ & $  2.18$ & $  3.82 \pm   1.35$ \\
51 & 2048 & $ -17.85$ & $ -18.57$ & 1.32 & \nodata & $  1.89$ & $  2.94 \pm   1.08$ \\
52 & 1871 & $ -17.31$ & $ -18.32$ & 0.79 & $1.43 \pm 0.44$ & $  1.36$ & $  2.26 \pm   0.58$ \\
53 &    9 & $ -18.04$ & $ -18.80$ & 1.57 & $1.64 \pm 0.65$ & $  2.17$ & $  3.08 \pm   0.61$ \\
54 &  575 & $ -18.42$ & $ -19.33$ & 1.34 & $5.30 \pm 0.61$ & $  3.56$ & $  5.00 \pm   1.75$ \\
55 & 1910 & $ -17.39$ & $ -18.33$ & 0.82 & $1.11 \pm 0.37$ & $  1.43$ & $  2.10 \pm   0.81$ \\
56 & 1049 & $ -16.69$ & $ -17.34$ & 1.26 & $1.43 \pm 0.53$ & $  0.60$ & $  0.52 \pm   0.40$ \\
57 &  856 & $ -17.57$ & $ -18.25$ & 0.76 & $0.76 \pm 0.47$ & $  1.47$ & $  2.22 \pm   0.85$ \\
58 &  140 & $ -17.51$ & $ -18.27$ & 1.24 & $1.30 \pm 0.46$ & $  1.37$ & $  2.33 \pm   0.71$ \\
59 & 1355 & $ -17.51$ & $ -18.16$ & 0.50 & $0.52 \pm 0.63$ & $  1.29$ & $  1.82 \pm   0.35$ \\
60 & 1087 & $ -17.79$ & $ -18.64$ & 0.25 & $0.34 \pm 0.46$ & $  1.84$ & $  3.29 \pm   1.07$ \\
61 & 1297 & $ -17.67$ & $ -18.75$ & 0.04 & $0.23 \pm 0.46$ & $  1.95$ & $  3.96 \pm   0.62$ \\
62 & 1861 & $ -17.60$ & $ -18.34$ & 0.80 & $1.15 \pm 0.45$ & $  1.63$ & $  2.88 \pm   1.00$ \\
63 &  543 & $ -17.41$ & $ -18.19$ & 0.91 & $1.44 \pm 0.44$ & $  1.27$ & $  2.19 \pm   0.65$ \\
64 & 1431 & $ -17.39$ & $ -18.25$ & 0.34 & $0.90 \pm 0.45$ & $  1.37$ & $  2.20 \pm   0.74$ \\
65 & 1528 & $ -17.16$ & $ -18.04$ & 0.34 & $0.70 \pm 0.45$ & $  1.06$ & $  1.63 \pm   0.48$ \\
66 & 1695 & $ -17.49$ & $ -18.32$ & 0.43 & $0.46 \pm 0.54$ & $  1.30$ & $  1.69 \pm   0.78$ \\
67 & 1833 & $ -17.13$ & $ -17.89$ & 1.22 & $1.33 \pm 0.45$ & $  0.97$ & $  0.86 \pm   0.48$ \\
68 &  437 & $ -17.82$ & $ -18.69$ & 1.62 & $1.65 \pm 0.48$ & $  1.92$ & $  2.80 \pm   1.25$ \\
69 & 2019 & $ -17.36$ & $ -18.18$ & 1.08 & $1.13 \pm 0.48$ & $  1.17$ & $  1.02 \pm   0.72$ \\
70 &   33 & $ -16.39$ & $ -17.01$ & 1.48 & $2.19 \pm 0.57$ & $  0.46$ & $  0.43 \pm   0.25$ \\
71 &  200 & $ -17.12$ & $ -17.80$ & 1.02 & $1.70 \pm 0.59$ & $  0.95$ & $  1.34 \pm   0.49$ \\
72 &  571 & $ -17.32$ & $ -18.29$ & 1.40 & $7.27 \pm 1.01$ & $  1.07$ & $  0.98 \pm   0.65$ \\
73 &   21 & $ -16.83$ & $ -17.38$ & 1.58 & \nodata & $  0.58$ & $  0.56 \pm   0.37$ \\
74 & 1488 & $ -16.78$ & $ -17.34$ & 0.88 & \nodata & $  0.57$ & $  0.41 \pm   0.38$ \\
75 & 1779 & $ -16.90$ & $ -17.38$ & 0.89 & \nodata & $  0.63$ & $  1.02 \pm   0.38$ \\
76 & 1895 & $ -16.60$ & $ -17.29$ & 1.16 & $1.47 \pm 0.37$ & $  0.57$ & $  0.76 \pm   0.39$ \\
77 & 1499 & $ -16.53$ & $ -16.95$ & 0.22 & \nodata & $  0.34$ & $  0.14 \pm   0.19$ \\
78 & 1545 & $ -16.91$ & $ -17.74$ & 0.26 & $0.30 \pm 0.54$ & $  0.87$ & $  1.41 \pm   0.42$ \\
79 & 1192 & $ -16.86$ & $ -18.14$ & 1.27 & \nodata & $  0.94$ & $  1.85 \pm   0.62$ \\
80 & 1857 & $ -16.64$ & $ -17.32$ & 0.90 & \nodata & $  0.53$ & $  0.73 \pm   0.14$ \\
81 & 1075 & $ -16.78$ & $ -17.55$ & 0.63 & $0.87 \pm 0.68$ & $  0.73$ & $  1.11 \pm   0.47$ \\
82 & 1948 & $ -16.06$ & $ -16.78$ & 0.99 & \nodata & $  0.35$ & $  0.50 \pm   0.09$ \\
83 & 1627 & $ -16.46$ & $ -17.39$ & 0.34 & $2.19 \pm 0.34$ & $  0.59$ & $  1.03 \pm   0.32$ \\
84 & 1440 & $ -16.86$ & $ -17.66$ & 0.88 & $1.21 \pm 0.45$ & $  0.78$ & $  1.19 \pm   0.44$ \\
85 &  230 & $ -16.21$ & $ -16.96$ & 0.96 & $1.41 \pm 0.84$ & $  0.43$ & $  0.69 \pm   0.29$ \\
86 & 2050 & $ -16.36$ & $ -17.15$ & 1.16 & $1.47 \pm 0.52$ & $  0.47$ & $  0.29 \pm   0.26$ \\
87 & 1993 & $ -16.30$ & $ -17.02$ & 0.95 & $1.03 \pm 0.38$ & $  0.48$ & $  0.75 \pm   0.27$ \\
88 &  751 & $ -16.97$ & $ -17.92$ & 1.72 & $2.05 \pm 0.44$ & $  0.92$ & $  1.41 \pm   0.51$ \\
89 & 1828 & $ -16.78$ & $ -17.47$ & 0.67 & $0.67 \pm 0.55$ & $  0.72$ & $  1.26 \pm   0.47$ \\
90 &  538 & $ -16.24$ & $ -17.10$ & 1.62 & $6.47 \pm 0.86$ & $  0.44$ & $  0.63 \pm   0.26$ \\
91 & 1407 & $ -16.72$ & $ -17.43$ & 0.17 & $0.23 \pm 0.39$ & $  0.68$ & $  1.24 \pm   0.42$ \\
92 & 1886 & $ -16.25$ & $ -16.93$ & 0.76 & \nodata & $  0.36$ & $  0.51 \pm   0.10$ \\
93 & 1199 & $ -15.47$ & $ -16.94$ & 1.25 & \nodata & $  0.29$ & $  0.58 \pm   0.16$ \\
94 & 1743 & $ -16.33$ & $ -16.95$ & 0.84 & $1.27 \pm 0.75$ & $  0.45$ & $  0.40 \pm   0.29$ \\
95 & 1539 & $ -16.05$ & $ -17.12$ & 0.25 & $0.26 \pm 0.87$ & $  0.37$ & $  0.52 \pm   0.11$ \\
96 & 1185 & $ -16.77$ & $ -17.37$ & 0.10 & $0.10 \pm 0.79$ & $  0.78$ & $  0.89 \pm   0.51$ \\
97 & 1826 & $ -16.03$ & $ -16.71$ & 0.98 & $1.11 \pm 0.61$ & $  0.35$ & $  0.61 \pm   0.22$ \\
98 & 1512 & $ -16.25$ & $ -16.54$ & 0.38 & $1.50 \pm 0.42$ & $  0.43$ & $  0.61 \pm   0.12$ \\
99 & 1489 & $ -15.61$ & $ -16.41$ & 0.45 & \nodata & $  0.22$ & $  0.32 \pm   0.06$ \\
100 & 1661 & $ -15.81$ & $ -17.40$ & 0.70 & $1.26 \pm 1.20$ & $  0.29$ & $  0.43 \pm   0.08$ \\
\enddata
\tablenotetext{1}{Running number, sorted by increasing $B_T$ magnitude}
\tablenotetext{2}{Number in Virgo Cluster Catalog}
\tablenotetext{3}{Absolute $V$ magnitude}
\tablenotetext{4}{Absolute $z$ magnitude}
\tablenotetext{5}{Projected distance from M87 (VCC~1316), in Mpc}
\tablenotetext{6}{3-dimensional distance from M87 (VCC~1316), in Mpc, using polynomial calibration from Paper~XIII}
\tablenotetext{7}{Stellar $z$ luminosity ($10^9 L_\odot$)}
\tablenotetext{8}{Stellar mass ($10^9 M_\odot$)}
\end{deluxetable}

%%%%%%%%%%%%%%%%%%%%%%%%%%%%%%%%%%

\clearpage
%%%%%%%%%%%%%%%%%%%%%%%%%%%%%%%%%%
\begin{landscape}
\LongTables
\tabletypesize{\tiny}
\begin{deluxetable}{crcrrrrrrrrrr}
%\rotate
\tablewidth{0pt}
\tablecaption{GC Formation Efficiencies in ACS Virgo Cluster Survey Galaxies\label{table:gctable}}
\tablehead{
\colhead{No.} & 
\colhead{VCC} & 
\colhead{$N_{GC}$} & 
\colhead{$S_N$} & 
\colhead{$S_{N,z}$} & 
\colhead{$T$} & 
\colhead{$S_L$} & 
\colhead{$S_M$} & 
\colhead{$S_{N,z,blue}$} & 
\colhead{$S_{N,z,red}$} & 
\colhead{$T_{blue}$} & 
\colhead{$T_{red}$} & 
\colhead{$f_{red}$} \\
\colhead{(1)} &
\colhead{(2)} &
\colhead{(3)} &
\colhead{(4)} &
\colhead{(5)} &
\colhead{(6)} &
\colhead{(7)} &
\colhead{(8)} &
\colhead{(9)} &
\colhead{(10)} &
\colhead{(11)} &
\colhead{(12)} &
\colhead{(13)}
}
\startdata
   1 & 1226 & $ 7813 \pm   830$ & $ 5.40 \pm 0.57$ & $ 2.20 \pm 0.23$ & $ 14.7 \pm   3.4$ & $ 1.23 \pm  0.01$ & $ 0.87 \pm  0.18$ & $ 1.57 \pm 0.19$ & $ 0.64 \pm 0.14$ & $ 10.4 \pm   2.5$ & $  4.3 \pm   1.3$ & $0.29$ \\ 
   2 & 1316 & $14660 \pm   891$ & $12.59 \pm 0.77$ & $ 4.19 \pm 0.25$ & $ 48.5 \pm  13.0$ & $ 2.33 \pm  0.02$ & $ 2.83 \pm  0.74$ & $ 3.04 \pm 0.21$ & $ 1.14 \pm 0.15$ & $ 35.2 \pm   9.5$ & $ 13.2 \pm   3.9$ & $0.27$ \\ 
   3 & 1978 & $ 4745 \pm  1099$ & $ 5.16 \pm 1.20$ & $ 2.03 \pm 0.47$ & $ 14.0 \pm   3.8$ & $ 1.15 \pm  0.02$ & $ 0.84 \pm  0.12$ & $ 1.28 \pm 0.38$ & $ 0.75 \pm 0.27$ & $  8.8 \pm   2.9$ & $  5.2 \pm   2.0$ & $0.37$ \\ 
   4 &  881 & $ 2660 \pm   129$ & $ 2.57 \pm 0.12$ & $ 1.03 \pm 0.05$ & $  9.2 \pm   1.9$ & $ 0.56 \pm  0.01$ & $ 0.49 \pm  0.10$ & $ 0.87 \pm 0.04$ & $ 0.16 \pm 0.03$ & $  7.7 \pm   1.6$ & $  1.5 \pm   0.4$ & $0.16$ \\ 
   5 &  798 & $ 1110 \pm   181$ & $ 1.29 \pm 0.21$ & $ 0.56 \pm 0.09$ & $  6.0 \pm   1.7$ & $ 0.20 \pm  0.01$ & $ 0.21 \pm  0.05$ & $ 0.33 \pm 0.07$ & $ 0.23 \pm 0.05$ & $  3.5 \pm   1.1$ & $  2.4 \pm   0.8$ & $0.41$ \\ 
   6 &  763 & $ 4301 \pm  1201$ & $ 5.20 \pm 1.45$ & $ 2.26 \pm 0.63$ & $ 18.2 \pm   6.9$ & $ 1.11 \pm  0.03$ & $ 0.89 \pm  0.23$ & $ 2.01 \pm 0.51$ & $ 0.24 \pm 0.37$ & $ 16.2 \pm   5.9$ & $  2.0 \pm   3.0$ & $0.11$ \\ 
   7 &  731 & $ 3246 \pm   598$ & $ 3.86 \pm 0.71$ & $ 1.45 \pm 0.27$ & $ 14.3 \pm   4.2$ & $ 0.85 \pm  0.02$ & $ 0.86 \pm  0.20$ & $ 0.80 \pm 0.22$ & $ 0.65 \pm 0.16$ & $  7.9 \pm   2.8$ & $  6.5 \pm   2.2$ & $0.45$ \\ 
   8 & 1535 & $  388 \pm   117$ & $ 1.09 \pm 0.33$ & $ 0.42 \pm 0.13$ & $  5.0 \pm   1.8$ & $ 0.20 \pm  0.00$ & $ 0.25 \pm  0.05$ & $ 0.26 \pm 0.10$ & $ 0.16 \pm 0.07$ & $  3.1 \pm   1.4$ & $  1.9 \pm   1.0$ & $0.38$ \\ 
   9 & 1903 & $  803 \pm   355$ & $ 2.70 \pm 1.19$ & $ 1.07 \pm 0.47$ & $  9.6 \pm   4.8$ & $ 0.68 \pm  0.01$ & $ 0.61 \pm  0.14$ & $ 0.55 \pm 0.38$ & $ 0.52 \pm 0.27$ & $  4.9 \pm   3.6$ & $  4.7 \pm   2.7$ & $0.49$ \\ 
  10 & 1632 & $  984 \pm   198$ & $ 2.82 \pm 0.57$ & $ 1.15 \pm 0.23$ & $ 10.3 \pm   2.8$ & $ 0.65 \pm  0.01$ & $ 0.61 \pm  0.11$ & $ 0.69 \pm 0.19$ & $ 0.46 \pm 0.14$ & $  6.2 \pm   2.0$ & $  4.1 \pm   1.4$ & $0.40$ \\ 
  11 & 1231 & $  376 \pm    97$ & $ 1.98 \pm 0.51$ & $ 0.88 \pm 0.23$ & $  7.0 \pm   2.4$ & $ 0.32 \pm  0.01$ & $ 0.28 \pm  0.06$ & $ 0.50 \pm 0.18$ & $ 0.38 \pm 0.13$ & $  4.0 \pm   1.7$ & $  3.0 \pm   1.3$ & $0.43$ \\ 
  12 & 2095 & $  211 \pm    34$ & $ 1.03 \pm 0.17$ & $ 0.88 \pm 0.14$ & $  4.0 \pm   1.0$ & $ 0.16 \pm  0.00$ & $ 0.14 \pm  0.03$ & $ 0.52 \pm 0.11$ & $ 0.35 \pm 0.08$ & $  2.4 \pm   0.7$ & $  1.6 \pm   0.5$ & $0.40$ \\ 
  13 & 1154 & $  218 \pm    28$ & $ 1.00 \pm 0.13$ & $ 0.42 \pm 0.05$ & $  2.8 \pm   0.6$ & $ 0.12 \pm  0.00$ & $ 0.09 \pm  0.02$ & $ 0.22 \pm 0.04$ & $ 0.20 \pm 0.03$ & $  1.5 \pm   0.4$ & $  1.3 \pm   0.3$ & $0.48$ \\ 
  14 & 1062 & $  178 \pm    30$ & $ 1.06 \pm 0.18$ & $ 0.52 \pm 0.09$ & $  3.4 \pm   0.8$ & $ 0.21 \pm  0.00$ & $ 0.16 \pm  0.03$ & $ 0.52 \pm 0.07$ & $ 0.00 \pm 0.05$ & $  3.4 \pm   0.7$ & $  0.0 \pm   0.3$ & $0.00$ \\ 
  15 & 2092 & $  103 \pm    17$ & $ 0.68 \pm 0.11$ & $ 0.23 \pm 0.04$ & $  2.2 \pm   0.5$ & $ 0.08 \pm  0.00$ & $ 0.07 \pm  0.01$ & $ 0.13 \pm 0.03$ & $ 0.09 \pm 0.02$ & $  1.3 \pm   0.4$ & $  0.9 \pm   0.3$ & $0.39$ \\ 
  16 &  369 & $  179 \pm    17$ & $ 1.51 \pm 0.14$ & $ 1.23 \pm 0.12$ & $  5.5 \pm   1.4$ & $ 0.28 \pm  0.01$ & $ 0.25 \pm  0.06$ & $ 0.58 \pm 0.09$ & $ 0.65 \pm 0.08$ & $  2.6 \pm   0.7$ & $  2.9 \pm   0.8$ & $0.53$ \\ 
  17 &  759 & $  200 \pm    41$ & $ 1.34 \pm 0.27$ & $ 0.54 \pm 0.11$ & $  4.4 \pm   1.3$ & $ 0.24 \pm  0.00$ & $ 0.20 \pm  0.05$ & $ 0.35 \pm 0.09$ & $ 0.19 \pm 0.07$ & $  2.9 \pm   1.0$ & $  1.6 \pm   0.6$ & $0.35$ \\ 
  18 & 1692 & $  139 \pm    23$ & $ 1.09 \pm 0.18$ & $ 0.53 \pm 0.09$ & $  4.1 \pm   1.0$ & $ 0.16 \pm  0.01$ & $ 0.16 \pm  0.03$ & $ 0.34 \pm 0.07$ & $ 0.19 \pm 0.05$ & $  2.6 \pm   0.7$ & $  1.5 \pm   0.5$ & $0.36$ \\ 
$\star$19 & 1030 & $  345 \pm    80$ & $ 2.27 \pm 0.53$ & $ 0.97 \pm 0.22$ & $ 25.9 \pm  18.6$ & $ 0.40 \pm  0.01$ & $ 1.06 \pm  0.72$ & $ 0.73 \pm 0.18$ & $ 0.24 \pm 0.13$ & $ 19.5 \pm  14.1$ & $  6.4 \pm   5.6$ & $0.25$ \\ 
  20 & 2000 & $  205 \pm    28$ & $ 2.97 \pm 0.41$ & $ 1.12 \pm 0.15$ & $  8.6 \pm   1.9$ & $ 0.53 \pm  0.01$ & $ 0.37 \pm  0.06$ & $ 0.97 \pm 0.13$ & $ 0.15 \pm 0.08$ & $  7.5 \pm   1.6$ & $  1.2 \pm   0.7$ & $0.13$ \\ 
  21 &  685 & $  196 \pm    60$ & $ 1.78 \pm 0.54$ & $ 0.75 \pm 0.23$ & $  6.3 \pm   2.3$ & $ 0.40 \pm  0.01$ & $ 0.36 \pm  0.08$ & $ 0.54 \pm 0.19$ & $ 0.22 \pm 0.13$ & $  4.5 \pm   1.8$ & $  1.8 \pm   1.2$ & $0.29$ \\ 
  22 & 1664 & $  213 \pm    31$ & $ 2.35 \pm 0.34$ & $ 0.87 \pm 0.13$ & $  8.2 \pm   2.3$ & $ 0.42 \pm  0.01$ & $ 0.37 \pm  0.09$ & $ 0.54 \pm 0.10$ & $ 0.33 \pm 0.07$ & $  5.1 \pm   1.5$ & $  3.1 \pm   1.0$ & $0.38$ \\ 
  23 &  654 & $   45 \pm    16$ & $ 0.50 \pm 0.18$ & $ 0.26 \pm 0.09$ & $  2.0 \pm   0.8$ & $ 0.05 \pm  0.00$ & $ 0.05 \pm  0.01$ & $ 0.23 \pm 0.08$ & $ 0.04 \pm 0.05$ & $  1.7 \pm   0.7$ & $  0.3 \pm   0.4$ & $0.15$ \\ 
  24 &  944 & $   72 \pm    10$ & $ 0.87 \pm 0.12$ & $ 0.39 \pm 0.05$ & $  2.5 \pm   0.5$ & $ 0.11 \pm  0.01$ & $ 0.08 \pm  0.01$ & $ 0.25 \pm 0.04$ & $ 0.14 \pm 0.03$ & $  1.6 \pm   0.4$ & $  0.9 \pm   0.3$ & $0.36$ \\ 
$\ddagger$25 & 1938 & $  59.8 \pm    9.2$ & $ 0.64 \pm 0.10$ & $ 0.27 \pm 0.04$ & $  2.4 \pm   0.6$ & $ 0.12 \pm  0.01$ & $ 0.11 \pm  0.02$ & $ 0.22 \pm 0.04$ & $ 0.05 \pm 0.02$ & $  1.9 \pm   0.5$ & $  0.5 \pm   0.2$ & $0.19$ \\ 
$\dagger$26 & 1279 & $   58 \pm    11$ & $ 0.72 \pm 0.15$ & $ 0.30 \pm 0.06$ & $  2.7 \pm   0.7$ & $ 0.13 \pm  0.01$ & $ 0.11 \pm  0.02$ & $ 0.21 \pm 0.05$ & $ 0.09 \pm 0.04$ & $  1.8 \pm   0.6$ & $  0.8 \pm   0.3$ & $0.30$ \\ 
  27 & 1720 & $   62 \pm    13$ & $ 0.87 \pm 0.18$ & $ 0.33 \pm 0.07$ & $  3.1 \pm   1.0$ & $ 0.12 \pm  0.00$ & $ 0.10 \pm  0.03$ & $ 0.22 \pm 0.06$ & $ 0.12 \pm 0.04$ & $  2.0 \pm   0.7$ & $  1.1 \pm   0.5$ & $0.36$ \\ 
  28 &  355 & $  100 \pm    31$ & $ 1.70 \pm 0.53$ & $ 0.69 \pm 0.21$ & $  6.3 \pm   2.4$ & $ 0.11 \pm  0.02$ & $ 0.11 \pm  0.03$ & $ 0.53 \pm 0.17$ & $ 0.16 \pm 0.12$ & $  4.9 \pm   2.0$ & $  1.4 \pm   1.2$ & $0.23$ \\ 
  29 & 1619 & $   84 \pm    19$ & $ 1.52 \pm 0.34$ & $ 0.70 \pm 0.16$ & $  4.9 \pm   1.4$ & $ 0.14 \pm  0.01$ & $ 0.10 \pm  0.02$ & $ 0.52 \pm 0.13$ & $ 0.19 \pm 0.09$ & $  3.6 \pm   1.1$ & $  1.3 \pm   0.7$ & $0.27$ \\ 
  30 & 1883 & $   83 \pm    25$ & $ 1.11 \pm 0.34$ & $ 0.42 \pm 0.13$ & $  5.0 \pm   1.9$ & $ 0.16 \pm  0.00$ & $ 0.16 \pm  0.04$ & $ 0.29 \pm 0.10$ & $ 0.13 \pm 0.07$ & $  3.4 \pm   1.5$ & $  1.6 \pm   0.9$ & $0.31$ \\ 
  31 & 1242 & $  116 \pm    24$ & $ 2.05 \pm 0.42$ & $ 0.88 \pm 0.18$ & $  7.6 \pm   2.2$ & $ 0.31 \pm  0.01$ & $ 0.26 \pm  0.05$ & $ 0.57 \pm 0.15$ & $ 0.31 \pm 0.11$ & $  4.9 \pm   1.6$ & $  2.7 \pm   1.1$ & $0.35$ \\ 
  32 &  784 & $   50 \pm    14$ & $ 0.95 \pm 0.27$ & $ 0.39 \pm 0.11$ & $  3.0 \pm   1.0$ & $ 0.07 \pm  0.01$ & $ 0.05 \pm  0.01$ & $ 0.23 \pm 0.09$ & $ 0.17 \pm 0.07$ & $  1.7 \pm   0.7$ & $  1.3 \pm   0.6$ & $0.44$ \\ 
  33 & 1537 & $  31.4 \pm    7.2$ & $ 0.80 \pm 0.18$ & $ 0.33 \pm 0.08$ & $  3.1 \pm   1.0$ & $ 0.00 \pm  0.00$ & $ 0.00 \pm  0.00$ & $ 0.29 \pm 0.07$ & $ 0.04 \pm 0.03$ & $  2.7 \pm   0.9$ & $  0.4 \pm   0.3$ & $0.12$ \\ 
  34 &  778 & $   74 \pm    32$ & $ 1.11 \pm 0.48$ & $ 0.72 \pm 0.31$ & $  4.1 \pm   2.0$ & $ 0.14 \pm  0.01$ & $ 0.12 \pm  0.03$ & $ 0.51 \pm 0.25$ & $ 0.21 \pm 0.18$ & $  2.9 \pm   1.6$ & $  1.2 \pm   1.1$ & $0.29$ \\ 
  35 & 1321 & $  31.0 \pm    9.0$ & $ 0.94 \pm 0.27$ & $ 0.33 \pm 0.10$ & $  4.4 \pm   1.7$ & $ 0.09 \pm  0.01$ & $ 0.08 \pm  0.02$ & $ 0.24 \pm 0.08$ & $ 0.09 \pm 0.05$ & $  3.3 \pm   1.3$ & $  1.2 \pm   0.8$ & $0.27$ \\ 
  36 &  828 & $  69.5 \pm    9.8$ & $ 1.52 \pm 0.21$ & $ 0.65 \pm 0.09$ & $  5.1 \pm   1.3$ & $ 0.00 \pm  0.00$ & $ 0.00 \pm  0.00$ & $ 0.50 \pm 0.08$ & $ 0.15 \pm 0.05$ & $  3.9 \pm   1.1$ & $  1.2 \pm   0.4$ & $0.23$ \\ 
$\dagger$37 & 1250 & $  20.1 \pm    7.3$ & $ 0.52 \pm 0.19$ & $ 0.18 \pm 0.07$ & $  5.5 \pm   3.6$ & $ 0.06 \pm  0.01$ & $ 0.14 \pm  0.08$ & $ 0.17 \pm 0.06$ & $ 0.01 \pm 0.03$ & $  5.0 \pm   3.3$ & $  0.4 \pm   1.0$ & $0.06$ \\ 
  38 & 1630 & $   47 \pm    11$ & $ 1.11 \pm 0.26$ & $ 0.44 \pm 0.10$ & $  4.0 \pm   1.3$ & $ 0.12 \pm  0.02$ & $ 0.11 \pm  0.03$ & $ 0.26 \pm 0.08$ & $ 0.18 \pm 0.06$ & $  2.4 \pm   0.9$ & $  1.7 \pm   0.7$ & $0.41$ \\ 
  39 & 1146 & $   72 \pm    12$ & $ 1.93 \pm 0.32$ & $ 1.00 \pm 0.17$ & $  8.3 \pm   2.5$ & $ 0.39 \pm  0.02$ & $ 0.40 \pm  0.10$ & $ 0.30 \pm 0.12$ & $ 0.70 \pm 0.12$ & $  2.5 \pm   1.1$ & $  5.8 \pm   1.8$ & $0.70$ \\ 
  40 & 1025 & $  141 \pm    34$ & $ 2.08 \pm 0.50$ & $ 0.84 \pm 0.20$ & $  6.6 \pm   2.1$ & $ 0.25 \pm  0.07$ & $ 0.19 \pm  0.08$ & $ 0.75 \pm 0.17$ & $ 0.09 \pm 0.11$ & $  5.9 \pm   1.8$ & $  0.7 \pm   0.9$ & $0.11$ \\ 
  41 & 1303 & $   72 \pm    18$ & $ 2.10 \pm 0.53$ & $ 0.87 \pm 0.22$ & $  6.9 \pm   2.3$ & $ 0.24 \pm  0.03$ & $ 0.18 \pm  0.05$ & $ 0.82 \pm 0.18$ & $ 0.05 \pm 0.11$ & $  6.5 \pm   2.1$ & $  0.4 \pm   0.9$ & $0.06$ \\ 
  42 & 1913 & $   71 \pm    14$ & $ 2.02 \pm 0.40$ & $ 0.90 \pm 0.18$ & $  6.6 \pm   2.1$ & $ 0.31 \pm  0.01$ & $ 0.23 \pm  0.06$ & $ 0.74 \pm 0.15$ & $ 0.16 \pm 0.09$ & $  5.4 \pm   1.8$ & $  1.2 \pm   0.8$ & $0.18$ \\ 
$\dagger$43 & 1327 & $   11 \pm    12$ & $ 0.26 \pm 0.29$ & $ 0.14 \pm 0.15$ & $  0.7 \pm   0.7$ & $ 0.07 \pm  0.00$ & $ 0.05 \pm  0.01$ & $ 0.09 \pm 0.12$ & $ 0.05 \pm 0.09$ & $  0.4 \pm   0.6$ & $  0.2 \pm   0.4$ & $0.36$ \\ 
  44 & 1125 & $  52.3 \pm    8.5$ & $ 1.23 \pm 0.20$ & $ 0.88 \pm 0.14$ & $  6.5 \pm   2.9$ & $ 0.16 \pm  0.01$ & $ 0.20 \pm  0.08$ & $ 0.81 \pm 0.13$ & $ 0.07 \pm 0.06$ & $  6.0 \pm   2.7$ & $  0.5 \pm   0.5$ & $0.08$ \\ 
  45 & 1475 & $   81 \pm    10$ & $ 3.05 \pm 0.40$ & $ 1.38 \pm 0.18$ & $ 10.5 \pm   2.5$ & $ 0.33 \pm  0.03$ & $ 0.23 \pm  0.05$ & $ 1.33 \pm 0.17$ & $ 0.06 \pm 0.06$ & $ 10.1 \pm   2.4$ & $  0.4 \pm   0.5$ & $0.04$ \\ 
$\dagger$46 & 1178 & $  25.3 \pm    9.2$ & $ 1.16 \pm 0.42$ & $ 0.46 \pm 0.17$ & $  3.6 \pm   1.5$ & $ 0.12 \pm  0.01$ & $ 0.09 \pm  0.02$ & $ 0.31 \pm 0.14$ & $ 0.15 \pm 0.10$ & $  2.4 \pm   1.2$ & $  1.2 \pm   0.8$ & $0.33$ \\ 
  47 & 1283 & $  58.6 \pm    9.3$ & $ 2.02 \pm 0.32$ & $ 0.77 \pm 0.12$ & $  6.5 \pm   1.6$ & $ 0.22 \pm  0.03$ & $ 0.16 \pm  0.04$ & $ 0.59 \pm 0.10$ & $ 0.18 \pm 0.06$ & $  5.0 \pm   1.3$ & $  1.5 \pm   0.6$ & $0.23$ \\ 
  48 & 1261 & $  35.1 \pm    7.6$ & $ 1.51 \pm 0.33$ & $ 0.72 \pm 0.16$ & $  7.2 \pm   2.9$ & $ 0.29 \pm  0.02$ & $ 0.28 \pm  0.10$ & $ 0.60 \pm 0.14$ & $ 0.12 \pm 0.08$ & $  6.0 \pm   2.5$ & $  1.2 \pm   0.9$ & $0.17$ \\ 
  49 &  698 & $  114 \pm    12$ & $ 3.53 \pm 0.38$ & $ 1.50 \pm 0.16$ & $ 12.0 \pm   3.5$ & $ 0.56 \pm  0.01$ & $ 0.41 \pm  0.11$ & $ 1.23 \pm 0.14$ & $ 0.27 \pm 0.07$ & $  9.8 \pm   2.9$ & $  2.2 \pm   0.8$ & $0.18$ \\ 
  50 & 1422 & $  24.9 \pm    6.0$ & $ 1.62 \pm 0.39$ & $ 0.80 \pm 0.19$ & $  6.5 \pm   2.8$ & $ 0.21 \pm  0.02$ & $ 0.17 \pm  0.06$ & $ 0.54 \pm 0.16$ & $ 0.26 \pm 0.11$ & $  4.4 \pm   2.0$ & $  2.1 \pm   1.2$ & $0.32$ \\ 
  51 & 2048 & $  17.2 \pm    5.4$ & $ 1.25 \pm 0.39$ & $ 0.64 \pm 0.20$ & $  5.9 \pm   2.8$ & $ 0.19 \pm  0.02$ & $ 0.19 \pm  0.07$ & $ 0.49 \pm 0.17$ & $ 0.15 \pm 0.11$ & $  4.5 \pm   2.3$ & $  1.3 \pm   1.1$ & $0.23$ \\ 
  52 & 1871 & $  10.4 \pm    5.0$ & $ 1.24 \pm 0.60$ & $ 0.49 \pm 0.24$ & $  4.6 \pm   2.5$ & $ 0.21 \pm  0.07$ & $ 0.17 \pm  0.07$ & $ 0.48 \pm 0.21$ & $ 0.01 \pm 0.11$ & $  4.5 \pm   2.3$ & $  0.1 \pm   1.0$ & $0.02$ \\ 
  53 &    9 & $  25.7 \pm    6.4$ & $ 1.57 \pm 0.39$ & $ 0.78 \pm 0.19$ & $  8.3 \pm   2.6$ & $ 0.21 \pm  0.02$ & $ 0.20 \pm  0.05$ & $ 0.68 \pm 0.17$ & $ 0.09 \pm 0.09$ & $  7.3 \pm   2.4$ & $  1.0 \pm   1.0$ & $0.12$ \\ 
  54 &  575 & $  18.0 \pm    6.1$ & $ 0.77 \pm 0.26$ & $ 0.33 \pm 0.11$ & $  3.6 \pm   1.7$ & $ 0.00 \pm  0.06$ & $-0.00 \pm  0.06$ & $ 0.26 \pm 0.10$ & $ 0.07 \pm 0.06$ & $  2.8 \pm   1.4$ & $  0.8 \pm   0.7$ & $0.21$ \\ 
  55 & 1910 & $  48.7 \pm    8.4$ & $ 5.38 \pm 0.93$ & $ 2.27 \pm 0.39$ & $ 23.2 \pm   9.8$ & $ 1.09 \pm  0.04$ & $ 1.12 \pm  0.43$ & $ 1.47 \pm 0.32$ & $ 0.79 \pm 0.23$ & $ 15.1 \pm   6.7$ & $  8.1 \pm   3.9$ & $0.35$ \\ 
  56 & 1049 & $   8.4 \pm    4.5$ & $ 1.77 \pm 0.95$ & $ 0.97 \pm 0.52$ & $ 16.2 \pm  15.1$ & $ 0.05 \pm  0.03$ & $ 0.09 \pm  0.08$ & $ 0.75 \pm 0.44$ & $ 0.22 \pm 0.28$ & $ 12.4 \pm  12.0$ & $  3.7 \pm   5.5$ & $0.23$ \\ 
  57 &  856 & $  43.4 \pm    7.9$ & $ 4.07 \pm 0.74$ & $ 2.17 \pm 0.39$ & $ 19.5 \pm   8.3$ & $ 0.55 \pm  0.03$ & $ 0.56 \pm  0.22$ & $ 1.80 \pm 0.35$ & $ 0.37 \pm 0.19$ & $ 16.2 \pm   6.9$ & $  3.3 \pm   2.1$ & $0.17$ \\ 
  58 &  140 & $  21.3 \pm    6.1$ & $ 2.10 \pm 0.60$ & $ 1.05 \pm 0.30$ & $  9.2 \pm   3.8$ & $ 0.18 \pm  0.03$ & $ 0.16 \pm  0.06$ & $ 0.79 \pm 0.25$ & $ 0.26 \pm 0.16$ & $  6.9 \pm   3.1$ & $  2.3 \pm   1.6$ & $0.25$ \\ 
  59 & 1355 & $  10.8 \pm    5.6$ & $ 1.07 \pm 0.56$ & $ 0.59 \pm 0.30$ & $  5.9 \pm   3.3$ & $ 0.05 \pm  0.01$ & $ 0.07 \pm  0.02$ & $ 0.53 \pm 0.26$ & $ 0.06 \pm 0.15$ & $  5.3 \pm   2.8$ & $  0.6 \pm   1.6$ & $0.10$ \\ 
  60 & 1087 & $  66.0 \pm    9.5$ & $ 5.07 \pm 0.73$ & $ 2.31 \pm 0.33$ & $ 20.0 \pm   7.1$ & $ 0.87 \pm  0.03$ & $ 0.73 \pm  0.24$ & $ 2.17 \pm 0.31$ & $ 0.14 \pm 0.12$ & $ 18.8 \pm   6.7$ & $  1.2 \pm   1.1$ & $0.06$ \\ 
$\dagger$61 & 1297 & $    4 \pm    11$ & $ 0.42 \pm 1.00$ & $ 0.16 \pm 0.37$ & $  1.2 \pm   3.0$ & $ 0.15 \pm  0.02$ & $ 0.12 \pm  0.02$ & $ 0.11 \pm 0.30$ & $ 0.05 \pm 0.22$ & $  0.9 \pm   2.4$ & $  0.4 \pm   1.7$ & $0.31$ \\ 
  62 & 1861 & $  37.6 \pm    7.4$ & $ 3.44 \pm 0.68$ & $ 1.73 \pm 0.34$ & $ 13.0 \pm   5.2$ & $ 0.69 \pm  0.07$ & $ 0.49 \pm  0.18$ & $ 1.26 \pm 0.29$ & $ 0.47 \pm 0.18$ & $  9.5 \pm   3.9$ & $  3.5 \pm   1.8$ & $0.27$ \\ 
  63 &  543 & $  18.1 \pm    5.5$ & $ 1.97 \pm 0.60$ & $ 0.96 \pm 0.29$ & $  8.3 \pm   3.5$ & $ 0.15 \pm  0.03$ & $ 0.14 \pm  0.05$ & $ 0.96 \pm 0.27$ & $ 0.00 \pm 0.11$ & $  8.3 \pm   3.4$ & $  0.0 \pm   0.9$ & $0.00$ \\ 
  64 & 1431 & $  60.6 \pm    9.3$ & $ 6.72 \pm 1.03$ & $ 3.03 \pm 0.46$ & $ 27.5 \pm  10.2$ & $ 0.78 \pm  0.04$ & $ 0.70 \pm  0.24$ & $ 2.57 \pm 0.41$ & $ 0.45 \pm 0.21$ & $ 23.4 \pm   8.7$ & $  4.1 \pm   2.4$ & $0.15$ \\ 
  65 & 1528 & $  40.7 \pm    7.6$ & $ 5.57 \pm 1.04$ & $ 2.46 \pm 0.46$ & $ 24.9 \pm   8.7$ & $ 0.66 \pm  0.04$ & $ 0.69 \pm  0.21$ & $ 2.24 \pm 0.42$ & $ 0.22 \pm 0.19$ & $ 22.7 \pm   7.9$ & $  2.2 \pm   2.0$ & $0.09$ \\ 
  66 & 1695 & $  14.4 \pm    5.7$ & $ 1.45 \pm 0.58$ & $ 0.68 \pm 0.27$ & $  8.5 \pm   5.2$ & $ 0.09 \pm  0.03$ & $ 0.10 \pm  0.06$ & $ 0.48 \pm 0.22$ & $ 0.20 \pm 0.15$ & $  6.0 \pm   3.9$ & $  2.5 \pm   2.2$ & $0.29$ \\ 
  67 & 1833 & $  18.1 \pm    5.5$ & $ 2.55 \pm 0.77$ & $ 1.26 \pm 0.38$ & $ 21.1 \pm  13.5$ & $ 0.15 \pm  0.04$ & $ 0.25 \pm  0.16$ & $ 0.87 \pm 0.32$ & $ 0.39 \pm 0.22$ & $ 14.6 \pm   9.8$ & $  6.5 \pm   5.2$ & $0.31$ \\ 
  68 &  437 & $  42.1 \pm    7.9$ & $ 3.13 \pm 0.59$ & $ 1.40 \pm 0.26$ & $ 15.0 \pm   7.2$ & $ 0.46 \pm  0.04$ & $ 0.34 \pm  0.15$ & $ 1.32 \pm 0.24$ & $ 0.08 \pm 0.10$ & $ 14.1 \pm   6.8$ & $  0.9 \pm   1.2$ & $0.06$ \\ 
  69 & 2019 & $  23.9 \pm    6.1$ & $ 2.72 \pm 0.69$ & $ 1.28 \pm 0.33$ & $ 23.5 \pm  17.8$ & $ 0.72 \pm  0.05$ & $ 1.19 \pm  0.85$ & $ 1.28 \pm 0.31$ & $ 0.00 \pm 0.11$ & $ 23.5 \pm  17.6$ & $  0.0 \pm   2.1$ & $0.00$ \\ 
  70 &   33 & $   2.2 \pm    4.2$ & $ 0.61 \pm 1.17$ & $ 0.34 \pm 0.66$ & $  5.1 \pm  10.1$ & $-0.13 \pm  0.07$ & $-0.21 \pm  0.16$ & $ 0.22 \pm 0.53$ & $ 0.12 \pm 0.38$ & $  3.3 \pm   8.1$ & $  1.8 \pm   5.7$ & $0.35$ \\ 
  71 &  200 & $  15.5 \pm    5.8$ & $ 2.21 \pm 0.83$ & $ 1.18 \pm 0.44$ & $ 11.5 \pm   6.0$ & $ 0.05 \pm  0.05$ & $ 0.05 \pm  0.06$ & $ 1.18 \pm 0.40$ & $ 0.00 \pm 0.19$ & $ 11.5 \pm   5.8$ & $  0.0 \pm   1.8$ & $0.00$ \\ 
  72 &  571 & $  10.9 \pm    5.6$ & $ 1.29 \pm 0.66$ & $ 0.53 \pm 0.27$ & $ 11.2 \pm   9.4$ & $ 0.14 \pm  0.07$ & $ 0.17 \pm  0.20$ & $ 0.53 \pm 0.24$ & $ 0.00 \pm 0.13$ & $ 11.2 \pm   9.0$ & $  0.0 \pm   2.7$ & $0.00$ \\ 
  73 &   21 & $  20.7 \pm    6.5$ & $ 3.85 \pm 1.21$ & $ 2.32 \pm 0.73$ & $ 37.0 \pm  27.2$ & $ 0.67 \pm  0.08$ & $ 0.97 \pm  0.65$ & $ 2.32 \pm 0.66$ & $ 0.00 \pm 0.30$ & $ 37.0 \pm  26.7$ & $  0.0 \pm   4.8$ & $0.00$ \\ 
  74 & 1488 & $   7.3 \pm    4.2$ & $ 1.42 \pm 0.82$ & $ 0.85 \pm 0.49$ & $ 17.9 \pm  19.5$ & $ 0.03 \pm  0.04$ & $ 0.08 \pm  0.12$ & $ 0.84 \pm 0.44$ & $ 0.01 \pm 0.22$ & $ 17.7 \pm  18.8$ & $  0.2 \pm   4.6$ & $0.01$ \\ 
  75 & 1779 & $   2.1 \pm    3.7$ & $ 0.37 \pm 0.64$ & $ 0.23 \pm 0.41$ & $  2.1 \pm   3.7$ & $-0.11 \pm  0.05$ & $-0.13 \pm  0.08$ & $ 0.23 \pm 0.35$ & $ 0.00 \pm 0.22$ & $  2.1 \pm   3.2$ & $  0.0 \pm   2.0$ & $0.00$ \\ 
  76 & 1895 & $   6.3 \pm    4.3$ & $ 1.45 \pm 0.99$ & $ 0.76 \pm 0.52$ & $  8.3 \pm   7.2$ & $ 0.07 \pm  0.06$ & $ 0.05 \pm  0.07$ & $ 0.65 \pm 0.44$ & $ 0.11 \pm 0.27$ & $  7.1 \pm   6.1$ & $  1.3 \pm   3.1$ & $0.14$ \\ 
  77 & 1499 & $  19.0 \pm    6.6$ & $ 4.65 \pm 1.62$ & $ 3.14 \pm 1.09$ & $133.8 \pm 185.1$ & $ 0.98 \pm  0.14$ & $ 4.07 \pm  5.47$ & $ 2.67 \pm 0.94$ & $ 0.47 \pm 0.55$ & $113.7 \pm 157.5$ & $ 20.1 \pm  35.7$ & $0.15$ \\ 
  78 & 1545 & $  54.2 \pm    8.8$ & $ 9.37 \pm 1.52$ & $ 4.34 \pm 0.71$ & $ 38.5 \pm  13.0$ & $ 1.02 \pm  0.07$ & $ 0.76 \pm  0.23$ & $ 3.91 \pm 0.64$ & $ 0.43 \pm 0.29$ & $ 34.6 \pm  11.8$ & $  3.8 \pm   2.8$ & $0.10$ \\ 
$\dagger$79 & 1192 & $   -6 \pm    13$ & $-1.12 \pm 2.42$ & $-0.34 \pm 0.74$ & $ -3.4 \pm   7.3$ & $-0.39 \pm  0.05$ & $-0.31 \pm  0.11$ & $-0.21 \pm 0.61$ & $-0.13 \pm 0.45$ & $ -2.1 \pm   5.9$ & $ -1.3 \pm   4.4$ & $0.00$ \\ 
  80 & 1857 & $  10.8 \pm    5.9$ & $ 2.39 \pm 1.31$ & $ 1.28 \pm 0.70$ & $ 14.8 \pm   8.6$ & $ 0.32 \pm  0.09$ & $ 0.29 \pm  0.12$ & $ 0.78 \pm 0.56$ & $ 0.50 \pm 0.42$ & $  9.1 \pm   6.7$ & $  5.8 \pm   5.0$ & $0.39$ \\ 
  81 & 1075 & $  16.5 \pm    5.2$ & $ 3.19 \pm 1.01$ & $ 1.58 \pm 0.50$ & $ 14.8 \pm   7.8$ & $ 0.33 \pm  0.06$ & $ 0.31 \pm  0.14$ & $ 1.58 \pm 0.46$ & $ 0.00 \pm 0.18$ & $ 14.8 \pm   7.6$ & $  0.0 \pm   1.7$ & $0.00$ \\ 
  82 & 1948 & $   2.5 \pm    3.0$ & $ 0.94 \pm 1.13$ & $ 0.48 \pm 0.58$ & $  5.0 \pm   6.1$ & $-0.28 \pm  0.17$ & $-0.22 \pm  0.12$ & $ 0.48 \pm 0.50$ & $ 0.00 \pm 0.29$ & $  5.0 \pm   5.3$ & $  0.0 \pm   3.0$ & $0.00$ \\ 
  83 & 1627 & $   3.6 \pm    3.7$ & $ 0.94 \pm 0.97$ & $ 0.40 \pm 0.41$ & $  3.5 \pm   3.8$ & $ 1.21 \pm  0.13$ & $ 1.05 \pm  0.34$ & $ 0.40 \pm 0.36$ & $ 0.00 \pm 0.21$ & $  3.5 \pm   3.3$ & $  0.0 \pm   1.8$ & $0.00$ \\ 
  84 & 1440 & $  26.7 \pm    6.8$ & $ 4.81 \pm 1.22$ & $ 2.30 \pm 0.58$ & $ 22.5 \pm  10.1$ & $ 0.29 \pm  0.05$ & $ 0.31 \pm  0.12$ & $ 1.93 \pm 0.51$ & $ 0.37 \pm 0.28$ & $ 18.9 \pm   8.6$ & $  3.6 \pm   3.1$ & $0.16$ \\ 
  85 &  230 & $  28.7 \pm    6.7$ & $ 9.39 \pm 2.19$ & $ 4.73 \pm 1.10$ & $ 41.4 \pm  20.0$ & $ 5.45 \pm  1.17$ & $ 3.68 \pm  1.73$ & $ 4.64 \pm 1.03$ & $ 0.09 \pm 0.41$ & $ 40.5 \pm  19.4$ & $  0.8 \pm   3.6$ & $0.02$ \\ 
  86 & 2050 & $   9.2 \pm    4.3$ & $ 2.63 \pm 1.23$ & $ 1.27 \pm 0.60$ & $ 31.5 \pm  32.0$ & $ 0.21 \pm  0.06$ & $ 0.43 \pm  0.41$ & $ 1.27 \pm 0.54$ & $ 0.00 \pm 0.25$ & $ 31.5 \pm  31.4$ & $  0.0 \pm   6.1$ & $0.00$ \\ 
  87 & 1993 & $  -1.6 \pm    2.1$ & $-0.48 \pm 0.63$ & $-0.25 \pm 0.33$ & $ -2.1 \pm   2.9$ & $-0.07 \pm  0.08$ & $ 0.29 \pm  0.13$ & $-0.03 \pm 0.30$ & $-0.22 \pm 0.29$ & $ -0.3 \pm   2.6$ & $ -1.9 \pm   2.6$ & $0.00$ \\ 
  88 &  751 & $   9.2 \pm    4.3$ & $ 1.49 \pm 0.70$ & $ 0.62 \pm 0.29$ & $  6.5 \pm   3.9$ & $ 0.13 \pm  0.04$ & $ 0.11 \pm  0.06$ & $ 0.62 \pm 0.27$ & $ 0.00 \pm 0.12$ & $  6.5 \pm   3.6$ & $  0.0 \pm   1.3$ & $0.00$ \\ 
  89 & 1828 & $  20.4 \pm    5.8$ & $ 3.94 \pm 1.12$ & $ 2.10 \pm 0.60$ & $ 16.3 \pm   7.7$ & $ 0.95 \pm  0.08$ & $ 0.78 \pm  0.30$ & $ 2.10 \pm 0.56$ & $ 0.00 \pm 0.22$ & $ 16.3 \pm   7.5$ & $  0.0 \pm   1.7$ & $0.00$ \\ 
  90 &  538 & $   0.6 \pm    3.5$ & $ 0.19 \pm 1.11$ & $ 0.09 \pm 0.51$ & $  0.9 \pm   5.5$ & $-0.31 \pm  0.22$ & $-0.25 \pm  0.19$ & $ 0.09 \pm 0.42$ & $ 0.00 \pm 0.29$ & $  0.9 \pm   4.6$ & $  0.0 \pm   3.1$ & $0.00$ \\ 
  91 & 1407 & $  49.7 \pm    8.6$ & $10.18 \pm 1.76$ & $ 5.30 \pm 0.92$ & $ 40.2 \pm  15.5$ & $ 0.69 \pm  0.07$ & $ 0.50 \pm  0.18$ & $ 4.61 \pm 0.82$ & $ 0.69 \pm 0.41$ & $ 35.0 \pm  13.5$ & $  5.2 \pm   3.6$ & $0.13$ \\ 
  92 & 1886 & $   3.9 \pm    2.6$ & $ 1.24 \pm 0.82$ & $ 0.66 \pm 0.44$ & $  7.7 \pm   5.3$ & $ 0.29 \pm  0.06$ & $ 0.41 \pm  0.11$ & $ 0.66 \pm 0.41$ & $ 0.00 \pm 0.17$ & $  7.7 \pm   5.0$ & $  0.0 \pm   1.9$ & $0.00$ \\ 
$\dagger$93 & 1199 & $   -9 \pm    14$ & $-6.24 \pm 9.23$ & $-1.60 \pm 2.37$ & $-16.6 \pm  24.9$ & $-3.39 \pm  0.24$ & $-2.55 \pm  0.74$ & $-0.90 \pm 1.93$ & $-0.70 \pm 1.45$ & $ -9.3 \pm  20.1$ & $ -7.3 \pm  15.2$ & $0.00$ \\ 
  94 & 1743 & $   9.8 \pm    6.4$ & $ 2.88 \pm 1.88$ & $ 1.63 \pm 1.06$ & $ 24.6 \pm  23.9$ & $-0.03 \pm  0.29$ & $-0.13 \pm  0.55$ & $ 1.11 \pm 0.87$ & $ 0.52 \pm 0.62$ & $ 16.7 \pm  17.8$ & $  7.9 \pm  10.9$ & $0.32$ \\ 
  95 & 1539 & $  31.0 \pm    7.0$ & $11.83 \pm 2.67$ & $ 4.40 \pm 0.99$ & $ 59.4 \pm  18.7$ & $ 2.41 \pm  0.33$ & $ 1.73 \pm  0.40$ & $ 4.40 \pm 0.93$ & $ 0.00 \pm 0.35$ & $ 59.4 \pm  18.1$ & $  0.0 \pm   4.7$ & $0.00$ \\ 
$\dagger$96 & 1185 & $  14.0 \pm    5.7$ & $ 2.73 \pm 1.11$ & $ 1.58 \pm 0.64$ & $ 15.7 \pm  11.0$ & $ 0.20 \pm  0.04$ & $ 0.26 \pm  0.16$ & $ 1.45 \pm 0.56$ & $ 0.13 \pm 0.31$ & $ 14.4 \pm   9.9$ & $  1.3 \pm   3.1$ & $0.08$ \\ 
  97 & 1826 & $   4.5 \pm    3.9$ & $ 1.74 \pm 1.51$ & $ 0.93 \pm 0.81$ & $  7.3 \pm   6.9$ & $ 0.08 \pm  0.11$ & $-0.01 \pm  0.12$ & $ 0.93 \pm 0.71$ & $ 0.00 \pm 0.40$ & $  7.3 \pm   6.1$ & $  0.0 \pm   3.1$ & $0.00$ \\ 
  98 & 1512 & $   4.7 \pm    3.8$ & $ 1.48 \pm 1.20$ & $ 1.14 \pm 0.92$ & $  7.7 \pm   6.5$ & $-0.40 \pm  0.58$ & $-0.42 \pm  0.42$ & $ 1.14 \pm 0.81$ & $ 0.00 \pm 0.44$ & $  7.7 \pm   5.7$ & $  0.0 \pm   3.0$ & $0.00$ \\ 
  99 & 1489 & $  11.7 \pm    4.8$ & $ 6.69 \pm 2.74$ & $ 3.19 \pm 1.31$ & $ 36.9 \pm  16.7$ & $ 0.46 \pm  0.16$ & $ 0.34 \pm  0.14$ & $ 2.20 \pm 1.08$ & $ 0.99 \pm 0.75$ & $ 25.4 \pm  13.3$ & $ 11.4 \pm   8.9$ & $0.31$ \\ 
 100 & 1661 & $  10.2 \pm    4.7$ & $ 4.83 \pm 2.23$ & $ 1.12 \pm 0.51$ & $ 23.9 \pm  12.0$ & $ 0.08 \pm  0.08$ & $ 0.06 \pm  0.14$ & $ 1.12 \pm 0.46$ & $ 0.00 \pm 0.22$ & $ 23.9 \pm  11.0$ & $  0.0 \pm   4.7$ & $0.00$ \\ 
\enddata
\tablenotetext{1}{Running number, sorted by increasing $B_T$ magnitude}
\tablenotetext{2}{Number in Virgo Cluster Catalog}
\tablenotetext{3}{Total number of globular clusters}
\tablenotetext{4}{Specific frequency}
\tablenotetext{5}{Specific frequency in $z$ bandpass}
\tablenotetext{6}{$N_{GC}$ normalized to stellar mass of $10^9 M_\odot$}
\tablenotetext{7}{Percentage of galaxy $z$ luminosity in GCs}
\tablenotetext{8}{Percentage of galaxy stellar mass in GCs}
\tablenotetext{9}{Specific frequency in $z$ for blue GCs}
\tablenotetext{10}{Specific frequency in $z$ for red GCs}
\tablenotetext{11}{$T$ for blue GCs}
\tablenotetext{12}{$T$ for red GCs}
\tablenotetext{13}{Fraction of red GCs}
\tablenotetext{$\dagger$}{Quantities calculated using an R=70\arcsec aperture around galaxy}
\tablenotetext{$\ddagger$}{Quantities possibly affected by nearby dE VCC 1941, which has been masked}
\tablenotetext{$\star$}{Mass-to-light ratio (i.e.\ optical-IR colors) suspect.}
\end{deluxetable}

\clearpage
\end{landscape}
%%%%%%%%%%%%%%%%%%%%%%%%%%%%%%%%%%

%\clearpage

%\clearpage

%\clearpage

%\clearpage

%% The following command ends your manuscript. LaTeX will ignore any text


\clearpage

\begin{thebibliography}{}

\bibitem[Adelman-McCarthy et al.(2007)]{2007ApJS..172..634A} 
Adelman-McCarthy, J.~K., et al.\ 2007, \apjs, 172, 634

\bibitem[Ashman \& Zepf(1992)]{1992ApJ...384...50A} Ashman, K.~M., \& Zepf, 
S.~E.\ 1992, \apj, 384, 50

\bibitem[Ashman \& Zepf(1998)]{1998gcs..book.....A} Ashman, K.~M., \& Zepf, 
S.~E.\ 1998, Globular cluster systems / Keith M.~Ashman, Stephen E.~Zepf.~ 
Cambridge, U.~K.~; New York : Cambridge University Press, 1998.~(Cambridge 
astrophysics series ; 30) QB853.5 .A84 1998

\bibitem[Ashman \& Zepf(2001)]{2001AJ....122.1888A} Ashman, K.~M., \& Zepf, 
S.~E.\ 2001, \aj, 122, 1888

\bibitem[Bagget et al.(2002)]{2002hst} Bagget, S.\ et al.\ 2002, in HST
  WFPC2 Data Handbook, v. 4.0, ed. B. Mobasher, Baltimore, STScI

\bibitem[Bastian et al.(2005)]{2005A&A...431..905B} Bastian, N., Gieles, 
M., Lamers, H.~J.~G.~L.~M., Scheepmaker, R.~A., \& de Grijs, R.\ 2005, 
\aap, 431, 905

\bibitem[Beasley et al.(2002)]{2002MNRAS.333..383B} Beasley, M.~A., Baugh, 
C.~M., Forbes, D.~A., Sharples, R.~M., \& Frenk, C.~S.\ 2002, \mnras, 333, 
383

\bibitem[Bekki et al.(2006)]{2006ApJ...645L..29B} Bekki, K., Yahagi, H., \& 
Forbes, D.~A.\ 2006, \apjl, 645, L29

\bibitem[Benson et al.(2002)]{2002MNRAS.333..177B} Benson, A.~J., Frenk, 
C.~S., Lacey, C.~G., Baugh, C.~M., \& Cole, S.\ 2002, \mnras, 333, 177

\bibitem[Berlind \& Weinberg(2002)]{2002ApJ...575..587B} Berlind, A.~A., \& 
Weinberg, D.~H.\ 2002, \apj, 575, 587 

\bibitem[Bertin \& Arnouts(1996)]{1996A&AS..117..393B} Bertin, E., \&
Arnouts, S.\ 1996, \aaps, 117, 393

\bibitem[Binggeli et al.(1985)]{1985AJ.....90.1681B} Binggeli, B., Sandage, 
A., \& Tammann, G.~A.\ 1985, \aj, 90, 1681

\bibitem[Blakeslee(1997)]{1997ApJ...481L..59B} Blakeslee, J.~P.\ 1997, 
\apjl, 481, L59

\bibitem[Blakeslee et al.(1997)]{1997AJ....114..482B} Blakeslee, J.~P., 
Tonry, J.~L., \& Metzger, M.~R.\ 1997, \aj, 114, 482

\bibitem[Blakeslee(1999)]{1999AJ....118.1506B} Blakeslee, J.~P.\ 1999, \aj, 
118, 1506

\bibitem[Blakeslee et al.(2003)]{2003IAUJD...6E..34B} Blakeslee, J.~P., 
Meurer, G.~R., Ford, H.~C., Benitez, N., White, R.~L., Zekser, K.~C., \& 
Sirianni, M.\ 2003, Extragalactic Globular Clusters and their Host 
Galaxies, 25th meeting of the IAU, Joint Discussion 6, 17 July 2003, 
Sydney, Australia, 6

\bibitem[Blakeslee et al.(2003b)]{2003ASPC..295..257B} Blakeslee, J.~P., 
Anderson, K.~R., Meurer, G.~R., Ben{\'{\i}}tez, N., \& Magee, D.\ 2003, 
Astronomical Data Analysis Software and Systems XII, 295, 257

\bibitem[Blanton \& Roweis(2007)]{2007AJ....133..734B} Blanton, M.~R., \& 
Roweis, S.\ 2007, \aj, 133, 734

\bibitem[Bohringer et al.(1994)]{1994Natur.368..828B} Bohringer, H., Briel, 
U.~G., Schwarz, R.~A., Voges, W., Hartner, G., \& Trumper, J.\ 1994, \nat, 
368, 828

\bibitem[Brodie \& Huchra(1991)]{1991ApJ...379..157B} Brodie, J.~P., \& 
Huchra, J.~P.\ 1991, \apj, 379, 157

\bibitem[Brodie \& Strader(2006)]{2006ARA&A..44..193B} Brodie, J.~P., \& 
Strader, J.\ 2006, \araa, 44, 193 

\bibitem[Bullock et al.(2000)]{2000ApJ...539..517B} Bullock, J.~S., 
Kravtsov, A.~V., \& Weinberg, D.~H.\ 2000, \apj, 539, 517 

\bibitem[Cantiello \& Blakeslee(2007)]{2007ApJ...669..982C} Cantiello, M., 
\& Blakeslee, J.~P.\ 2007, \apj, 669, 982

\bibitem[Cen(2001)]{2001ApJ...560..592C} Cen, R.\ 2001, \apj, 560, 592 

\bibitem[Cappellari et al.(2006)]{2006MNRAS.366.1126C} Cappellari, M., et 
al.\ 2006, \mnras, 366, 1126 

\bibitem[Chaboyer et al.(1996)]{1996Sci...271..957C} Chaboyer, B., 
Demarque, P., Kernan, P.~J., \& Krauss, L.~M.\ 1996, Science, 271, 957 

\bibitem[Chabrier(2003)]{2003PASP..115..763C} Chabrier, G.\ 2003, \pasp, 
  115, 763

\bibitem[Chandar et al.(2004)]{2004ApJ...611..220C} Chandar, R., Whitmore, 
B., \& Lee, M.~G.\ 2004, \apj, 611, 220

\bibitem[Cote et al.(1998)]{1998ApJ...501..554C} C{\^ o}t{\' e}, P., 
Marzke, R.~O., \& West, M.~J.\ 1998, \apj, 501, 554

\bibitem[C{\^ o}t{\' e} et al.(2000)]{2000ApJ...533..869C} C{\^ o}t{\' e}, 
P., Marzke, R.~O., West, M.~J., \& Minniti, D.\ 2000, \apj, 533, 869

\bibitem[C{\^ o}t{\' e} et al.(2001)]{2001ApJ...559..828C} C{\^ o}t{\' e}, 
P., et al.\ 2001, \apj, 559, 828

\bibitem[C{\^ o}t{\' e} et al.(2003)]{2003ApJ...591..850C} C{\^ o}t{\' e}, 
P., McLaughlin, D.~E., Cohen, J.~G., \& Blakeslee, J.~P.\ 2003, \apj, 591, 
850

\bibitem[C{\^ o}t{\' e} et al.(2004)]{2004ApJS..153..223C} C{\^ o}t{\' e}, 
P., et al.\ 2004, \apjs, 153, 223 (Paper I)

\bibitem[C{\^o}t{\'e} et al.(2006)]{2006ApJS..165...57C} C{\^o}t{\'e}, P., 
et al.\ 2006, \apjs, 165, 57

\bibitem[Croton et al.(2006)]{2006MNRAS.365...11C} Croton, D.~J., et al.\ 
2006, \mnras, 365, 11

\bibitem[C{\^o}t{\'e} et al.(2007)]{2007ApJ..} C{\^o}t{\'e}, P., 
et al.\ 2007, \apj, submitted

\bibitem[De Lucia et al.(2006)]{2006MNRAS.366..499D} De Lucia, G., 
Springel, V., White, S.~D.~M., Croton, D., \& Kauffmann, G.\ 2006, \mnras, 
366, 499

\bibitem[De Lucia \& Blaizot(2007)]{2007MNRAS.375....2D} De Lucia, G., \& 
Blaizot, J.\ 2007, \mnras, 375, 2 

\bibitem[de Vaucouleurs et al.(1991)]{1991trcb.book.....D} de Vaucouleurs, 
G., de Vaucouleurs, A., Corwin, H.~G., Jr., Buta, R.~J., Paturel, G., \& 
Fouque, P.\ 1991, Volume 1-3, XII, 2069 pp.~7 figs..~ Springer-Verlag 
Berlin Heidelberg New York

\bibitem[Dekel \& Birnboim(2006)]{2006MNRAS.368....2D} Dekel, A., \& 
Birnboim, Y.\ 2006, \mnras, 368, 2

\bibitem[Diemand et al.(2005)]{2005MNRAS.364..367D} Diemand, J., Madau, P., 
\& Moore, B.\ 2005, \mnras, 364, 367

\bibitem[Dirsch et al.(2003)]{2003AJ....125.1908D} Dirsch, B., Richtler, 
T., Geisler, D., Forte, J.~C., Bassino, L.~P., \& Gieren, W.~P.\ 2003a, \aj, 
125, 1908

\bibitem[Dirsch et al.(2003)]{2003A&A...408..929D} Dirsch, B., Richtler, 
T., \& Bassino, L.~P.\ 2003b, \aap, 408, 929

\bibitem[Dirsch et al.(2005)]{2005A&A...433...43D} Dirsch, B., Schuberth, 
Y., \& Richtler, T.\ 2005, \aap, 433, 43

\bibitem[Durrell et al.(1996)]{1996AJ....112..972D} Durrell, P.~R., Harris, 
W.~E., Geisler, D., \& Pudritz, R.~E.\ 1996a, \aj, 112, 972

\bibitem[Durrell et al.(1996)]{1996ApJ...463..543D} Durrell, P.~R., 
McLaughlin, D.~E., Harris, W.~E., \& Hanes, D.~A.\ 1996b, \apj, 463, 543 

\bibitem[Elmegreen \& Efremov(1997)]{1997ApJ...480..235E} Elmegreen, B.~G., 
\& Efremov, Y.~N.\ 1997, \apj, 480, 235 

\bibitem[Fall \& Rees(1977)]{1977MNRAS.181P..37F} Fall, S.~M., \& Rees, 
M.~J.\ 1977, \mnras, 181, 37P

\bibitem[Fall \& Zhang(2001)]{2001ApJ...561..751F} Fall, S.~M., \& Zhang, 
Q.\ 2001, \apj, 561, 751

\bibitem[Fall et al.(2005)]{2005ApJ...631L.133F} Fall, S.~M., Chandar, R., 
\& Whitmore, B.~C.\ 2005, \apjl, 631, L133 

\bibitem[Ferrarese et al.(2006a)]{2006ApJS..164..334F} Ferrarese, L., et 
al.\ 2006a, \apjs, 164, 334

\bibitem[Ferrarese et al.(2006b)]{2006ApJ...644L..21F} Ferrarese, L., et 
al.\ 2006b, \apjl, 644, L21

\bibitem[Forbes et al.(1996)]{1996AJ....112.2448F} Forbes, D.~A., Brodie, 
J.~P., \& Huchra, J.\ 1996, \aj, 112, 2448

\bibitem[Forbes(2005)]{2005ApJ...635L.137F} Forbes, D.~A.\ 2005, \apjl, 
635, L137

\bibitem[Ford et al.(1998)]{1998SPIE.3356..234F} Ford, H.~C., et al.\ 1998, 
\procspie, 3356, 234

\bibitem[Freedman et al.(2001)]{2001ApJ...553...47F} Freedman, W.~L., et 
al.\ 2001, \apj, 553, 47

\bibitem[Gao et al.(2005)]{2005MNRAS.363L..66G} Gao, L., Springel, V., \& 
White, S.~D.~M.\ 2005, \mnras, 363, L66 

\bibitem[Geha et al.(2003)]{2003AJ....126.1794G} Geha, M., Guhathakurta, 
P., \& van der Marel, R.~P.\ 2003, \aj, 126, 1794

\bibitem[Goudfrooij et al.(2003)]{2003MNRAS.343..665G} Goudfrooij, P., 
Strader, J., Brenneman, L., Kissler-Patig, M., Minniti, D., \& Edwin 
Huizinga, J.\ 2003, \mnras, 343, 665

\bibitem[G{\'o}mez \& Richtler(2004)]{2004A&A...415..499G} G{\'o}mez, M., 
\& Richtler, T.\ 2004, \aap, 415, 499

\bibitem[Ha{\c s}egan et al.(2005)]{2005ApJ...627..203H} Ha{\c s}egan, M., 
et al.\ 2005, \apj, 627, 203 (Paper VII)

\bibitem[Harris et al.(2004)]{2004AJ....128..723H} Harris, G.~L.~H., 
Harris, W.~E., \& Geisler, D.\ 2004, \aj, 128, 723 

\bibitem[Harris \& van den Bergh(1981)]{1981AJ.....86.1627H} Harris, W.~E., 
\& van den Bergh, S.\ 1981, \aj, 86, 1627

\bibitem[Harris(1991)]{1991ARA&A..29..543H} Harris, W.~E.\ 1991, \araa, 29, 
543

\bibitem[Harris \& Pudritz(1994)]{1994ApJ...429..177H} Harris, W.~E., \& 
Pudritz, R.~E.\ 1994, \apj, 429, 177

\bibitem[Harris et al.(1998)]{1998AJ....115.1801H} Harris, W.~E., Harris, 
G.~L.~H., \& McLaughlin, D.~E.\ 1998, \aj, 115, 1801

\bibitem[Harris \& Harris(2002)]{2002AJ....123.3108H} Harris, W.~E., \& 
Harris, G.~L.~H.\ 2002, \aj, 123, 3108 

\bibitem[Hoekstra et al.(2005)]{2005ApJ...635...73H} Hoekstra, H., Hsieh, 
B.~C., Yee, H.~K.~C., Lin, H., \& Gladders, M.~D.\ 2005, \apj, 635, 73

\bibitem[Jarrett et al.(2003)]{2003AJ....125..525J} Jarrett, T.~H., 
Chester, T., Cutri, R., Schneider, S.~E., \& Huchra, J.~P.\ 2003, \aj, 125, 
525

\bibitem[Jord{\'a}n et al.(2003)]{2003AJ....125.1642J} Jord{\'a}n, A., 
West, M.~J., C{\^o}t{\'e}, P., \& Marzke, R.~O.\ 2003, \aj, 125, 1642 

\bibitem[Jord{\' a}n et al.(2004a)]{2004AJ....127...24J} Jord{\' a}n, A., 
C{\^ o}t{\' e}, P., West, M.~J., Marzke, R.~O., Minniti, D., \& Rejkuba, 
M.\ 2004a, \aj, 127, 24

\bibitem[Jord{\' a}n et al.(2004b)]{2004ApJS..154..509J} Jord{\' a}n, A., et 
al.\ 2004a, \apjs, 154, 509 (Paper II)

\bibitem[Jord{\' a}n et al.(2004c)]{2004ApJ...613..279J} Jord{\' a}n, A., et 
al.\ 2004b, \apj, 613, 279 (Paper III)

\bibitem[Jord{\'a}n et al.(2005)]{2005ApJ...634.1002J} Jord{\'a}n, A., et 
al.\ 2005, \apj, 634, 1002 (Paper X)

\bibitem[Jord{\'a}n et al.(2006)]{2006ApJ...651L..25J} Jord{\'a}n, A., et 
al.\ 2006, \apjl, 651, L25

\bibitem[Jord{\'a}n et al.(2007)]{2007ApJS..171..101J} Jord{\'a}n, A., et 
al.\ 2007, \apjs, 171, 101 (Paper XII)

\bibitem[Kauffmann et al.(2003)]{2003MNRAS.341...54K} Kauffmann, G., et
al.\ 2003, \mnras, 341, 54

\bibitem[Kenney et al.(2004)]{2004AJ....127.3361K} Kenney, J.~D.~P., van 
Gorkom, J.~H., \& Vollmer, B.\ 2004, \aj, 127, 3361 

\bibitem[King(1966)]{1966AJ.....71...64K} King, I.~R.\ 1966, \aj, 71, 64

\bibitem[Kodama et al.(1998)]{1998A&A...334...99K} Kodama, T., Arimoto, N., 
Barger, A.~J., \& Arag'on-Salamanca, A.\ 1998, \aap, 334, 99

\bibitem[Koekemoer et al.(2002)]{2002hstc.conf..339K} Koekemoer, A.~M., 
Fruchter, A.~S., Hook, R.~N., \& Hack, W.\ 2002, The 2002 HST Calibration 
Workshop : Hubble after the Installation of the ACS and the NICMOS Cooling 
System, Proceedings of a Workshop held at the Space Telescope Science 
Institute, Baltimore, Maryland, October 17 and 18, 2002.~ Edited by 
Santiago Arribas, Anton Koekemoer, and Brad Whitmore.~Baltimore, MD: Space 
Telescope Science Institute, 2002., p.339, 339

\bibitem[Kundu \& Whitmore(2001a)]{2001AJ....121.2950K} Kundu, A., \& 
Whitmore, B.~C.\ 2001a, \aj, 121, 2950 

\bibitem[Kundu \& Whitmore(2001b)]{2001AJ....122.1251K} Kundu, A., \& 
Whitmore, B.~C.\ 2001b, \aj, 122, 1251 

\bibitem[Kravtsov \& Gnedin(2005)]{2005ApJ...623..650K} Kravtsov, A.~V., \& 
Gnedin, O.~Y.\ 2005, \apj, 623, 650

\bibitem[Kronawitter et al.(2000)]{2000A&AS..144...53K} Kronawitter, A., 
Saglia, R.~P., Gerhard, O., \& Bender, R.\ 2000, \aaps, 144, 53 

\bibitem[Larsen \& Richtler(2000)]{2000A&A...354..836L} Larsen, S.~S., \& 
Richtler, T.\ 2000, \aap, 354, 836

\bibitem[Larsen et al.(2001)]{2001AJ....121.2974L} Larsen, S.~S., Brodie, 
J.~P., Huchra, J.~P., Forbes, D.~A., \& Grillmair, C.~J.\ 2001, \aj, 121, 
2974

\bibitem[Lotz et al.(2004)]{2004ApJ...613..262L} Lotz, J.~M., Miller, 
B.~W., \& Ferguson, H.~C.\ 2004, \apj, 613, 262

\bibitem[Mandelbaum et al.(2006)]{2006MNRAS.368..715M} Mandelbaum, R., 
Seljak, U., Kauffmann, G., Hirata, C.~M., \& Brinkmann, J.\ 2006, \mnras, 
368, 715

\bibitem[Mayer et al.(2007)]{2007Natur.445..738M} Mayer, L., Kazantzidis, 
S., Mastropietro, C., \& Wadsley, J.\ 2007, \nat, 445, 738 

\bibitem[McLaughlin(1999a)]{1999AJ....117.2398M} McLaughlin, D.~E.\ 1999a,
\aj, 117, 2398

\bibitem[McLaughlin(1999b)]{1999ApJ...512L...9M} McLaughlin, D.~E.\ 1999b, 
\apjl, 512, L9

\bibitem[Mei et al.(2005a)]{2005ApJS..156..113M} Mei, S., et al.\ 2005a, 
\apjs, 156, 113

\bibitem[Mei et al.(2005b)]{2005ApJ...625..121M} Mei, S., et al.\ 2005b, 
\apj, 625, 121 (Paper V)

\bibitem[Mei et al.(2007)]{2007ApJ...655..144M} Mei, S., et al.\ 2007, 
\apj, 655, 144

\bibitem[Mieske et al.(2006)]{2006ApJ...653..193M} Mieske, S., et al.\ 
2006, \apj, 653, 193 (Paper XIV)

\bibitem[Miller et al.(1998)]{1998ApJ...508L.133M} Miller, B.~W., Lotz, 
J.~M., Ferguson, H.~C., Stiavelli, M., \& Whitmore, B.~C.\ 1998, \apjl, 
508, L133

\bibitem[Miller \& Lotz(2007)]{2007ApJ...670.1074M} Miller, B.~W., \& Lotz, J.~M.\ 2007, \apj, 670, 1074 

\bibitem[Mo et al.(1998)]{1998MNRAS.295..319M} Mo, H.~J., Mao, S., \& 
White, S.~D.~M.\ 1998, \mnras, 295, 319 

\bibitem[Moore et al.(1998)]{1998ApJ...495..139M} Moore, B., Lake, G., \& 
Katz, N.\ 1998, \apj, 495, 139 

\bibitem[Moore et al.(2006)]{2006MNRAS.368..563M} Moore, B., Diemand, J., 
Madau, P., Zemp, M., \& Stadel, J.\ 2006, \mnras, 368, 563

\bibitem[Napolitano et al.(2005)]{2005MNRAS.357..691N} Napolitano, N.~R., 
et al.\ 2005, \mnras, 357, 691

\bibitem[Peng et al.(2004)]{2004ApJ...602..685P} Peng, E.~W., Ford, H.~C., 
\& Freeman, K.~C.\ 2004, \apj, 602, 685

\bibitem[Peng et al.(2006a)]{2006ApJ...639...95P} Peng, E.~W., et al.\ 2006a, 
\apj, 639, 95 (Paper IX)

\bibitem[Peng et al.(2006b)]{2006ApJ...639..838P} Peng, E.~W., et al.\ 2006b, 
\apj, 639, 838 (Paper XI)

\bibitem[Puzia et al.(2006)]{2006ApJ...648..383P} Puzia, T.~H., 
Kissler-Patig, M., \& Goudfrooij, P.\ 2006, \apj, 648, 383 

\bibitem[Puzia \& Sharina(2007)]{2007arXiv0710.1550P} Puzia, T.~H., \& Sharina,
M.~E.\ 2007, ArXiv e-prints, 710, arXiv:0710.1550

\bibitem[Rhode \& Zepf(2001)]{2001AJ....121..210R} Rhode, K.~L., \& Zepf, 
S.~E.\ 2001, \aj, 121, 210

\bibitem[Rhode \& Zepf(2004)]{2004AJ....127..302R} Rhode, K.~L., \& Zepf,
S.~E.\ 2004, \aj, 127, 302

\bibitem[Rhode et al.(2005)]{2005ApJ...630L..21R} Rhode, K.~L., Zepf, 
S.~E., \& Santos, M.~R.\ 2005, \apjl, 630, L21

\bibitem[Ricotti(2004)]{2004ASPC..322..509R} Ricotti, M.\ 2004, The 
Formation and Evolution of Massive Young Star Clusters, 322, 509

\bibitem[Romanowsky et al.(2003)]{2003Sci...301.1696R} Romanowsky, A.~J., 
Douglas, N.~G., Arnaboldi, M., Kuijken, K., Merrifield, M.~R., Napolitano, 
N.~R., Capaccioli, M., \& Freeman, K.~C.\ 2003, Science, 301, 1696 

\bibitem[Salpeter(1955)]{1955ApJ...121..161S} Salpeter, E.~E.\ 1955, \apj, 
  121, 161

\bibitem[Guzik \& Seljak(2002)]{2002MNRAS.335..311G} Guzik, J., \& Seljak, 
U.\ 2002, \mnras, 335, 311

\bibitem[Santos(2003)]{2003egcs.conf..348S} Santos, M.~R.\ 2003, 
Extragalactic Globular Cluster Systems, 348

\bibitem[Schlegel et al.(1998)]{1998ApJ...500..525S} Schlegel, D.~J., 
Finkbeiner, D.~P., \& Davis, M.\ 1998, \apj, 500, 525

\bibitem[Searle \& Zinn(1978)]{1978ApJ...225..357S} Searle, L., \& Zinn, 
R.\ 1978, \apj, 225, 357

\bibitem[Sersic(1968)]{1968adga.book.....S} Sersic, J.~L.\ 1968, Cordoba, 
Argentina: Observatorio Astronomico, 1968

\bibitem[Seth et al.(2004)]{2004AJ....127..798S} Seth, A., Olsen, K., 
Miller, B., Lotz, J., \& Telford, R.\ 2004, \aj, 127, 798

\bibitem[Sirianni et al.(2005)]{2005PASP...Sirianni} Sirianni, M.,
  Jee, M.J., Benítez, N., Blakeslee, J.P., Martel, A.R., Meurer, G.,
  Clampin, M., De Marchi, G., Ford, H.C., Gilliland, R., Hartig, G.F.,
  Illingworth, G.D., Mack, J., \& McCann, W.J. 2005, PASP, accepted

\bibitem[Sivakoff et al.(2007)]{2007ApJ...660.1246S} Sivakoff, G.~R., et 
al.\ 2007, \apj, 660, 1246

\bibitem[Skrutskie et al.(2006)]{2006AJ....131.1163S} Skrutskie, M.~F., et 
al.\ 2006, \aj, 131, 1163

\bibitem[Spergel et al.(2007)]{2007ApJS..170..377S} Spergel, D.~N., et al.\ 
2007, \apjs, 170, 377

\bibitem[Spitler et al.(2007)]{2007arXiv0712.1382S} Spitler, L.~R., Forbes, 
D.~A., Strader, J., Brodie, J.~P., \& Gallagher, J.~S., III 2007, ArXiv 
e-prints, 712, arXiv:0712.1382

\bibitem[Springel et al.(2005)]{2005Natur.435..629S} Springel, V., et al.\ 
2005, \nat, 435, 629 

\bibitem[Strader et al.(2006)]{2006AJ....132.2333S} Strader, J., Brodie, 
J.~P., Spitler, L., \& Beasley, M.~A.\ 2006, \aj, 132, 2333

\bibitem[Thomas et al.(2005)]{2005ApJ...621..673T} Thomas, D., Maraston, 
C., Bender, R., \& Mendes de Oliveira, C.\ 2005, \apj, 621, 673 

\bibitem[Tonry et al.(2001)]{2001ApJ...546..681T} Tonry, J.~L., Dressler, 
A., Blakeslee, J.~P., Ajhar, E.~A., Fletcher, A.~B., Luppino, G.~A., 
Metzger, M.~R., \& Moore, C.~B.\ 2001, \apj, 546, 681\

\bibitem[Tremonti et al.(2004)]{2004ApJ...613..898T} Tremonti, C.~A., et 
al.\ 2004, \apj, 613, 898

\bibitem[Vale \& Ostriker(2007)]{2007astro.ph..1096V} Vale, A., \& 
Ostriker, J.~P.\ 2007, ArXiv Astrophysics e-prints,
arXiv:astro-ph/0701096

\bibitem[van den Bergh(1975)]{1975ARA&A..13..217V} van den Bergh, S.\ 1975, 
\araa, 13, 217

\bibitem[van den Bosch et al.(2003)]{2003MNRAS.340..771V} van den Bosch, 
F.~C., Yang, X., \& Mo, H.~J.\ 2003, \mnras, 340, 771

\bibitem[van Dokkum(2001)]{2001PASP..113.1420V} van Dokkum, P.~G.\ 2001, 
\pasp, 113, 1420

\bibitem[Vesperini(1998)]{1998MNRAS.299.1019V} Vesperini, E.\ 1998, \mnras, 
299, 1019

\bibitem[West(1993)]{1993MNRAS.265..755W} West, M.~J.\ 1993, \mnras, 265, 
755

\bibitem[West et al.(1995)]{1995ApJ...453L..77W} West, M.~J., Cote, P., 
Jones, C., Forman, W., \& Marzke, R.~O.\ 1995, \apjl, 453, L77

\bibitem[West et al.(2004)]{2004Natur.427...31W} West, M.~J., C{\^ o}t{\' 
e}, P., Marzke, R.~O., \& Jord{\' a}n, A.\ 2004, \nat, 427, 31

\bibitem[West et al.(2007)]{2007AJ..} West, A.~A., et al.\ 2007, AJ, submitted

\bibitem[Whitmore et al.(2007)]{2007AJ....133.1067W} Whitmore, B.~C., 
Chandar, R., \& Fall, S.~M.\ 2007, \aj, 133, 1067 

\bibitem[Williams et al.(2007)]{2007ApJ...654..835W} Williams, B.~F., et 
al.\ 2007, \apj, 654, 835

\bibitem[Yoon et al.(2006)]{2006Sci...311.1129Y} Yoon, S.-J., Yi, S.~K., \& 
Lee, Y.-W.\ 2006, Science, 311, 1129 

\bibitem[Zepf et al.(1995)]{1995ApJ...443..570Z} Zepf, S.~E., Ashman, 
K.~M., \& Geisler, D.\ 1995, \apj, 443, 570

\end{thebibliography}
\end{document}